\documentclass[useAMS,usenatbib,usegraphicx]{mn2e}

\title[Massive young stellar objects in G333.2$-$0.4]{A {\it Spitzer Space Telescope} survey of massive young stellar objects in the G333.2$-$0.4 giant molecular cloud}
\author[J. P. Simpson et al.]
{Janet P. Simpson$^{1}$\thanks{E-mail: jsimpson@seti.org},
Angela S. Cotera$^{1}$, Michael G. Burton$^{2}$, 
\newauthor 
Maria R. Cunningham$^{2}$, Nadia Lo$^{3}$, and Indra Bains$^{4}$\\
$^{1}$SETI Institute, 189 Bernardo Ave, Mountain View, CA 94043, USA\\
$^{2}$School of Physics, University of New South Wales, Sydney, NSW 2052, Australia\\
$^{3}$Departamento de Astronom\'ia, Universidad de Chile, Camino el Observatorio 1515, Las Condes, Santiago, Chile\\
$^{4}$Centre for Astrophysics and Supercomputing, Swinburne University of Technology, PO Box 218, Hawthorn, VIC 3122, Australia}
\begin{document}
\date{}
\pagerange{\pageref{firstpage}--\pageref{lastpage}} \pubyear{2011}

\maketitle

\label{firstpage}

\begin{abstract}

The G333 giant molecular cloud contains a few star clusters and \mbox{H\,{\sc ii}} regions,
plus a number of condensations currently forming stars.
We have mapped 13 of these sources with the appearance of young stellar objects (YSOs) 
with the {\it Spitzer} Infrared Spectrograph
in the Short-Low, Short-High, and Long-High modules (5-36~\micron).
We use these spectra plus available photometry and images to characterize the YSOs.
The spectral energy distributions (SEDs) of all sources peak between 35 and 110~\micron,
thereby showing their young age.
The objects are divided into two groups:
YSOs associated with extended emission in Infrared Array Camera (IRAC) band 2 at 4.5~\micron\ (`outflow sources')
and YSOs that have extended emission in all IRAC bands peaking at the longest wavelengths
(`red sources').
The two groups of objects have distinctly different spectra:
All the YSOs associated with outflows show evidence of massive envelopes surrounding
the protostar because the spectra show deep silicate absorption features
and absorption by ices at 6.0, 6.8, and 15.2~\micron.
We identify these YSOs with massive envelopes cool enough to contain ice-coated grains 
as the `bloated' protostars in the models of Hosokawa et al.
All spectral maps show ionized forbidden lines and
polycyclic aromatic hydrocarbon emission features.
For four of the red sources, these lines are concentrated
to the centres of the maps, from which we infer that these YSOs are
the source of ionizing photons.
Both types of objects show evidence of shocks, with most of the 
outflow sources showing a line of neutral sulphur in the outflows and 
two of the red sources showing the more highly excited \mbox{[Ne\,{\sc iii}]} and \mbox{[S\,{\sc iv}]} lines 
in outflow regions at some distance from the YSOs.
The 4.5-\micron\ emission seen in the IRAC band 2 images of the outflow sources 
is not due to H$_2$ lines,
which are too faint in the 5 -- 10~\micron\ wavelength region 
to be as strong as is needed to account for the IRAC band 2 emission.
\end{abstract}


\section{Introduction}

Massive stars form from gas and dust condensations in molecular clouds. 
Although there have been significant advances in our understanding of 
massive star formation, outstanding questions remain.
No one at this time is able to compute the evolution of a star 
from molecular cloud clump all the way to the main sequence 
because of the huge range of scales involved. 
Numerical models of molecular clouds collapsing under the force of gravity 
assume that when enough mass gets into the smallest grid cube or cubes, 
a star is formed.
These models, with fragmentation according to the local Jeans mass, 
produce both small and massive stars in agreement with observations 
(e.g., Bonnell \& Bate 2006; Peters et al. 2010; Wang et al. 2010).

A serious problem in understanding the formation of massive stars 
had long been that the radiation pressure at the surface 
of any star greater than about 8 M$_\odot$ 
should prevent the further addition of any mass, and yet 
stars an order of magnitude more massive are known.
However, it has now been shown through  
three-dimensional models of the collapse of a rotating core 
that the actual accretion can occur through an optically-thick accretion disc 
and the radiation escapes along the axes (e.g., Krumholz et al. 2009),
thus solving the problem of the excessive radiation pressure 
keeping the infalling material off the star.
Computations of low-mass stars (Vorobyov \& Basu 2005, 2006, 2010)
show that the disc fragments due to gravitational instabilities 
and the fragments lose angular momentum through viscosity 
and traverse through the disc until very close to the star, 
at which time it is assumed they are accreted.
Although such episodes of increased mass accretion are observed for low-mass stars 
(e.g., the FU Orionis stars, Hartmann \& Kenyon 2006),
such episodic accretion has not been observed for massive stars, 
probably because most mass is accreted while the star is still completely 
obscured optically and it is difficult to distinguish the star itself 
at mid-infrared (MIR) and far-infrared (FIR) wavelengths from the warm dust of the natal cloud.
Other computations indicate that such fragmentation and resulting accretion 
are also likely for high-mass stars (Kratter, Matzner, \& Krumholz 2008).
On the other hand, the mechanism of accretion onto the surface of the star 
from the disc is very complex and is still not understood 
(see the review of McKee \& Ostriker 2007).

The internal structure and evolution of an accreting massive protostar 
has been modelled by Hosokawa, Yorke, \& Omukai (2010), who find that 
when a model that is accreting at a rate of $10^{-3}$ M$_\odot$ yr$^{-1}$ approaches 
$\sim 10$ M$_\odot$, it `bloats up' to a radius $\sim 100$ R$_\odot$.
They predict that protostars at that stage are unable to ionize any 
\mbox{H\,{\sc ii}} region, in spite of their high luminosities,
and that this stage should be observable as a high-luminosity 
young stellar object (YSO) with no \mbox{H\,{\sc ii}} region. 

A molecular cloud collapsing into stars is by its nature turbulent.
It has been suggested that supersonic turbulence is crucial
in both preventing precipitous collapse on large scales and promoting
the density enhancements needed at smaller scales to actually form stars
(e.g., Zinnecker \& Yorke 2007; McKee \& Ostriker 2007).
Turbulence is induced into molecular clouds both 
on the large scale by rotation through a Galactic spiral arm 
and on the small scale by outflows from newly formed stars (e.g., Wang et al. 2010;
but see Padoan et al. 2009).
Given enough time, supernovae are major contributors. 

The giant molecular cloud (GMC) complex centred at $l=333.2$, $b=-0.4$ 
(the G333 cloud)
is ideal for testing theories of star formation.
We have been studying the effect of turbulence on giant molecular cloud structure
and on star formation through molecular line mapping of the G333 cloud 
(Bains et al. 2006; Wong et al. 2008; Lo et al. 2009).
This GMC contains the very luminous \mbox{H\,{\sc ii}} region and cluster G333.6$-$0.2,
plus a number of other, smaller star clusters.
The major star clusters, visible optically or in the near-infrared (NIR),
all lie along the major axis of the cloud. 
The FIR (Karnik et al. 2001) and mm (Mookerjea et al. 2004) maps
and 6.7 GHz Class II methanol maser surveys (Walsh et al. 1998)
also locate regions currently forming stars in the G333 GMC.

This paper discusses twelve additional massive star forming regions as observed 
with the {\it Spitzer Space Telescope} (Werner et al. 2004).
These are seen by the presence of very red, extended sources 
in images taken for the {\it Spitzer} Legacy programmes GLIMPSE (Churchwell et al. 2009) 
and MIPSGAL (Carey et al. 2009) with 
the {\it Spitzer} Infrared Array Camera (IRAC, Fazio et al. 2004)
and the Multiband Imaging Photometer for {\it Spitzer} (MIPS, Rieke et al. 2004), 
respectively.
Although the usual expectation is that star formation occurs in the centres
of GMCs where the density is highest, most of the star-forming regions
newly detected in the MIR {\it Spitzer} images are isolated and lie in
the outer regions of the cloud.

Our observed sources were chosen from two groups: 
(1) outflow sources as identified by either the presence of diffuse emission 
in the IRAC 4.5~\micron\ band 
(these have been described as `extended green objects', hereinafter EGOs, by Cyganowski et al. 2008, or
`green fuzzies' by Chambers et al. 2009)
or the presence of 7 mm SiO emission (Lo et al. 2007), 
both thought to arise in shocks;
and (2) sources with diffuse red emission in the IRAC 8.0-\micron\ images.
All objects are compact and bright in the MIPS 24-\micron\ images.
All but one of the outflow sources also produce 6.7-GHz Class II methanol masers. 
Fig.~1 shows these objects plotted on the IRAC 8.0-\micron\ image 
of the G333 cloud (all images in this paper are log scale). 
Their coordinates and estimated distances (all kinematic assuming 
the solar distance to the Galactic Center, $R_0 = 8.3$ kpc, from Brunthaler et al. 2011 
and the Galactic rotation curve of Levine, Heiles, \& Blitz 2008) are given in Table~1 
(note that a few sources lie outside the G333 cloud and appear to be associated 
with a more distant spiral arm).  
In this paper we use the word `sources' to refer to 
the whole extended emission region that we mapped, 
`YSOs' as the approximately point source including the envelope and disc 
that is the exciting source of the outflow and the MIR and FIR emission, 
and `protostar' that is the core object that is accreting mass 
and will eventually become a star.

Our goal for this {\it Spitzer} survey is to characterize these YSOs according 
to masses, outflow parameters, and ages. 
We also hope to illuminate observationally some of the theoretical studies 
mentioned in the previous paragraphs. 
In Section 2 we describe the observations, 
and in Section 3 we describe the results.
The spectra of the sources also fall into two distinct groups: 
objects with outflows and ice-absorption features at the central YSO, 
and objects with localized forbidden lines indicating the 
presence of sources of ionizing photons. 
In Section 3 we also discuss individual objects and the frequent multiple sources 
in a single map,
and in Section 4 we discuss the possible evolutionary stages
as indicated by the presence or absence of various spectral features.
Finally, in Section 5 we present our summary and conclusions.

\section{Observations}

%Figure 1
%\clearpage
\begin{figure*}
\includegraphics[width=165mm]{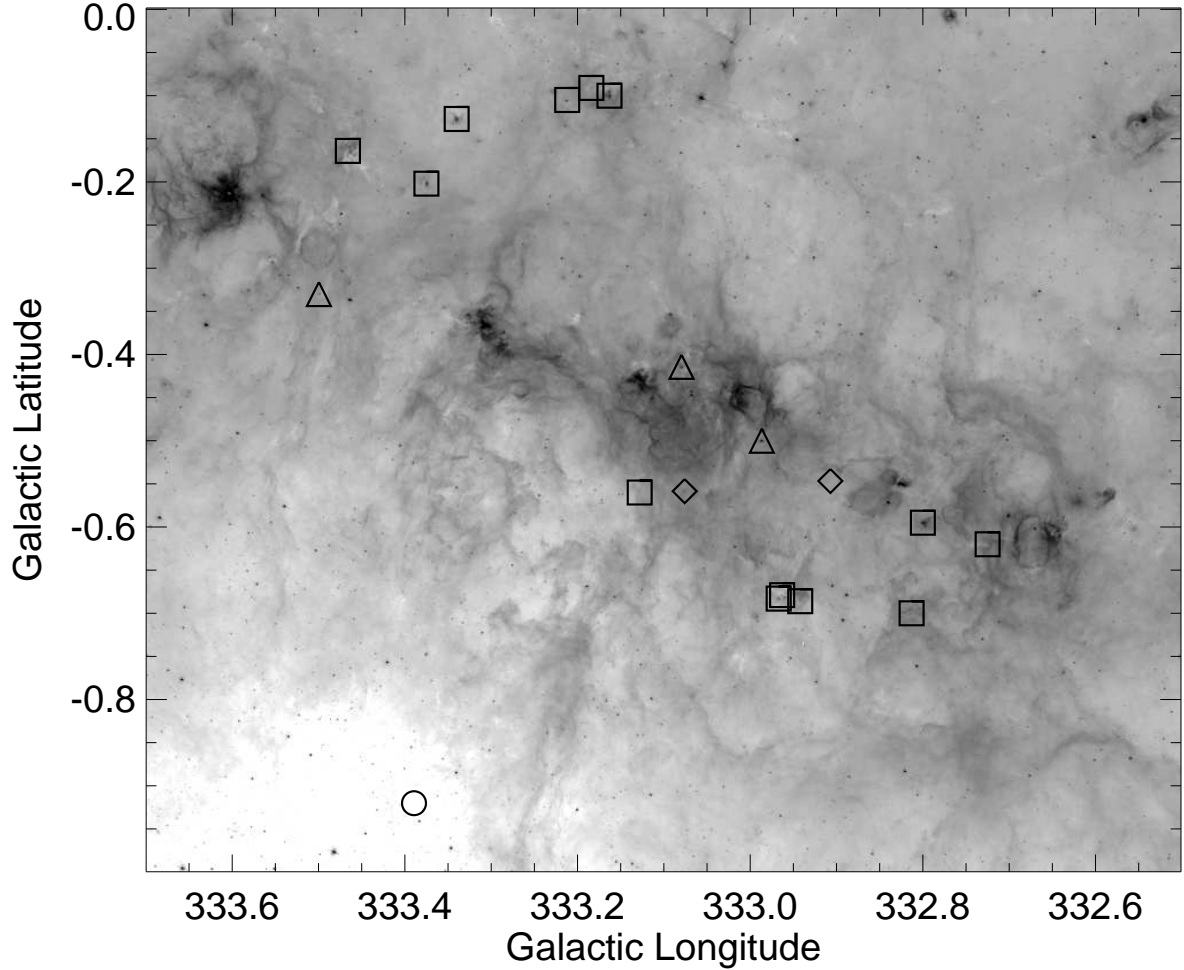}
\caption{IRAC band 4 (8.0~\micron) image of the G333 cloud in logarithmic scaling.
The sources in this survey are marked by square boxes.
The open circle marks our background position.
The outflow sources from the catalog of Cyganowski et al. (2008) that were not 
observed in our survey are marked by diamonds. 
Additional red sources that were not observed in our survey are marked by triangles.
}
\end{figure*}

\setcounter{table}{0}
\begin{table*}
\centering
\begin{minipage}{155mm}
\caption{Observational parameters.
The table gives for each source the RA and Dec., the date of the observation 
(end of cryogen was 2009 May 15), the AOR identification number, 
the source type (red or outflow, see text), name of any co-located source in the {\it IRAS} Point Source Catalog,
the kinematic distance estimated from an observed $V_{\rm LSR}$, and references:  
(a) All sources in the G333 cloud are assumed to have the cloud $V_{\rm LSR} = -50$ km s$^{-1}$ 
measured by Bains et al. (2006).
(b) The three sources north of the G333 cloud are assumed to have the $V_{\rm LSR} = -91$ km s$^{-1}$ 
measured for G333.168-0.081 by Caswell \& Haynes (1987).
The masers observed in all these sources have velocities consistent with these assumptions 
(Caswell 2009; Urquhart et al. 2009).
 }
\begin{tabular}{@{}lcccclccc@{}}
\hline
Source   &           RA     &     Dec  & Obs. Date &  AORKEY   &    Source Type & {\it IRAS} name & Distance & Ref. \\
       &         (J2000)  &   (J2000) &           &        &                &           &    (kpc)   &  \\

\hline
G332.725$-$0.620 & 16:20:02.8 &  $-$51:00:32 &  2009-04-30 & 25913088 &   Outflow    &             &  3.6 & a \\
G332.800$-$0.595 & 16:20:16.4 &  $-$50:56:19 &  2009-05-03 & 25915392 &   Red        &             &  3.6 & a \\
G332.813$-$0.700 & 16:20:48.1 &  $-$51:00:15 &  2009-04-30 & 25912832 &   Outflow   &   16170-5053 &  3.6 & a \\
G332.942$-$0.686 & 16:21:18.9 &  $-$50:54:10 &  2009-04-30 & 25912576 &   Outflow   &   16175-5046 &  3.6 & a \\
G332.963$-$0.679 & 16:21:22.9 &  $-$50:52:59 &  2009-04-30 & 25912064 &   Outflow   &              &  3.6 & a \\
G332.967$-$0.683 & 16:21:25.0 &  $-$50:53:01 &  2009-04-30 & 25916160 &   Red       &              &  3.6 & a \\
G333.131$-$0.560 & 16:21:36.1 &  $-$50:40:49 &  2009-05-03 & 25916416 &   Outflow   &              &  3.6 & a \\
G333.163$-$0.100 & 16:19:42.6 &  $-$50:19:53 &  2009-04-30 & 25915648 &   Red       &              &  5.8 & b \\
G333.184$-$0.091 & 16:19:45.6 &  $-$50:18:35 &  2009-05-03 & 25913344 &   Outflow   &              &  5.8 & b \\
G333.212$-$0.105 & 16:19:56.9 &  $-$50:18:01 &  2009-04-30 & 25915904 &   Red       &              &  5.8 & b \\
G333.340$-$0.127 & 16:20:36.9 &  $-$50:13:35 &  2009-05-03 & 25914624 &   Red       &   16168-5006  & 3.6 & a \\
G333.375$-$0.202 & 16:21:06.0 &  $-$50:15:15 &  2009-04-30 & 25915136 &   Red       &              &  3.6 & a \\
G333.466$-$0.164 & 16:21:20.2 &  $-$50:09:49 &  2009-05-07 & 25911552 &   Outflow  &    16175-5002 &  3.6 & a \\
\hline
\end{tabular}
\end{minipage}
\end{table*}

We mapped each of our sources with the Infrared Spectrograph (IRS, 
Houck et al. 2004) 
with the Short-Low (SL, 5.2 -- 14.5~\micron), Short-High (SH, 10 -- 19.5~\micron), 
and Long-High (LH, 19.5 -- 38~\micron) modules
on 2009 April 29 -- May 7 (a week before end of cryogen).
The observations were split into two groups according to date 
(2009 April 29--30 and May 3--7 May) because the IRS campaign was split by 
an IRAC mini-campaign.
Fig. 2 gives an example of the individual IRAC and MIPS images of a source 
with overlays of the IRS slit positions; 
the exposure identifier number (FITS keyword: EXPID) is given for the LH module.
Similar figures are given for the other sources in the online 
Supporting Information.
A background position was observed for each source; the background position  
%is at G333.3894$-$0.9202, RA 16$^{\rm h}$ 2$^{\rm m}$ 22\fs0 Dec $-$50\degr 45\arcmin 2\arcsec.
is at G333.3894$-$0.9202, or RA 16$^{\rmn{h}}~24^{\rmn{m}}~22\fs0$, Dec. $-50\degr~45\arcmin~2\arcsec$.

%Figure 2
%\clearpage
\begin{figure}
\includegraphics[width=84mm]{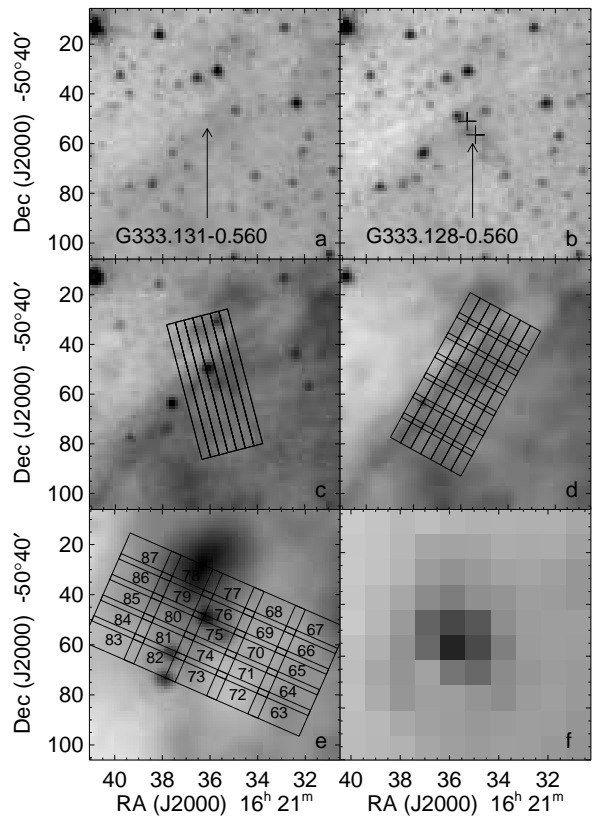}
\caption{IRAC and MIPS images of G333.131$-$0.560 and G333.128$-$0.560.
Panel (a): IRAC band 1 (3.6~\micron) image. 
Panel (b): IRAC band 2 (4.5~\micron) image. 
The two 6.7 GHz methanol masers (Caswell 2009) are plotted as black crosses.
Panel (c): IRAC band 3 (5.8~\micron) image with overlays of the SL slit positions.
Panel (d): IRAC band 4 (8.0~\micron) image with overlays of the SH slit positions.
Panel (e): MIPS 24~\micron\ image with overlays of the LH slit positions
on which is printed the LH exposure identifier numbers.
Panel (f): MIPS 70~\micron\ image.
G333.128$-$0.560 is the brightest source in the MIPS 70~\micron\ image (panel f) 
and is best located in the MIPS 24~\micron\ image (panel e).
}
\end{figure}

The data were reduced and calibrated with the S18.7 pipeline at the
Spitzer Science Center (SSC).
The basic calibrated data (BCD) from the {\it Spitzer} pipeline consist 
of 128 by 128 pixel arrays showing 10 spectral orders for the high resolution modules 
and 2 spectral orders for the low resolution modules.
For each telescope pointing and for each module, 
we took four spectra (resulting in four BCD images) such that 
the total integration time per pointing per module was 24 s.
These four BCD images of each detector array
for each telescope pointing
were median-combined and the median of all the background BCD images 
in the appropriate time group was subtracted. 
The final backgrounds have less noise than any of the source BCD images.
This process of subtracting fairly low-noise background BCD images 
greatly reduces the number 
of `rogue' pixels\footnote{http://irsa.ipac.caltech.edu/data/SPITZER/docs/irs/features/}
(pixels with temporary extra dark current or non-linear response as a result of 
hits by cosmic rays or solar protons). 
The resulting background-subtracted BCD images were further cleaned of rogue pixels
using the interactive version of 
{\sc irsclean}\footnote{http://irsa.ipac.caltech.edu/data/SPITZER/docs/dataanalysistools/tools/irsclean/}. 
The noisy order edges (the ends of the slits) were cleaned with our own software.

Two tools employing the Interactive Data Language ({\sc idl}) 
from the SSC were used to obtain spectra from the BCD images:
the Spectroscopy Modeling, Analysis and Reduction Tool 
({\sc smart}, Higdon et al. 2004) and 
the CUbe Builder for IRS Spectral Mapping ({\sc cubism}, Smith et al. 2007a).

\subsection{Spectra and line fluxes -- {\sevensize\bf SMART}}

{\sc smart} can be used to extract spectra from individual BCD images, or sections of BCD images 
for long slit spectra. 
We used {\sc smart} to extract spectra from the SH and LH BCD images.
Because the IRS was calibrated using point sources but all our sources 
are extended, a correction for the telescope diffraction pattern as a function
of wavelength is necessary; these `slitloss' correction factors were taken
from tables provided by the 
SSC\footnote{http://irsa.ipac.caltech.edu/data/SPITZER/docs/dataanalysistools/tools/spice/}.
{\sc smart} returns separate spectra for each of the ten SH or LH spectral orders.
Many of the spectra in certain orders, particularly LH, are contaminated 
by spectral fringing of order 3--6 per cent.
An advantage of the use of {\sc smart} to measure line fluxes is that it includes 
a tool ({\sc irsfringe}) to fit and subtract the fringes, order by order.
 
Line and continuum fluxes were estimated by fitting
Gaussian profiles plus a linear continuum to each of the line profiles
in the spectra using both {\sc idl}'s {\sc curvefit} procedure and the procedures in {\sc smart}.
For the faintest lines, the fluxes were estimated by integrating 
over the line profile. 
Errors for all fluxes were estimated
by computing the root-mean-squared deviation of the data from the fit;
because of the systematic effects of low-level, uncorrectable rogue pixels on the spectra,
these errors are consistently larger than errors estimated any other way,
such as the statistics of median averaging or the counting statistics plus read noise
from the SSC's pipeline.

The lines discussed in this paper are listed in Table~2,
along with references to the atomic or molecular physics parameters
(wavelengths, transition probabilities, collisional cross sections)
needed for further analysis.

%\documentclass[useAMS,usenatbib,usegraphicx]{mn2e}
%\begin{document}

\setcounter{table}{1}
\begin{table*}
\centering
\begin{minipage}{120mm}
\caption{Line parameters.
For each line the table gives the wavelength, 
the ionization potential ($IP$, the energy required to produce the ground state of the molecule or ion),
the energy of the upper level of the transition, the extinction law for that wavelength as
a fraction of the extinction law for 9.6 \micron, and relevant references.
The references in the last column are as follows:
(1) Bautista \& Pradhan (1996),
(2) Dufton \& Kingston (1991),
(3) Erickson et al. (1989),
(4) Feuchtgruber et al. (1997),
(5) Griffin, Mitnik, \& Badnell (2001),
(6) Kelly \& Lacy (1995),
(7) McLaughlin \& Bell (2000),
(8) Nussbaumer \& Storey (1988),
(9) Pelen \& Berrington (1995),
(10) Quinet (1996),
(11) Quinet, Le Dorneuf, \& Zeippen (1996),
(12) Ramsbottom et al. (2005),
(13) Saraph \& Storey (1999),
(14) Storey \& Hummer (1995),
(15) Tayal (2004),
(16) Tayal (2006),
(17) Tayal \& Gupta (1999),
(18) Wolniewicz, Simbotin, \& Dalgarno (1998),
(19) Zhang (1996).
}
\begin{tabular}{@{}lccccl@{}}
\hline
Line & Wavelength & $IP$ & $E_{upper}$ & $\tau_\lambda/\tau_{9.6}$ & References \\
     & (\micron)  & (eV) & (cm$^{-1}$) &                        &  \\
\hline
H$_2$ S(0) & 28.219  & 0 & 354.37 & 0.369 & 18 \\ %   Wolniewicz et al. (1998)
H$_2$ S(1) & 17.035  & 0 & 705.69 & 0.502 & 18 \\ %    ``
H$_2$ S(2) & 12.279  & 0 & 1168.80 & 0.344 & 18 \\ %   ``
H$_2$ S(3) & 9.665  & 0 & 1740.36 & 0.997 & 18 \\ %   ``
H$_2$ S(5) & 6.909  & 0 & 3187.64 & 0.273 & 18 \\ %   ``
H$_2$ S(7) & 5.512  & 0 & 5002.04 & 0.290 & 18 \\ %   ``
\mbox{H\,{\sc i}} 7-6    & 12.372  & 13.60 & 107440.45 & 0.332 & 14 \\ %   Storey \& Hummer 1995
\mbox{[O\,{\sc iv}]}  $^2$P$_{3/2}-^2$P$_{1/2}$   & 25.890 & 54.94 & 386.245  & 0.404 & 4, 16 \\ %   F97 T06
\mbox{[Ne\,{\sc ii}]}  $^2$P$_{1/2}-^2$P$_{3/2}$  & 12.814 & 21.56 & 780.42 & 0.313 & 5 \\ %   Griffin et al.
\mbox{[Ne\,{\sc iii}]} $^3$P$_1-^3$P$_2$       & 15.555 & 40.96 & 642.88 & 0.417 & 4, 7 \\ %  F97 McLauBell
\mbox{[Si\,{\sc ii}]} $^2$P$_{3/2}-^2$P$_{1/2}$  & 34.815 & 8.15 & 287.23 & 0.306 & 2, 4 \\ %  F97 DK91
\mbox{[S\,{\sc iii}]} $^3$P$_1-^3$P$_0$         & 33.481 & 23.34 & 298.68 & 0.311 & 4, 17 \\ %   F97 TG99
\mbox{[S\,{\sc iii}]} $^3$P$_2-^3$P$_1$         & 18.713 & 23.34 & 833.06 & 0.548 & 6, 17\\ %  KL95 TG99 
\mbox{[S\,{\sc iv}]}  $^2$P$_{3/2}-^2$P$_{1/2}$  & 10.511 & 34.79 & 951.43 & 0.777 & 13 \\ %   SS99
\mbox{[Ar\,{\sc ii}]}  $^2$P$_{1/2}-^2$P$_{3/2}$  & 6.985 & 15.76 & 1431.58 & 0.272 & 9 \\ % Pelen & Berrington 1995
%Ar III
\mbox{[Fe\,{\sc ii}]} $^6$D$_{7/2}-^6$D$_{9/2}$ & 25.988 & 7.90 & 384.79 & 0.403 & 1, 8, 11, 12 \\ %   NS88 Qetal96 BP96 R05
\mbox{[Fe\,{\sc iii}]} $^5$D$_3-^5$D$_4$       &  22.925 & 16.19 & 436.21 & 0.448  & 3, 10, 19 \\ %   Q96, Zhang 1996
\hline
\end{tabular}
\end{minipage}
\end{table*}

%\end{document}

\subsection{Spectra and line fluxes -- {\sevensize\bf CUBISM}}

{\sc cubism} (Smith et al. 2007a) combines all the spectra from a single or multiple Astronomical 
Observation Requests (AORs) for a given module if SH or LH or a given order 
if SL (SL order 1 covers 7.5 -- 14.2~\micron\ and order 2 covers 5.2 -- 7.5~\micron)
or LL (Long-Low, 14 -- 38~\micron) and 
produces a three-dimensional cube of $x$, $y$, and wavelength, 
where $x$ and $y$ refer to coordinates parallel and perpendicular to the module slit.
It was designed for spectral maps of galaxies 
from the SIRTF Nearby Galaxies Survey (SINGS, Kennicutt et al. 2003) 
and the calibration assumes that the source is extended.
The user selects an area of the sky in $x$ and $y$ to extract a spectrum.
Since the output spectrum header contains the coordinates on the sky of 
the corners of the $x$,$y$ rectangle, there is an option to extract a spectrum from 
a different module of exactly the same area of the sky.

We used {\sc cubism} to first extract the spectrum of every telescope pointing 
observed by the SH module (e.g., panel `d' of Fig.~2) 
and then extracted spectra from the LH and SL cubes of all the same positions 
on the sky (skipping those map corners that were not observed by all three modules).
To increase the signal/noise ratio (S/N), we also extracted spectra from regions 
that looked especially interesting, such as large areas seen to have \mbox{H\,{\sc ii}} region lines,
appropriate areas for background subtraction,  
and point sources where the observed grid did not exactly overlay the YSO.
Because the spectra covered exactly the same regions of the sky in all modules, 
they could be easily stitched together to make a complete 5.2 -- 35~\micron\ spectrum 
with very little flux adjustment at the wavelengths where the modules overlap 
(e.g., Smith et al. 2007b).

%Figure 3
%\clearpage
\begin{figure}
\includegraphics[width=84mm]{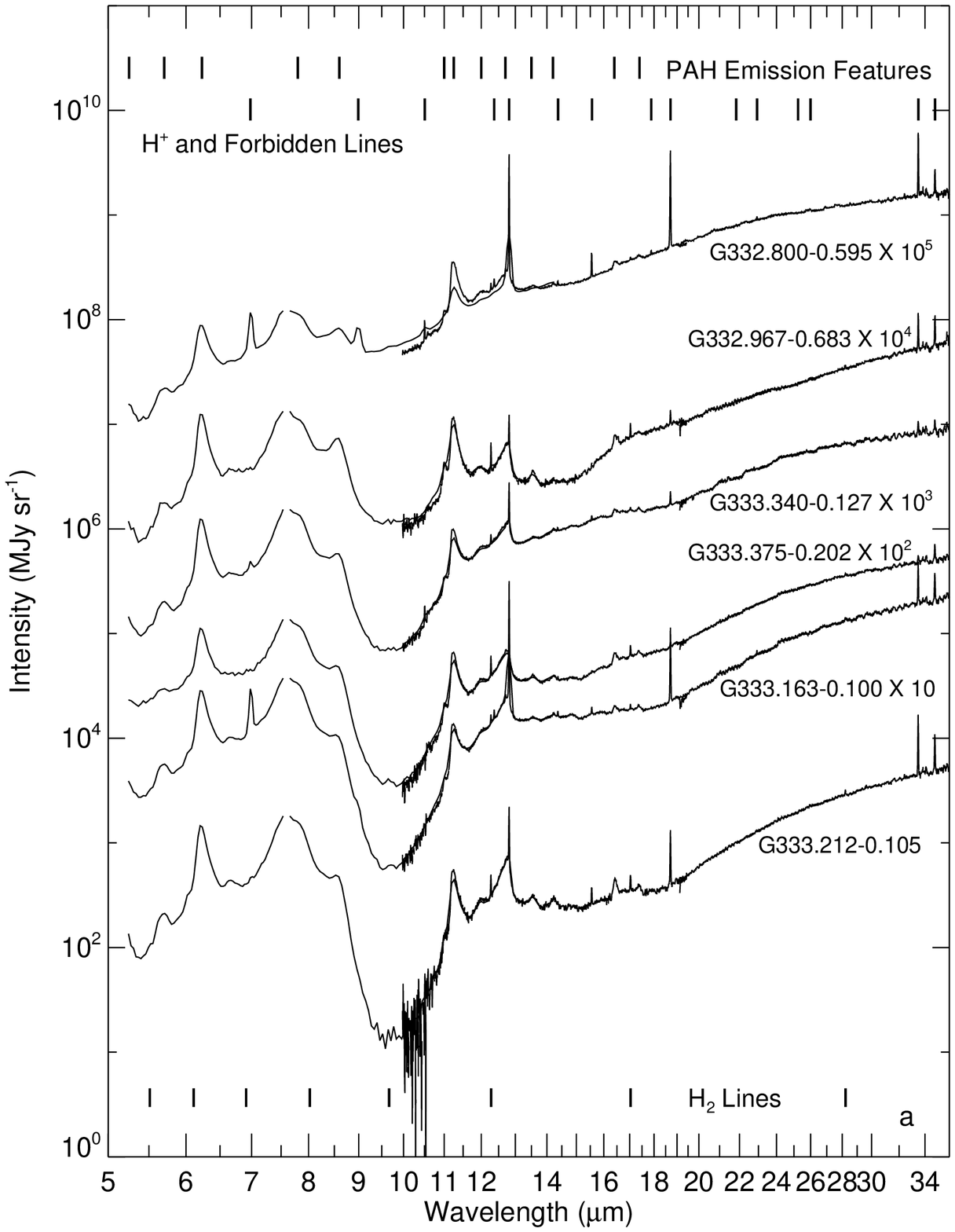}
\includegraphics[width=84mm]{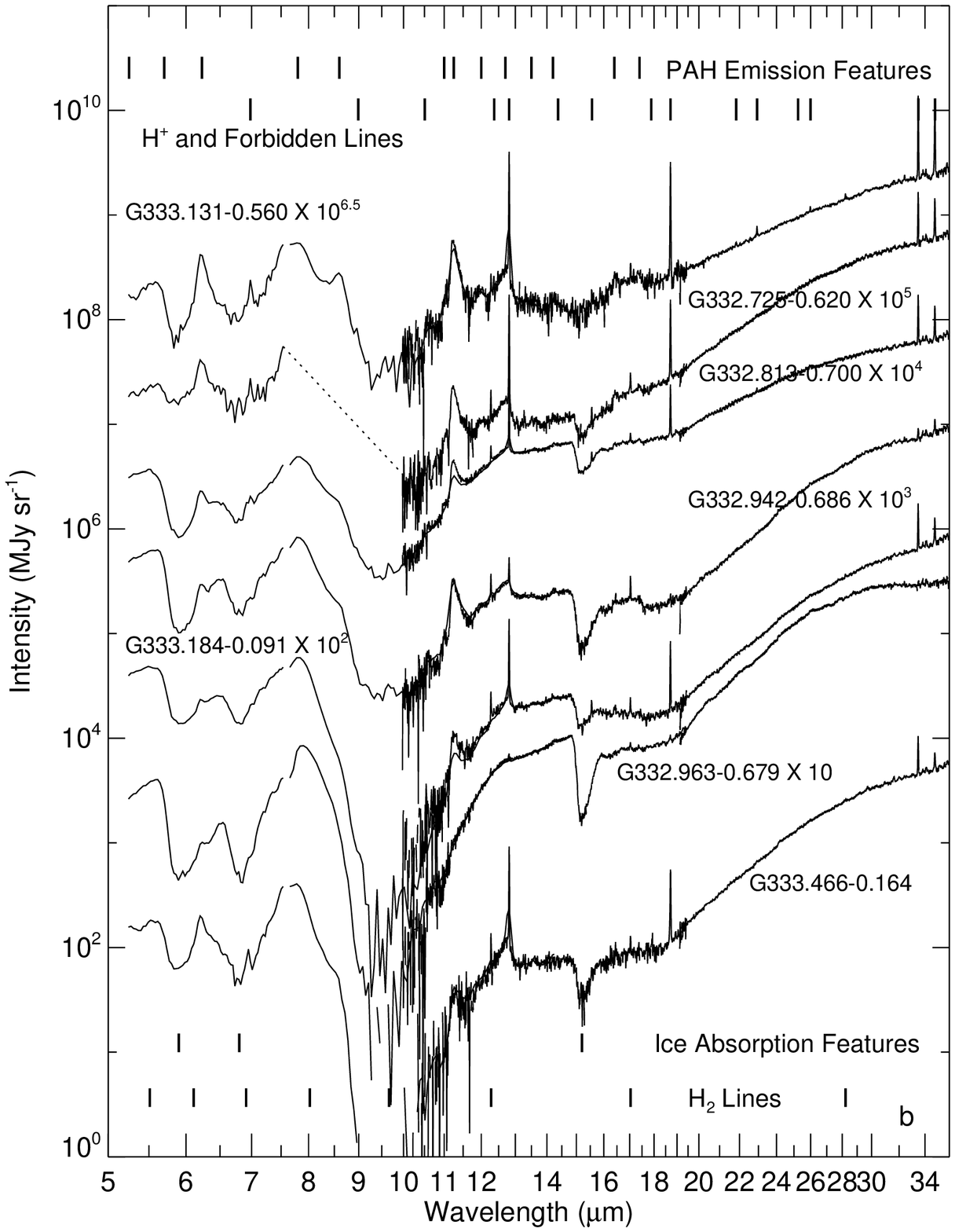}
\caption{
Panel (a): spectra of `red' sources with identification of features of interest.
Panel (b): spectra of `outflow' sources with identification of features of interest.
}
\end{figure}

The spectra of the YSOs are shown in Fig. 3.
These spectra will be discussed further in Section 3, but we wish to point out 
the following features:
All spectra are dominated by a warm continuum that increases steeply 
to longer wavelengths. 
The sharp lines are the \mbox{H\,{\sc ii}} region and photodissociation region (PDR) lines 
listed in Table 2; these are present in almost every spectrum,
including the `background' spectrum, and are probably due to the same 
diffuse interstellar medium (ISM) that is seen in the molecular cloud image of Fig. 1. 
The broad emission features are interstellar polycyclic aromatic 
hydrocarbon molecules (PAHs) at 6.2, 7.7, 8.6, 11.2, 12.0, and 17~\micron\ 
plus additional wavelengths as indicated in the figure;
they are also seen in almost all spectra, including the background.
The absorption features include silicates at 9.6 and 18~\micron\ 
and ices at 6.0 (H$_2$O and HCOOH), 6.8 (CH$_3$OH and NH$_4^+$) and 15.2~\micron\ (CO$_2$)
(e.g., Boogert et al. 2008).

The advantage of {\sc cubism} over {\sc smart} is that a complete spectrum can be produced 
for a given region of the sky 
and there are no `jumps' in the spectrum at the SH and LH order boundaries
(which jumps are due to {\sc smart}'s use of point sources for calibration). 
The advantages of {\sc smart} over {\sc cubism} is that 
it can be run as a script for all SH and LH spectra and 
the LH spectrum that was extracted by {\sc smart} can be defringed 
whereas the 19 -- 35~\micron\ LH spectrum that was extracted by {\sc cubism} can not.
We will give an example of the need for defringing in Section 3.6.

\section{Results}

We start with the basic results for all sources (extinction, luminosity) 
and then detail the results from the ice features and the forbidden line analysis.
We end with a brief discussion of each YSO.

\subsection{Extinction -- {\sevensize\bf PAHFIT}}  

The MIR extinction was determined by fitting the observed spectra 
with the program {\sc pahfit}, developed by Smith et al. (2007b) 
and available online.
We use the fitted 9.6~\micron\ optical depth and MIR extinction law 
to correct the measured line intensities, 
both hydrogen and forbidden lines, for extinction
and to determine the variations in extinction between the sources.

The extinction in the MIR ($> 8$~\micron) is due to absorption by silicates. 
This is most notable in the depth of the 9.6~\micron\ silicate feature in 
some of the spectra in Fig. 3, although in many of the spectra this feature 
is obscured by the gap in the PAH spectrum from 9 -- 11~\micron.
Consequently, the whole spectrum needs to be fitted at the same time: 
MIR continuum, PAHs, and the 9.6 and 18~\micron\ silicate absorption features.

Smith et al. (2007b) developed a program, called {\sc pahfit},  
to fit all these components by non-linear least-squares 
to the low-resolution SINGS galaxy spectra. 
This {\sc idl} program, with modifications, was used to estimate 
the MIR extinction affecting all our spectra. 
Although {\sc pahfit} could produce good fits for all our spectra 
that are most like \mbox{H\,{\sc ii}} region spectra (the red sources and the outskirts of all the maps,
e.g., Fig. 3a), as might be expected since the SINGS spiral galaxy  
spectra are dominated by \mbox{H\,{\sc ii}} regions, 
the spectra of the YSOs (e.g., Fig. 3b) 
that are dominated by absorption features often gave poor fits. 
{\sc pahfit} uses a Drude profile to fit each of the 24 PAH features in its database;
however, Drude profiles have very broad wings, and the least-squares fitting routine 
sometimes fitted the 5 -- 10~\micron\ continuum with exceptionally strong 
PAH features and no 5 -- 10~\micron\ continuum.
This was particularly likely for the sources with the 6.0 and 6.8~\micron\ ice features.
(A  non-physical result is something all users of least-squares fitting routines 
must always watch out for.)

Because most of the PAHs in our spectra arise in the diffuse ISM, 
we expect, in fact, that the PAH contribution should be similar 
for all the spectra (there can be substantial variations in the PAH spectrum 
between types of sources; see e.g., the review of Tielens 2008). 
Consequently, in order to improve the fit and to fit all sources consistently, 
we modified the {\sc pahfit} program to use a template for the PAH features 
instead of the Drude profiles for each individual feature.
(We kept the structure of the {\sc pahfit} program rather than write a new program
because of all its other useful features: 
silicate absorption, continuum fits, line fits, graphics, etc.)
Although it would be preferable to use a PAH template from the G333 region, 
none is available that does not include at least some extinction, since 
the G333 cloud is at a distance $\sim 3.6$ kpc (Table 1).
An unextincted PAH spectrum of high S/N that has minimal forbidden lines 
taken with {\it Spitzer}'s IRS SL module is needed for the template --
we found such a spectrum in {\it Spitzer} program 0045 (PI: T. Roellig) 
just south of the Orion Nebula bar at 5$^{\rm h}$ 35$^{\rm m}$ 21\fs3 $-$5\degr 25\arcmin 6\arcsec.
We extracted this spectrum with {\sc cubism} over a 72-arcsec$^2$ aperture 
from SL and SH AORs 4117760 and 4118016, stitched on a longer wavelength contribution 
from the {\it Midcourse Space Experiment (MSX)} spectrum of the whole Orion Nebula 
(Simpson et al. 1998), 
and estimated the continuum with {\sc pahfit}.
Fig. 4 shows the spectrum and fit; the PAH template consists of 
the data (square boxes) minus the fitted continuum (solid black line) 
over intervals between 5.2 and 17.6~\micron\ (see Smith et al. 2007b for a description of
the other fit components).
This fit found zero silicate absorption.

%Figure 4
%\clearpage
\begin{figure}
\includegraphics[width=84mm]{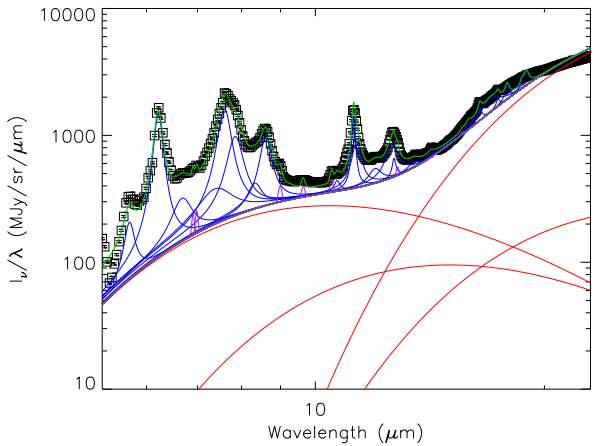}
\caption{Spectrum of the Orion Bar from  Spitzer program 0045 with results from {\sc pahfit} overplotted.
The data, marked by square boxes with error bars, consist of 
IRS SL spectra from 5.24 to 12.44~\micron, SH spectra smoothed to the SL resolution 
from 12.44 to 19.05~\micron, and spectra from the full Orion Nebula measured by {\it MSX} 
(Simpson et al. 1998) from 19.05 to 25.65~\micron.
Lines of \mbox{[Ne\,{\sc ii}]} 12.8~\micron, \mbox{[Cl\,{\sc ii}]} 14.4~\micron, \mbox{[Ne\,{\sc iii}]} 15.6~\micron, and 
\mbox{[S\,{\sc iii}]} 18.7 were excised from the spectrum before fitting.
The results from {\sc pahfit} are as follows: the fit is plotted in green, 
narrow lines are violet, 
PAH features are blue, blackbody continua are red, and the total continuum is plotted in black.
The PAH template consists of the Orion data minus the total continuum from 
5.24 - 8.85~\micron, 10.77 - 14.41~\micron, and 16.29 - 17.59~\micron;
from 8.85 -- 10.77~\micron\ and 14.41 -- 16.29~\micron\  
the template consists of the wings of the {\sc pahfit} PAH features,
which present a much less noisy continuum than the Orion data.
}
\end{figure}

%Figure 5
%\clearpage
\begin{figure}
\includegraphics[width=84mm]{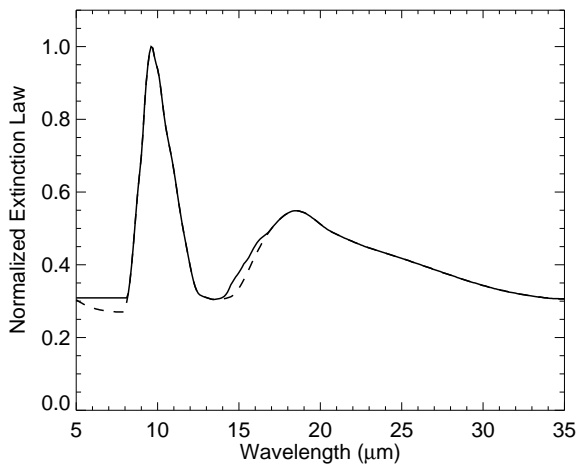}
\caption{Extinction law.
The solid line is the extinction law as was used for the G333 cloud;
it was taken originally from the Galactic Centre law of Chiar \& Tielens (2006)
but modified in the 13.6 -- 16.8~\micron\ region (see text) 
and at wavelengths short of 8~\micron\ where we use the extinction law of
Indebetouw et al. (2005).
The original law  of Chiar \& Tielens is plotted as the dashed line.
}
\end{figure}

For the MIR extinction law, we used the Galactic Centre law of Chiar \& Tielens (2006),
slightly modified to include a little more extinction at 15~\micron.
This is shown in Fig. 5, normalized to 1.0 at 9.6~\micron.
The reason for the modification is that otherwise, our very heavily extincted 
positions would have a little bump in the fitted spectrum at 15~\micron\ that is not observed.
The modification flattened the bump; however, a larger increase in the extinction 
might actually be the case but the exact shape of the extinction curve
at this wavelength would be difficult to determine because 
the objects with the largest extinction also exhibit the 15.2~\micron\ CO$_2$ ice feature. 
The Chiar \& Tielens (2006) Galactic Centre extinction has 
a larger 18/9.6~\micron\ ratio than the extinction laws of Draine (2003a,b); 
we prefer it because it gives a better fit to the H$_2$ S(0), S(1), and S(2) 
line ratios, which lines arise in the diffuse ISM along the line of sight 
to the Galactic Centre (Simpson et al. 2007).

The final modification was to add a component to the fitting procedure 
for the ice features.
To keep to the structure of the {\sc pahfit} program, the ice features 
were added as Gaussian absorption lines at 5.93, 6.80, and 15.30~\micron\ 
with full width at half maximum (FWHM) of 0.429, 0.843, and 0.602~\micron.
Gaussian line profiles are certainly not appropriate, but all that is needed 
for this modification is to make some accounting for the strength of the 
absorption relative to the continuum and the PAH emission, 
since the purpose of our using {\sc pahfit} is to estimate the extinction
and the strength of the PAHs. 
The feature central wavelengths and FWHM listed above 
were estimated using the line fitting procedure in {\sc smart}. 

%Figure 6
%\clearpage
\begin{figure}
\includegraphics[width=84mm]{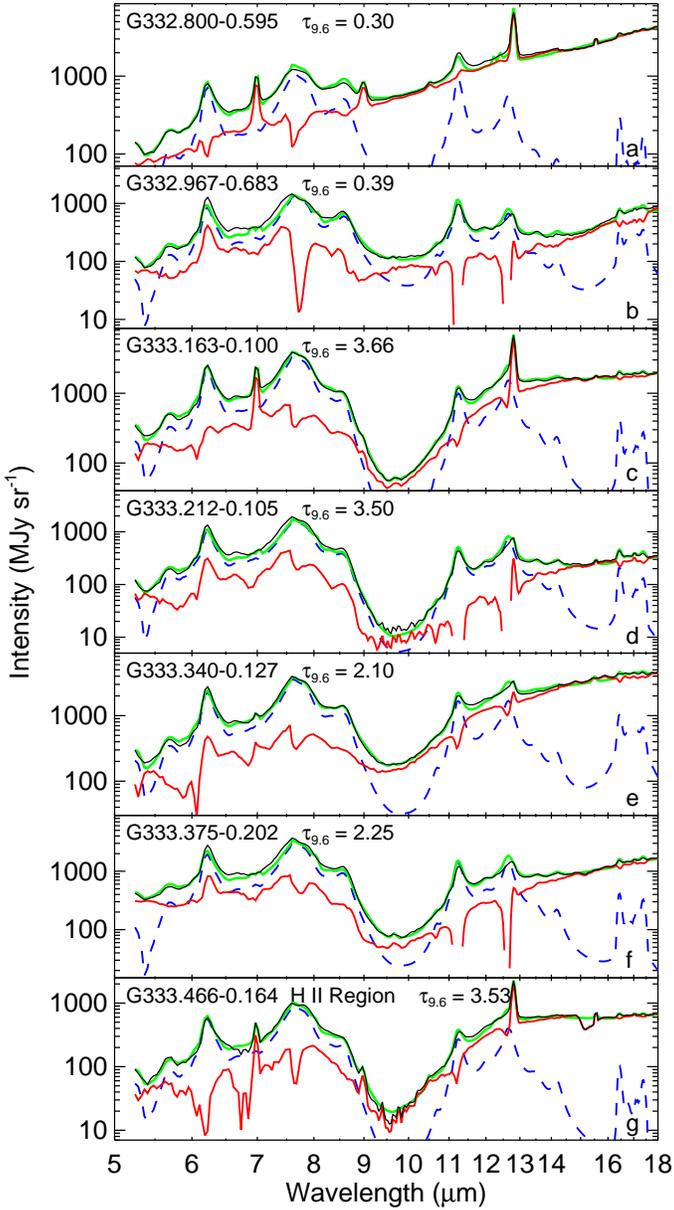}
\caption{Fit results for the SL and smoothed SH data for the position that includes the YSO
for the red sources plus the G333.466 position that has the maximum forbidden line emission.
The data are plotted in black, the PAH template, as absorbed by the extinction 
curve multiplied by the $\tau_{9.6}$ as given in each panel, is plotted in blue (dashed), 
the total fit is plotted in green (see text). 
The data minus the extincted PAH template are plotted in red.
Note that PAH template is never a really good fit, but the deviations vary with 
the different sources, indicating that there is no one single PAH spectrum,
even for very similar sources.
Panel (g): G333.466 is an outflow source that has an \mbox{H\,{\sc ii}} region component to the west 
of the YSO and outflow.  
The plotted position contains the peak of the \mbox{H\,{\sc ii}} region component,
and shows a typical \mbox{H\,{\sc ii}} region spectrum as seen in panels (a) to (f), 
except that it also has a substantial CO$_2$ 15-\micron\ ice absorption feature 
from the accreting dust and gas.
}
\end{figure}

%Figure 7
%\clearpage
\begin{figure}
\includegraphics[width=84mm]{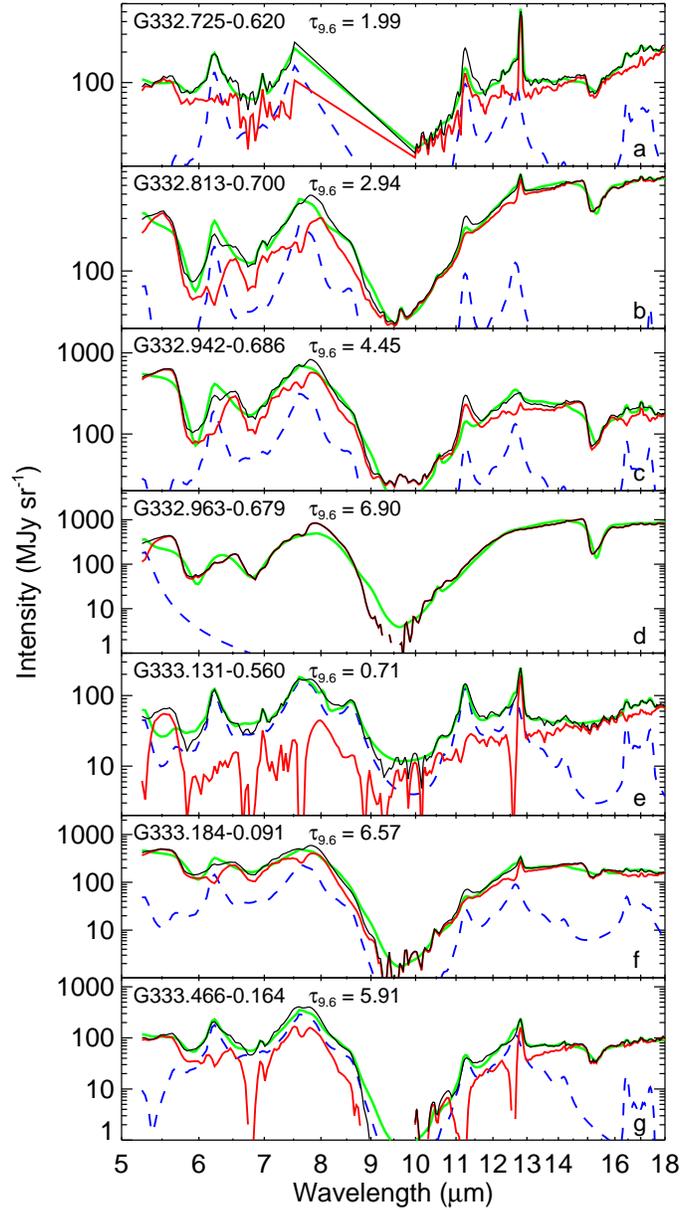}
\caption{Fit results for the SL and smoothed SH data for the position that includes the YSO
for the outflow sources.
The colours are described in Fig. 6.
The ice features were represented by Gaussians in the fit process for lack of 
a good template and thus their fits are not accurate.
Panel (d): note that even though the fitting program found no PAH components for G332.963,
there are small bumps at 6.2, 7.7, and 11.3~\micron\ in the spectrum,
showing that some PAH emission (probably foreground) is present, even here.
}
\end{figure}

The {\sc pahfit} deconvolutions are given in Figs. 6 and 7 
for the spectra at the YSO positions of all the sources.
The optical depths at 9.6~\micron, $\tau_{9.6}$, are at the YSO positions.
Note that compared to the spectra in Fig. 3, we used the SL spectra 
for the 5.2-14~\micron\ wavelength region 
and degraded the resolution of the rest of the SH and LH spectra 
when we stitched all spectra together in order to better match the 
line widths in {\sc pahfit}, which assumes all spectra are from the low resolution modules. 
We remark here that the PAH template is never a perfect fit to the 
observed spectra, and that each source deviates in a different way,
regarding both feature strengths and shapes.
See the review of Tielens (2008) for more discussion of the possible components 
to the PAH spectrum.

%Figure 8
%\clearpage
\begin{figure}
\includegraphics[width=84mm]{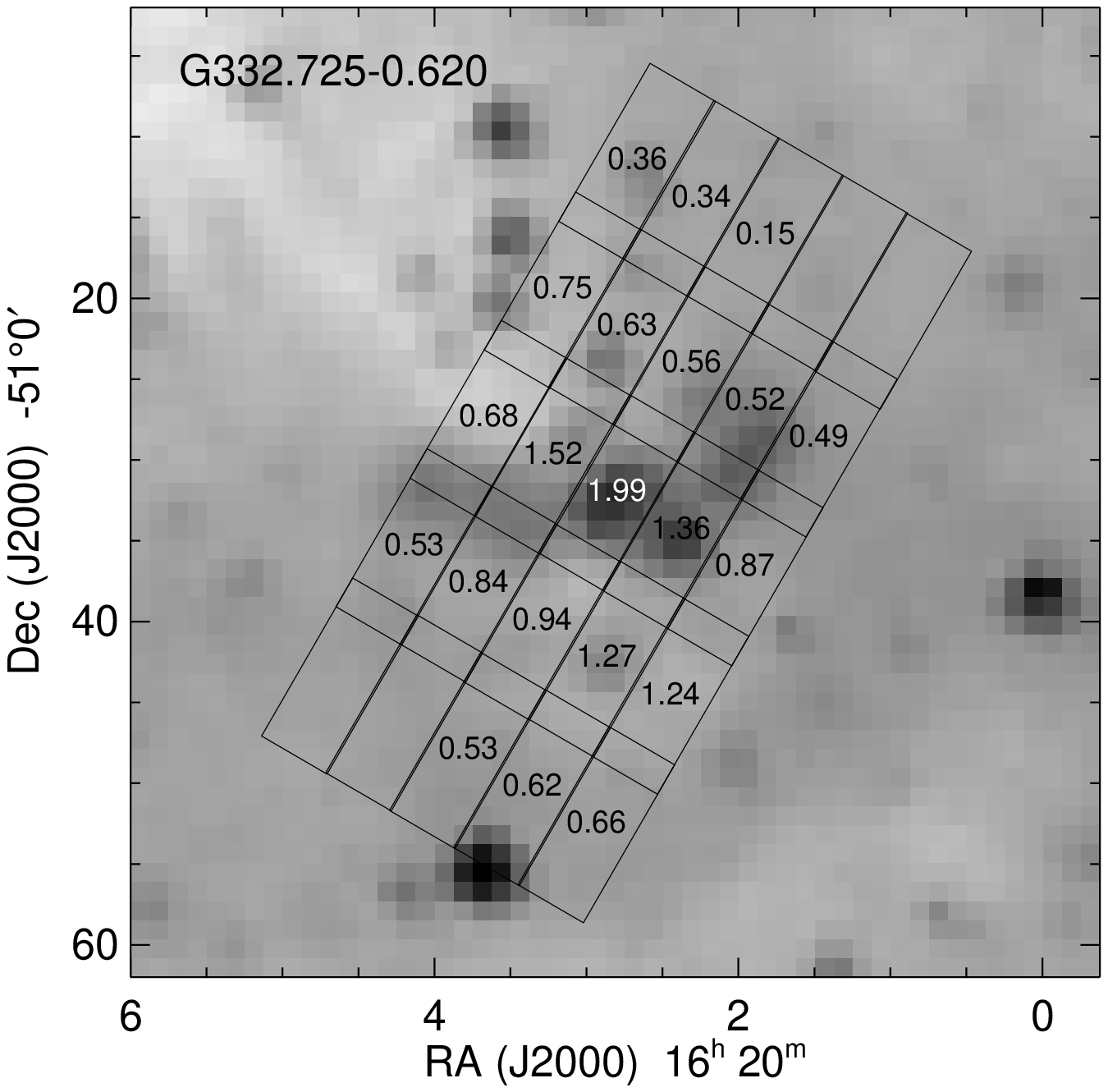}
\caption{Extinction map and SH slit positions for G332.725$-$0.620 plotted on the IRAC band 2 (4.5~\micron) image.
Positions without a value for the optical depth do not have SL spectra.
}
\end{figure}

%Figure 9
%\clearpage
\begin{figure}
\includegraphics[width=84mm]{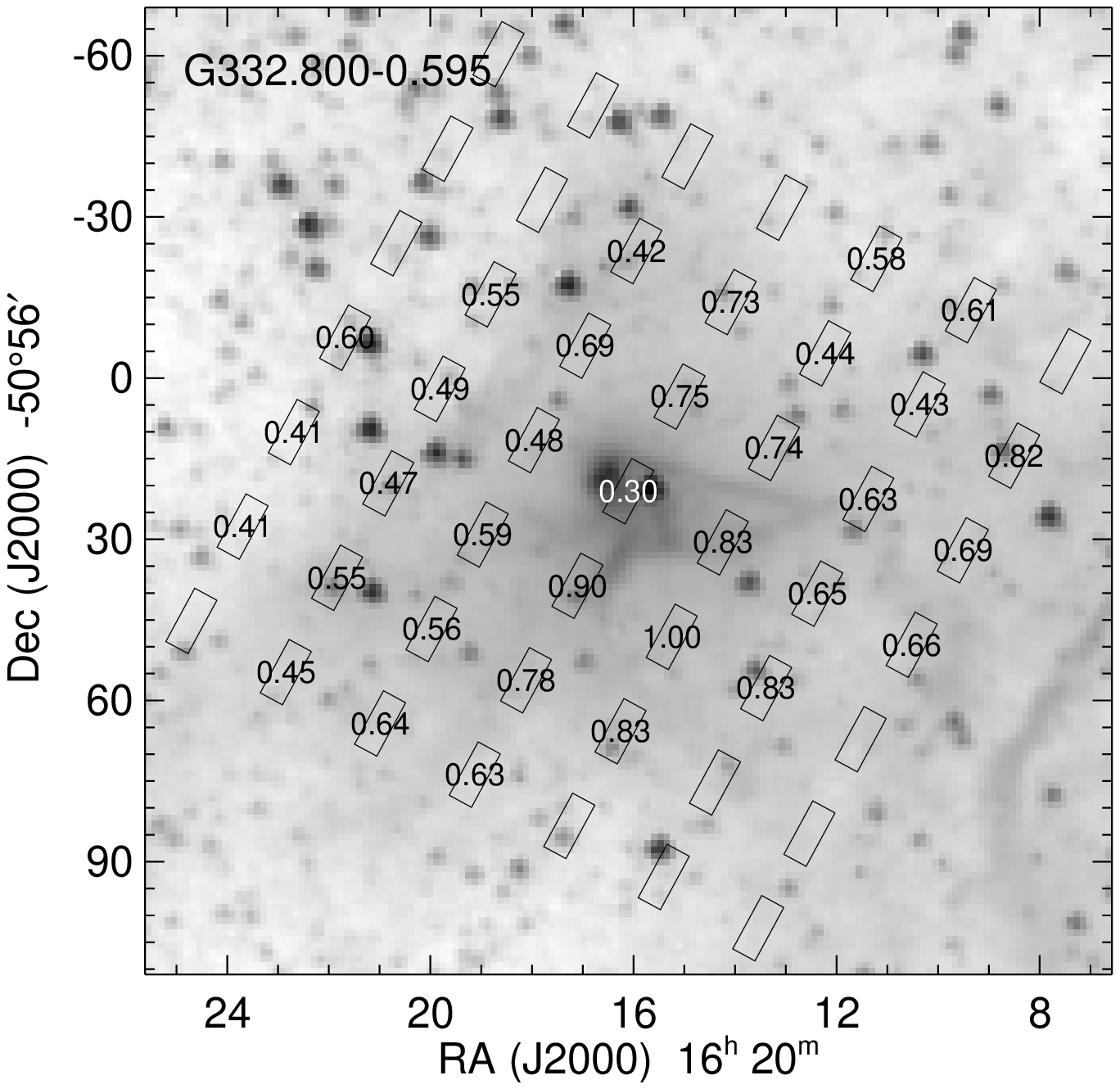}
\caption{Extinction map and SH slit positions for G332.800$-$0.595 plotted on the IRAC band 2 (4.5~\micron) image.
}
\end{figure}

%Figure 10
%\clearpage
\begin{figure}
\includegraphics[width=84mm]{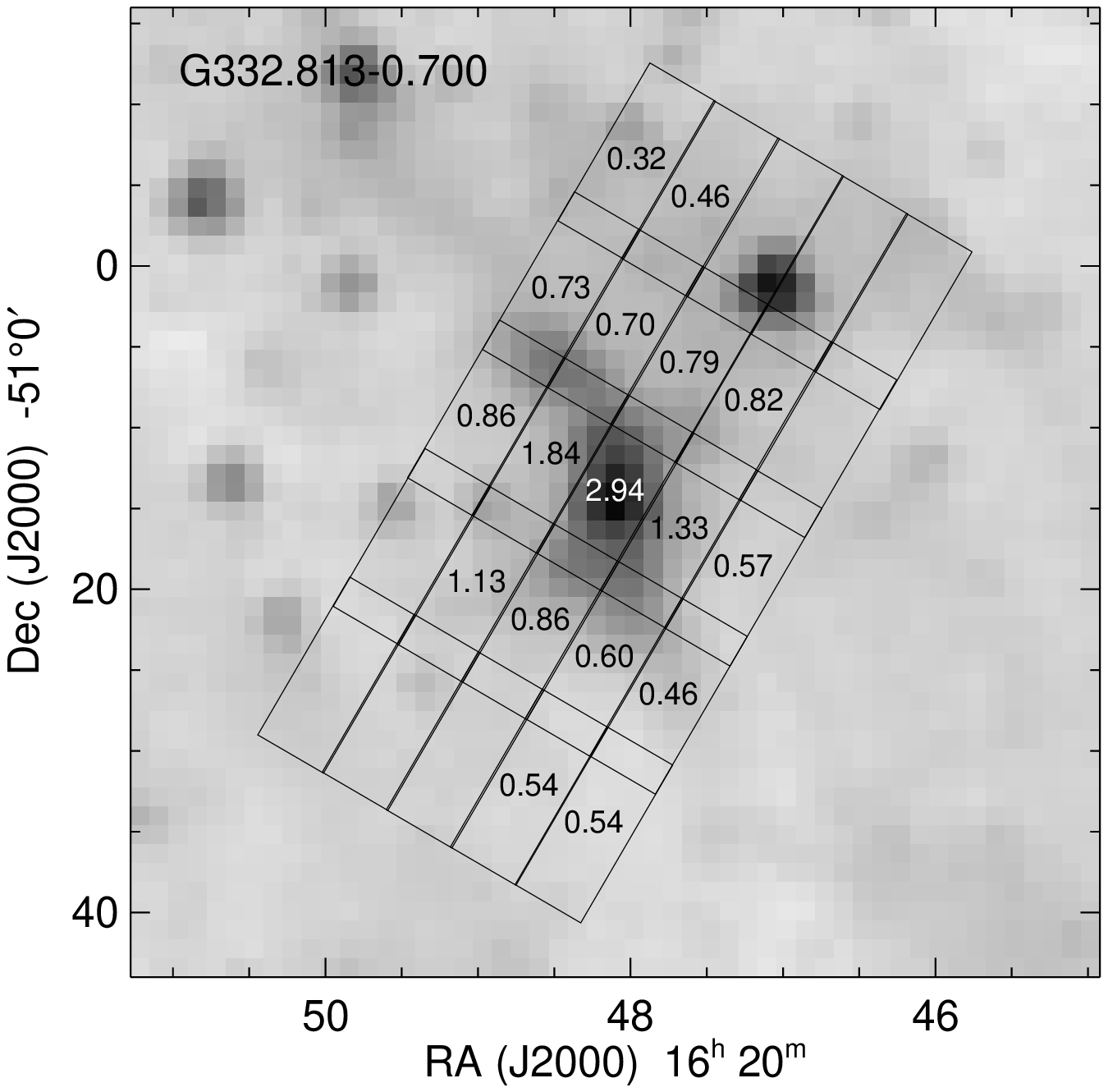}
\caption{Extinction map and SH slit positions for G332.813$-$0.700 plotted on the IRAC band 2 (4.5~\micron) image.
}
\end{figure}

%Figure 11
%\clearpage
\begin{figure}
\includegraphics[width=84mm]{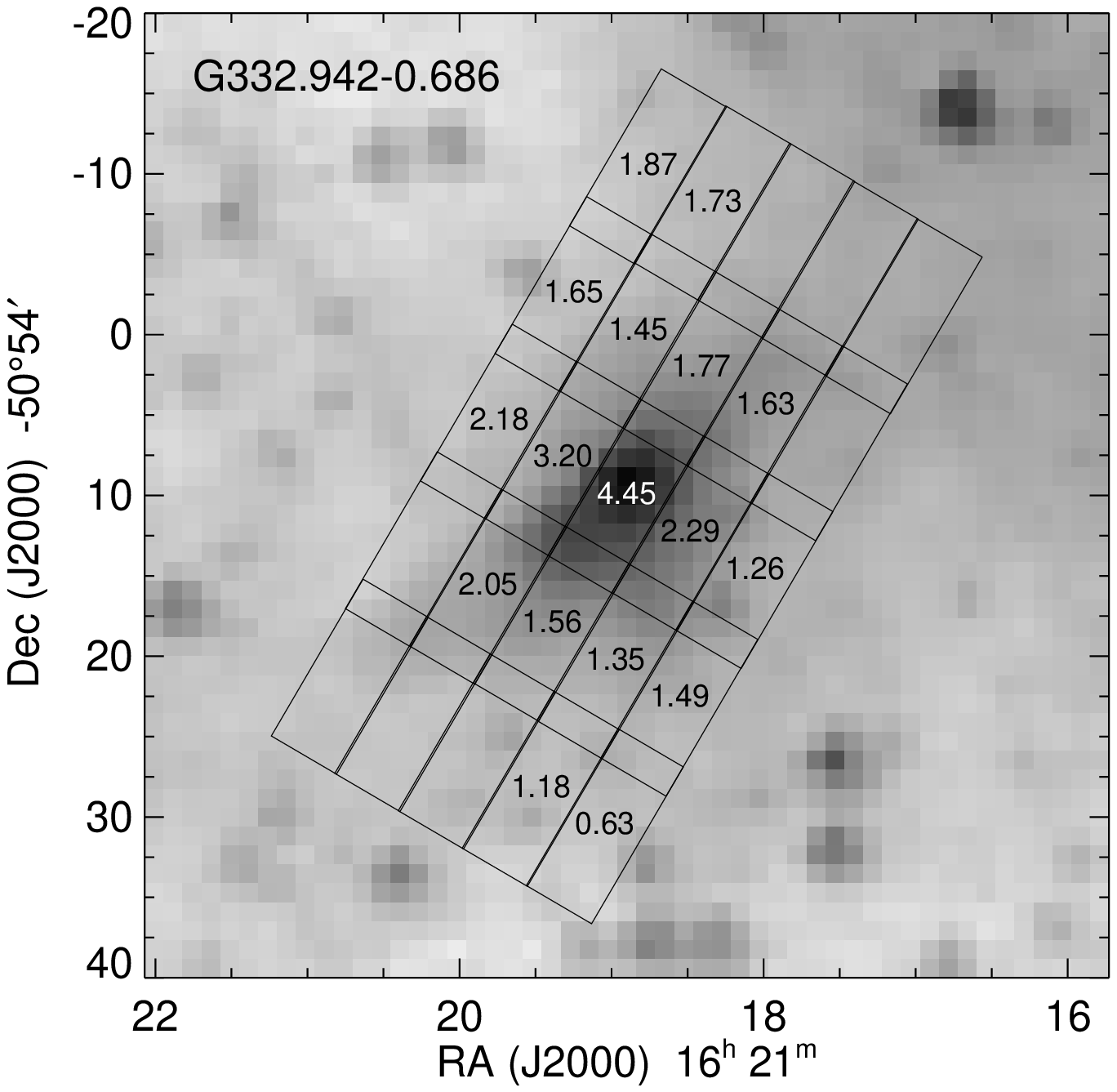}
\caption{Extinction map and SH slit positions for G332.942$-$0.686 plotted on the IRAC band 2 (4.5~\micron) image.
}
\end{figure}

%Figure 12
%\clearpage
\begin{figure}
\includegraphics[width=84mm]{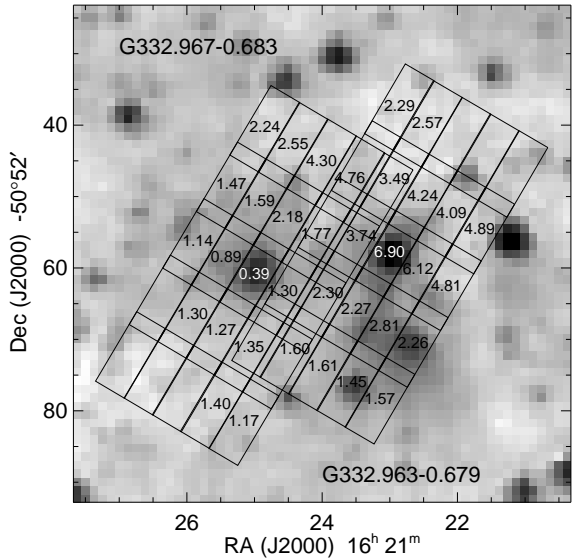}
\caption{Extinction map and SH slit positions for G332.963$-$0.679 and G332.967$-$0.683 plotted on the IRAC band 2 (4.5~\micron) image.
}
\end{figure}

%Figure 13
%\clearpage
\begin{figure}
\includegraphics[width=84mm]{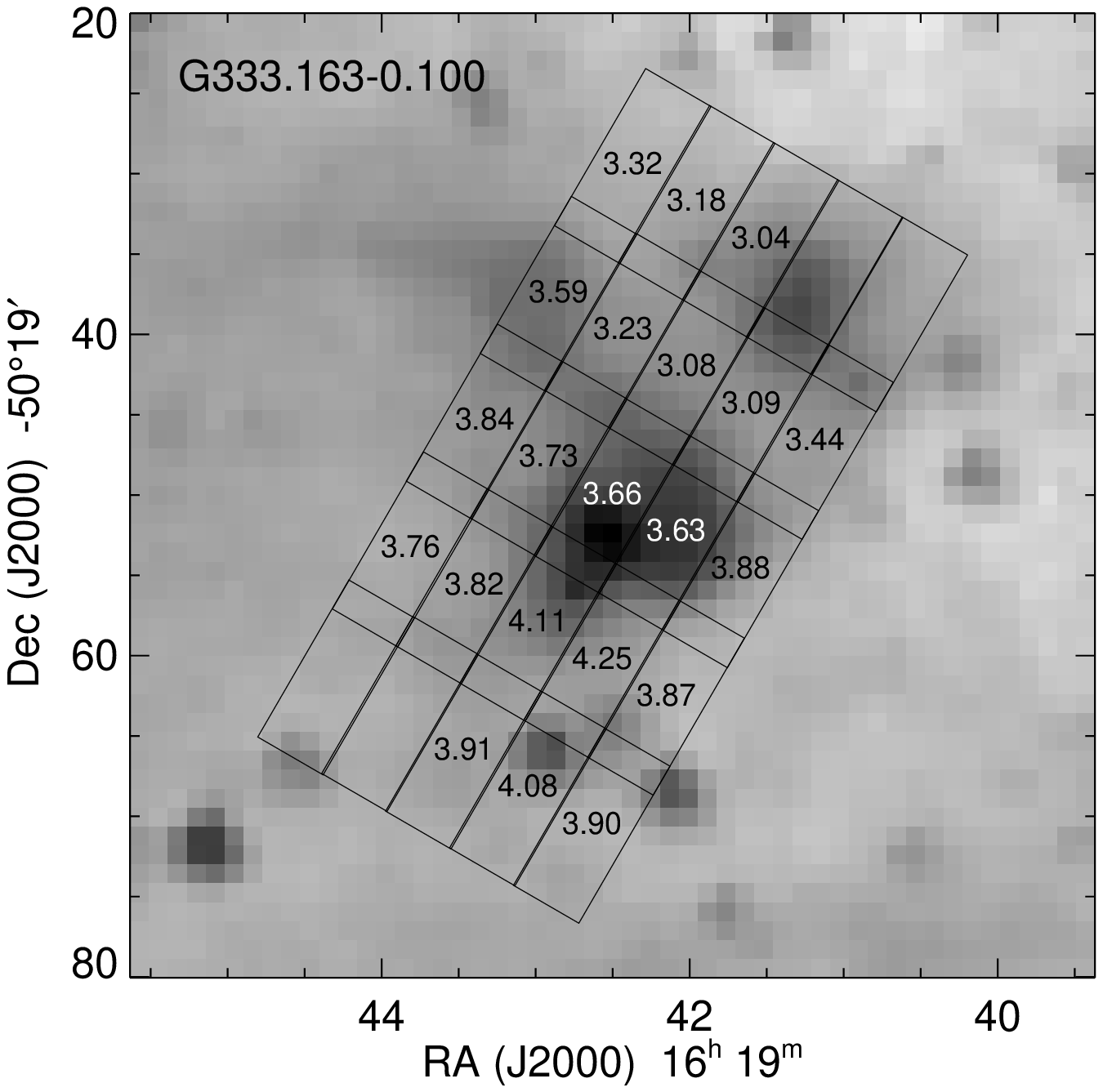}
\caption{Extinction map and SH slit positions for G333.163$-$0.100 plotted on the IRAC band 2 (4.5~\micron) image.
}
\end{figure}

%Figure 14
%\clearpage
\begin{figure}
\includegraphics[width=84mm]{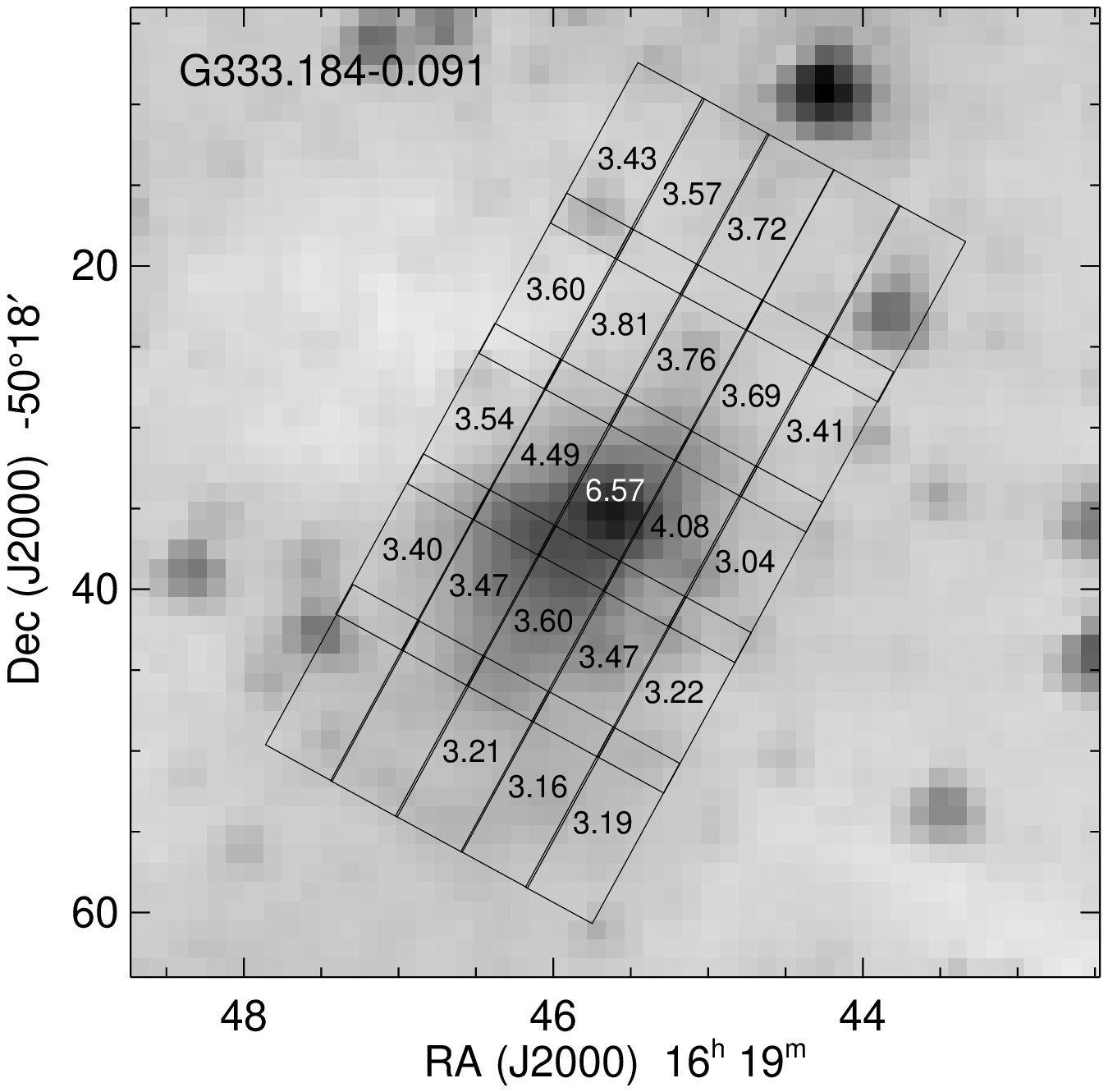}
\caption{Extinction map and SH slit positions for G333.184$-$0.091 plotted on the IRAC band 2 (4.5~\micron) image.
}
\end{figure}

%Figure 15
%\clearpage
\begin{figure}
\includegraphics[width=84mm]{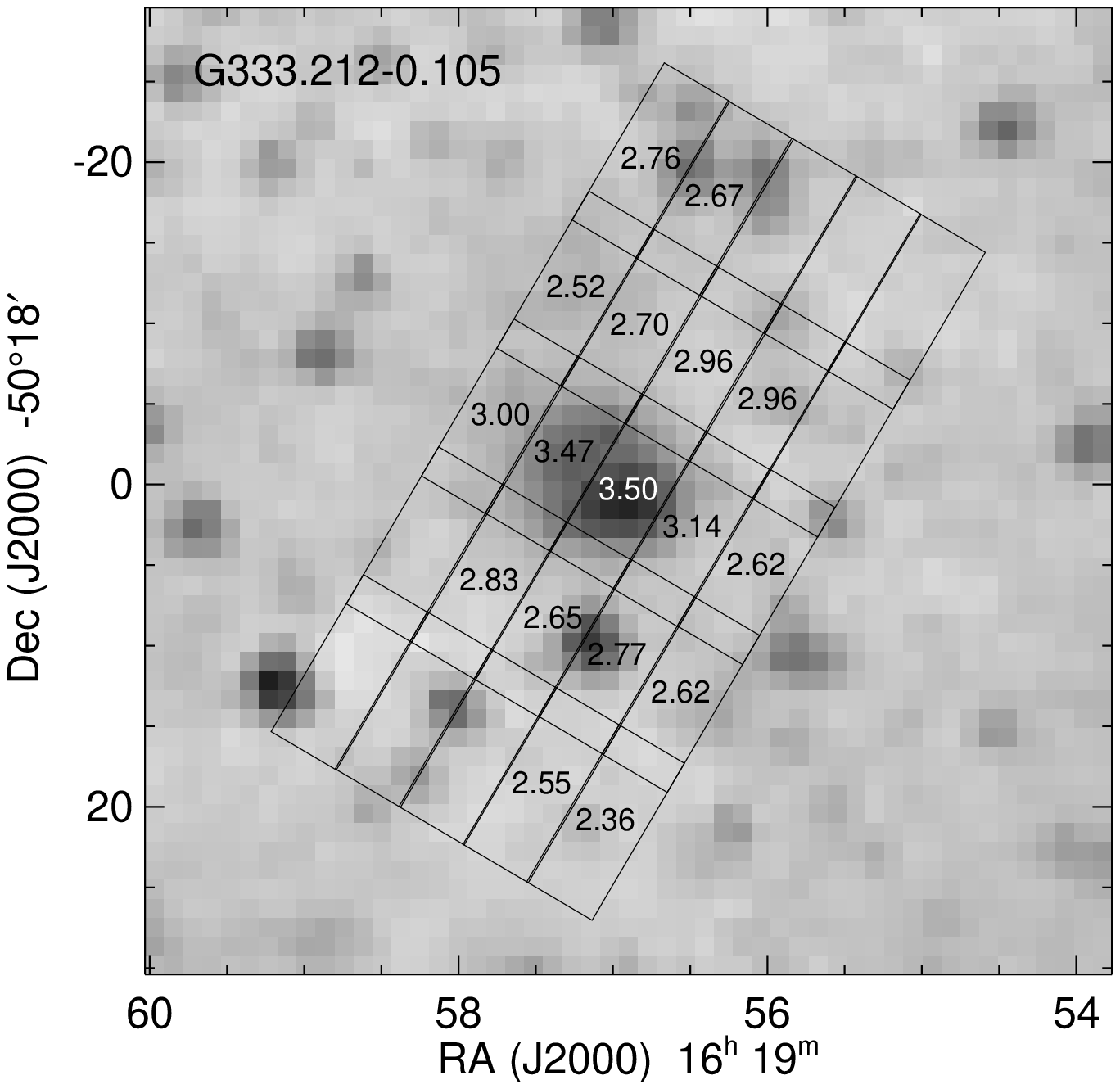}
\caption{Extinction map and SH slit positions for G333.212$-$0.105 plotted on the IRAC band 2 (4.5~\micron) image.
}
\end{figure}

%Figure 16
%\clearpage
\begin{figure}
\includegraphics[width=84mm]{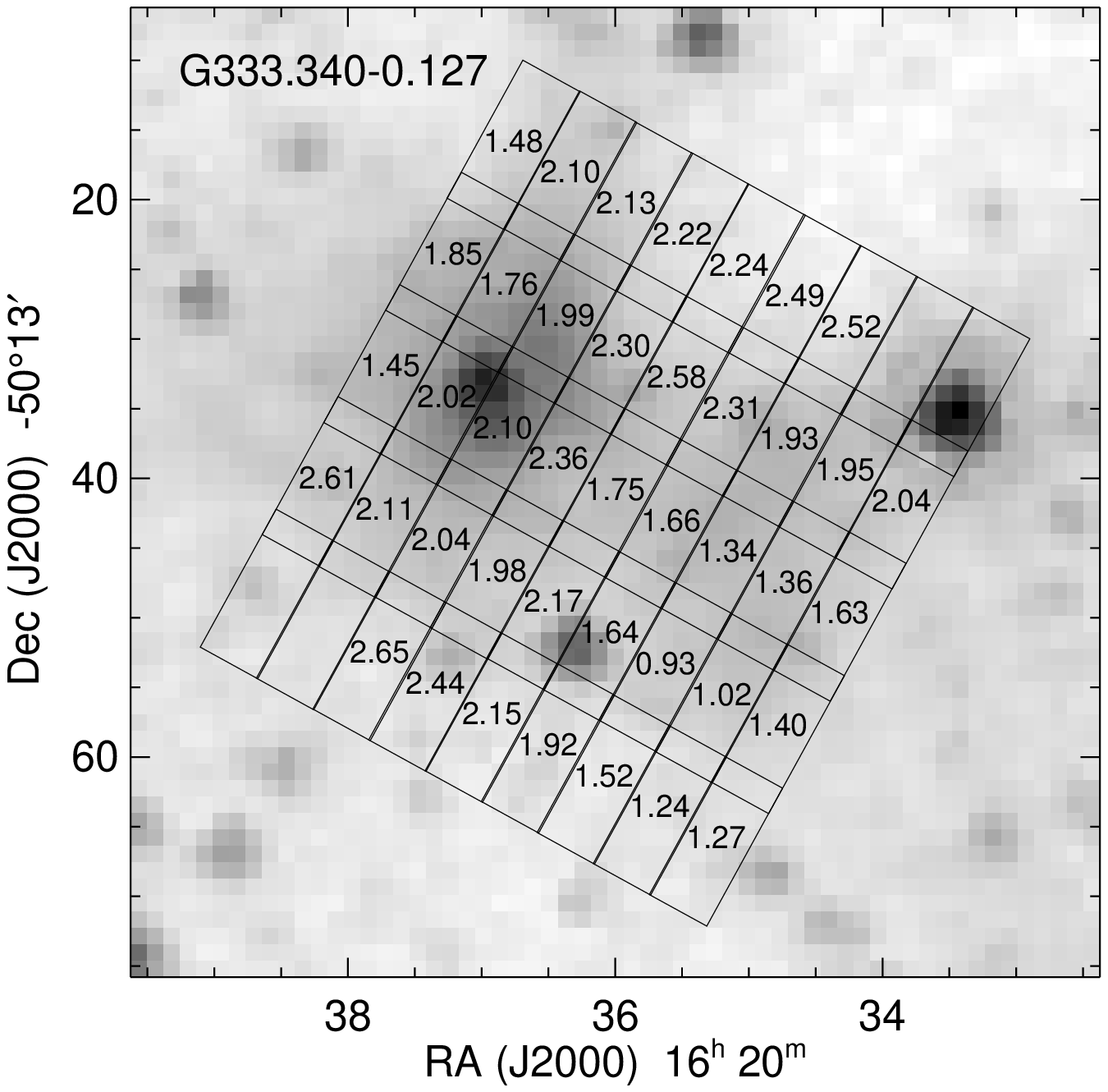}
\caption{Extinction map and SH slit positions for G333.340$-$0.127 plotted on the IRAC band 2 (4.5~\micron) image.
}
\end{figure}

%Figure 17
%\clearpage
\begin{figure}
\includegraphics[width=84mm]{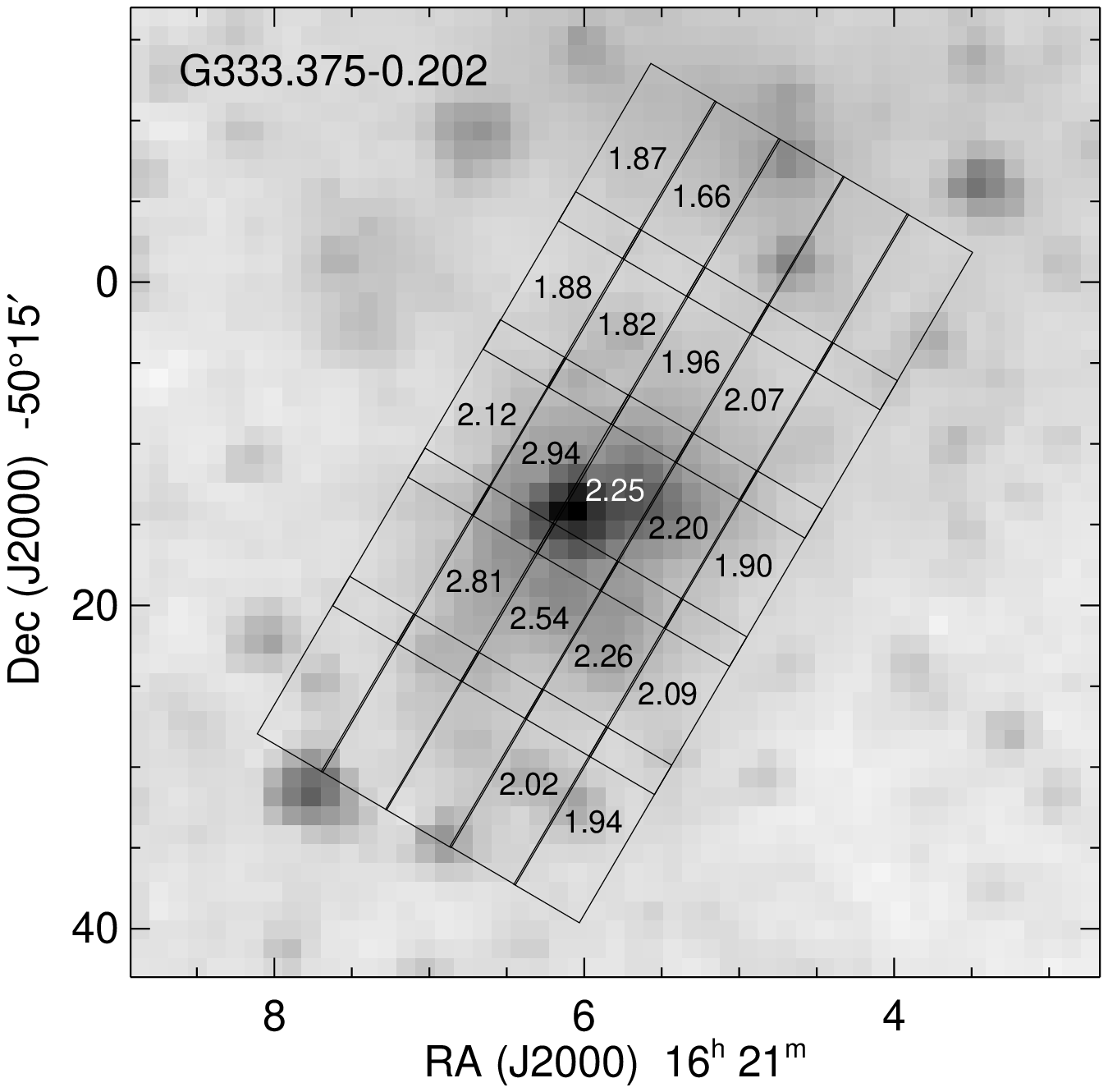}
\caption{Extinction map and SH slit positions for G333.375$-$0.202 plotted on the IRAC band 2 (4.5~\micron) image.
}
\end{figure}

%Figure 18
%\clearpage
\begin{figure}
\includegraphics[width=84mm]{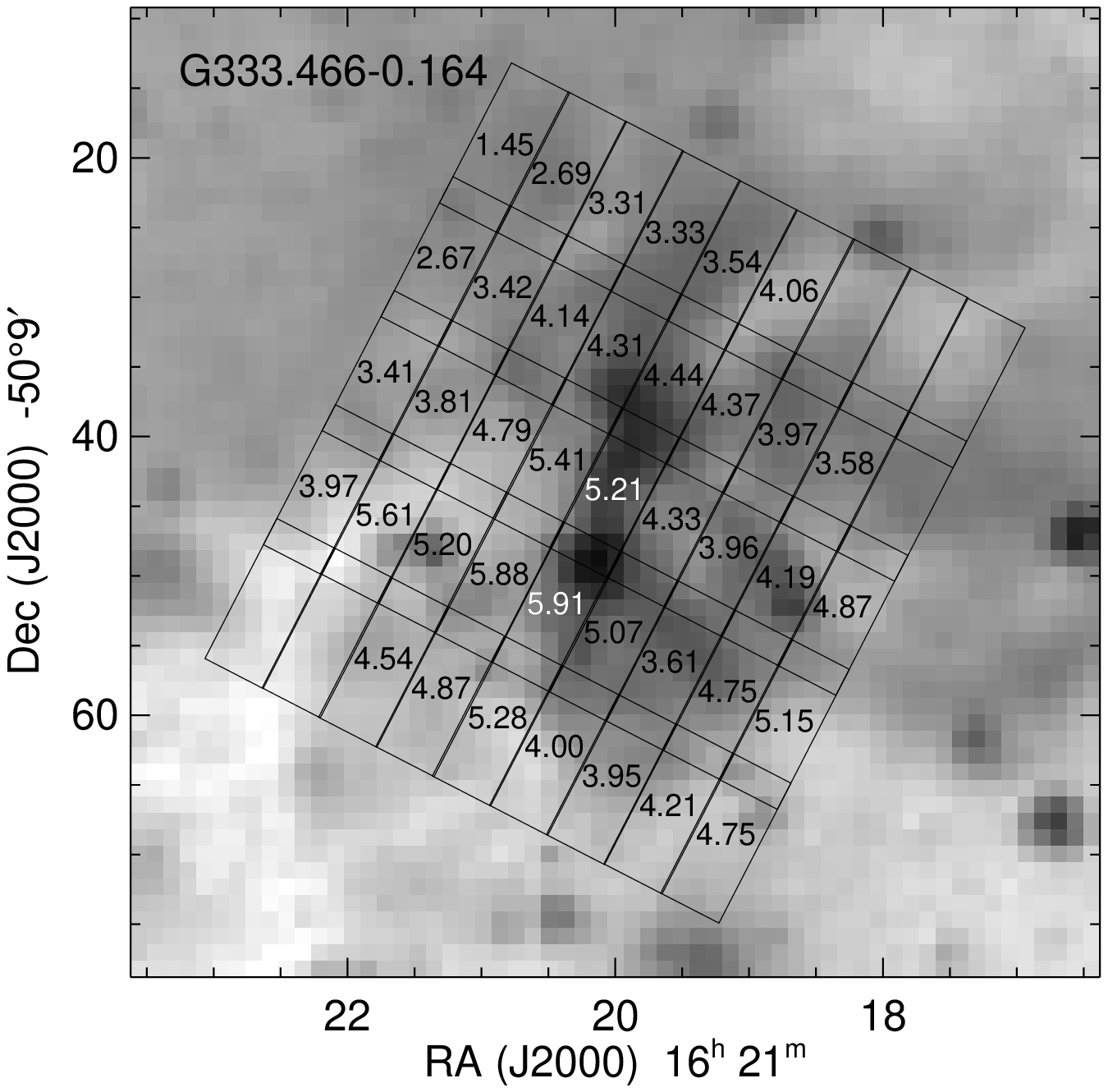}
\caption{Extinction map and SH slit positions for G333.466$-$0.164 plotted on the IRAC band 2 (4.5~\micron) image.
}
\end{figure}

The computed values of $\tau_{9.6}$ plotted as functions of position 
are given in Figs. 8 to 18 for all maps except G333.131 (Fig.~2), which has too little
S/N for production of stitched spectra from all three modules.
The values of $\tau_{9.6}$ are also given in Table~3 for each source,  
both the value of $\tau_{9.6}$ at the central YSO (this is usually the deepest absorption) 
and a value of $\tau_{9.6}$ from the outskirts of the spectral map  
for an estimate of the background extinction.
The sources with the largest distances from the Sun (Table 1) have the largest background extinctions
as might be expected if the background extinction is due to the diffuse ISM.
We also give the difference in extinction of source minus background.
Here we see that for the most part the outflow YSOs have larger extinction 
than the red sources.
We note that we assumed in the {\sc pahfit} computation 
that the unextincted continuum consists of a smooth black body with no silicate emission.
This should be appropriate for a very optically thick YSO core, but 
it is probably not correct for the diffuse ISM of the background positions.
Including silicate emission as well as silicate absorption could produce 
significantly larger estimates for the background extinction (Simpson \& Rubin 1990).

\subsection{Luminosities and associated parameters}

The source luminosities can be estimated from the total spectral energy distribution 
(SED) and the distance.
However, the luminosity cannot be estimated by simply integrating 
over the observed fluxes as a function of wavelength because YSOs are not spherically symmetric.  
Robitaille et al. (2006) computed a `grid' of 20,000 YSO models,  
each calculated with 10 inclination angles ranging from cos~$i = 0.05$ to cos~$i = 0.95$ 
($i = 87^\circ$ to $i = 18^\circ$).
The models consisted of a protostar, a disc, and an envelope with cavities 
along the axes due to outflows. 
They found that the optical depth to the central protostar can be quite variable 
depending on the axis inclination, $i$. 
For pole-on views ($i = 0$), the central protostar would be visible and 
there can be substantial observed NIR and even visible flux, whereas for $i = 90$, 
the star can be completely obscured by an edge-on, optically thick disc or envelope
(see also Whitney et al. 2004).

Consequently, to estimate the luminosities we used the SED fitter of 
Robitaille et al. (2007)\footnote{http://caravan.astro.wisc.edu/protostars}.
This online application takes input fluxes as a function of wavelength,   
estimated distances (Table~1), and visible extinction
and fits models from the Robitaille et al. (2006) grid.
For the estimated extinction, we used either the estimated background extinction from Table~3 
or we used a very large value so that the extinction is a completely free parameter.
As would be expected, the fits with the constrained extinction have much larger $\chi^2$; 
they also exhibit a larger range of fitted parameters, particularly 
the inclination angle and the model age.

The ranges of the fitted parameters are listed in Table~3 for luminosity, $L/L_{\odot}$, 
protostar mass, $M_*/M_{\odot}$, envelope mass, $M_{\rm env}/M_{\odot}$, 
protostar effective temperature, $T_{\rm eff}$, inclination, 
and model age for the best-fitting models. 
The parameters in Table~3 are combined from fits made with constrained and unconstrained extinction.
From the fits, which are shown in Fig.~19 for the fits with the unconstrained extinction, 
we also estimate and give in Table~3 the wavelength of the maximum of the fitted SED 
as plotted in $\lambda F_\lambda$ space.

%Figure 19
%\clearpage
\begin{figure*}
\includegraphics[width=150mm]{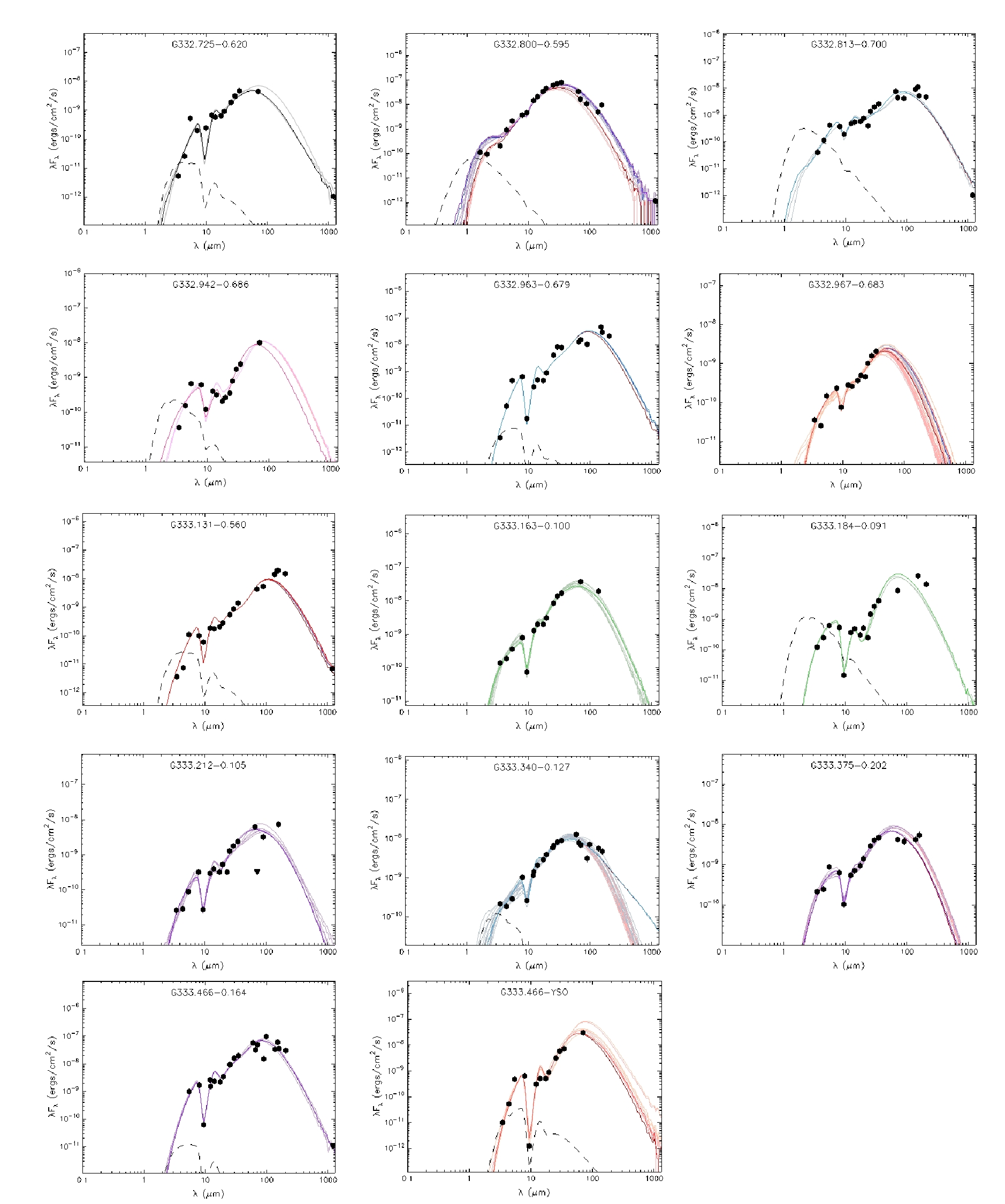}
\caption{Spectral energy distributions for all sources with fits from the online SED fitter 
of Robitaille et al. (2007).
The plotted points are our data from 5 to 35~\micron, fluxes measured from the IRAC and MIPS,
and fluxes taken from the literature (see text). 
The multiple solid lines are the SEDs of the top ten fitting models 
with unconstrained extinction from Robitaille et al. (2006).
(A figure showing the models with constrained extinction is very similar.)
The dashed line is the SED of the central source of the best-fitting model
including interstellar extinction but not the extinction from the envelope and disc.
There are two SEDs for G333.466: the whole source including the fluxes of IRAS 16175-5002 
and the SED for the YSO with the ice absorption features.
}
\end{figure*}

\setcounter{table}{2}
\begin{table*}
\centering
\begin{minipage}{170mm}
\caption{Results from fitting models to the spectra and to the SEDs.
The optical depth, $\tau_{9.6}$, is given for the spectrum of the source maximum ($\tau_{9.6}$ Src), 
minimum values near the corners of the map which should represent the diffuse ISM or background ($\tau_{9.6}$ Bkgr), 
and the difference of source maximum minus background ($\tau_{9.6}$ Src$-$Bkgr).
The values of model luminosity $L/L_{\odot}$, protostar mass $M_*/M_{\odot}$, envelope mass $M_{\rm env}/M_{\odot}$, 
protostar effective temperature $T_{\rm eff}$, inclination (Incl.), 
and model age are the range of the best fitting models from the online SED model fitter of Robitaille et al. (2007) (see text). 
$\lambda$(SEDmax) is the wavelength at which the FIR SED is maximum in Fig. 19.
Exponents are written as follows: 2-5e3 equals $2 \times 10^3 - 5 \times 10^3$.
}
\begin{tabular}{@{}lcccccccccc@{}}
\hline
Source   &  $\tau_{9.6}$  &  $\tau_{9.6}$  &  $\tau_{9.6}$ & $L$ &  $M_*$ &  $M_{\rm env}$ & $T_{\rm eff}$ & Incl. &   Model Age & $\lambda$(SEDmax) \\ % Comment
             & Src  & Bkgr  & Src$-$Bkgr  & ($L_\odot$)  &  ($M_\odot$) &  ($M_\odot$) &  (K)  &   (degrees)  &    (years)   &  (\micron) \\ %
\hline
Outflow Sources &&&&&&&&& \\ %
G332.725$-$0.620\footnote{Optical depths are less certain because wavelengths from 7.5 to 10 \micron\ were not observed.} &    1.99 &  0.4  &  1.6    &     2-5e3 &  9-12  & 4-158e1 &  4-12e3 &  18-57 &     2-17e3  & 60 \\ %  No SL1
G332.813$-$0.700 &    2.94 &  0.5  &  2.4    &     2-6e3 & 10-14  & 6-28e2 &  4-7e3 &  18      &     1-15e3  & 80 \\ %  Not much PAHs at YSO
G332.942$-$0.686 &    4.45 &  1.0  &  3.5    &     2-6e3 & 8-16  & 1-16e2 &  4-11e3 &  18-49   &     1-15e3   & 70 \\ %  tau10 too big, not much PAHs
G332.963$-$0.679 &    6.90 &  1.6  &  5.3    &     1-2e4 &  10-20 & 6-47e2 &  4-38e3 &  18-63  &    1-174e3  & 100 \\ %  No PAHs at YSO
G333.131$-$0.560 &    1.15 &  0.6  &  0.5    &     1-9e3 &  8-18 & 6-39e2 &  4-11e3  & 18-41   &     1-61e3  & 110  \\ %  
G333.184$-$0.091 &    6.57 &  3.2  &  3.4    &     1-5e4 &  20-32 & 8-18e2 &  4-7e3  & 18      &     1-5e3  &  70  \\ %   
G333.466$-$0.164\footnote{Parameters derived from the \mbox{H\,{\sc ii}} region peak and the SED of the whole cluster, IRAS 16175-5002.} &    3.97 &  1.6  &  2.4    &     2-5e4 &  12-30 & 13-36e2 &  4-36e3 & 18-57  &  1-96e3 &  80  \\ %  Whole cluster, IRAS 16175-5002, tau10 at MIPS24 saturated pixel
G333.466$-$0.164\footnote{Parameters derived from the outflow YSO $\tau_{9.6}$ and SED.} &    5.91 &  1.6  &  4.5    &     1-4e4 &  12-28  & 2-36e2 & 4-34e3 & 18-63  &  1-62e3 &  70  \\ %  IRAC3 outflow source
&&&&&&&&& \\
Red Sources \\ %
G332.800$-$0.595\footnote{Luminosity may be underestimated because the map coverage is incomplete.} &    0.30 &  0.7 &   0       &     5-7e4 &  21-23 & 6-15e1 & 35-37e3 & 81-87  &      8-35e4 &  35 \\ %  Incomplete coverage
G332.967$-$0.683 &    0.39 &  1.2 &   0       &     1-4e3 &  7-9   & 4-42 &  9-24e3  & 18-81  &      1-12e5  &  50 \\ % Extinction very variable 
G333.163$-$0.100 &    3.66 &  3.1 &   0.6    &     3-7e4 &  15-23 & 8-25e2 & 16-37e3 & 18-49   &     3-55e4 &  60  \\ %     
G333.212$-$0.105 &    3.50 &  2.5 &   1.0    &     4-10e3 & 10-18  & 8-25e2 &  4-18e3  & 18-32   &     1-78e3 &  70  \\ %     
G333.340$-$0.127 &    2.10 &  1.3 &   0.8    &     3-11e3 & 9-14 & 1-6e2 &  4-28e3 & 18-41   &     3-180e3 &  50  \\ %      
G333.375$-$0.202 &    2.25 &  1.9 &   0.4    &     4-7e3  & 8-10 & 2-8e2 & 21-24e3 & 31-56   &    15-22e4 &  60  \\ %       
\hline
\end{tabular}
\end{minipage}
\end{table*}

Other parameters from the fitted models are of less use in characterizing the YSOs:
The models with the larger ages usually have discs but the disc may or may not 
contribute substantial flux to the SED at the shortest wavelengths; 
the models with the shortest ages and lowest $T_{\rm eff}$ (the reddest SEDs) 
are the least likely to have discs. 
The total visible extinction, $A_V$, for the SED is the sum of 
the circumstellar and interstellar values -- 
the largest unconstrained interstellar extinction is $A_V \sim 102$ (in G333.466 YSO), 
for which the model circumstellar $A_V \sim 463$.
We constrained the interstellar $A_V$ to be $< 50$ for the YSOs with distance 5.8~kpc 
(this is effectively unconstrained) 
and $A_V < 10$ to 30 for the YSOs with distance 3.6~kpc (Table~1).
In general, for a given SED the model circumstellar $A_V$ varies much more widely 
than the interstellar $A_V$,
even by over two orders of magnitude for some of the red sources.

The model inclination is also not well-determined. 
For the fits with unconstrained extinction, 
all the outflow YSOs have $i = 18^\circ$, the minimum 
of the limited number (10) of inclination angles available to the SED fitter. 
For these fits, the best fits (lowest $\chi^2$) have large interstellar $A_V$; 
however, since all the outflow YSOs are readily visible at the shorter wavelengths 
(IRAC bands 2 and 3), the YSOs must be observed almost pole-on 
where the envelope extinction is smallest. 
On the other hand, when the interstellar extinction is constrained to be smaller ($A_V < 25$),
the observed depth of the 10~\micron\ silicate feature and the small NIR fluxes 
force the fit to have a large optical depth to the protostar via the envelope and disc 
(hence a larger inclination angle) -- 
but then the YSO must have a hotter star and a larger model age to 
have the hot disc that produces the NIR flux. 
Because the background extinction could be significantly larger (Section 3.1),
we regard the differences between fits using the constrained and unconstrained extinction 
as not significant. 
As emphasized by Robitaille et al. (2007), the fits are not unique but serve to 
demonstrate the range of possible parameters.
These parameters can almost certainly be better constrained by detailed modelling 
(e.g., Sewi{\l}o et al. 2010).

Because the range of fitted parameters is greatly decreased as more wavelengths 
are included (Robitaille et al. 2007), we increased the fitted wavelength range as follows:
No source is optically visible and only G332.800 has useable 2MASS 
(Skrutskie et al. 2006) $J$, $H$, and $Ks$ magnitudes. 
Some sources appear to have extended emission at $Ks$, particularly the outflow regions.
Since this is probably scattered light and the YSO itself either is not or is barely visible,
we do not use these wavelengths.
We measured fluxes from the IRAC band 1 (3.6~\micron) and band 2 (4.5~\micron) images
(only about half the sources have GLIMPSE catalog band 1 or band 2 fluxes) 
and used our own spectra instead of IRAC bands 3 and 4 (5.8 and 8.0~\micron, respectively).
Since our spectra are greatly contaminated by PAHs (and so would also be the photometry 
in IRAC bands 3 and 4),
we first subtracted the PAH templates fitted by {\sc pahfit} as described in the previous section.
This enabled us to measure much more reliable fluxes 
at 5.48, 7.99, 9.59, 12.02, 14.02, 18.02, 20.01, 25.03, 30.00, and 35.02~\micron\ 
for all YSOs except G332.725, for which we were not able to 
obtain a SL order 1 spectrum (7.5 -- 14~\micron) and so substituted 7.40 and 10.18~\micron\ fluxes 
for the 7.99 and 9.59~\micron\ fluxes.
These wavelengths miss the ice absorption features 
and most of the PAHs except at 7.99~\micron,
but the template fit and consequent subtraction is not bad at that wavelength.
Avoiding these emission and absorption features is important because they are not 
modelled by Robitaille et al. (2006).
Examples of spectra with the PAH template subtracted are given in Figs. 6 and 7
and the resulting continuum fluxes are given in Table A1.

For longer wavelengths we measured the flux at 70~\micron\ on the MIPS images 
and used {\it Infrared Astronomical Satellite (IRAS)} fluxes where available (Table 1).
We used fluxes from the {\it AKARI} (Murakami et al. 2007) 
point source catalog (Yamamura et al. 2010) at 65, 90, 140, and 160~\micron, 
fluxes from Karnik et al. (2001) at 150 and 210~\micron, 
and fluxes from Mookerjea et al. (2004) at 1.2 mm.
In the future we will add the 150~\micron\ to 600~\micron\ fluxes 
estimated from the images taken by the {\it Herschel Space Observatory} (Pilbratt et al. 2010)
as part of the Hi-GAL Open Time Key Project (Molinari et al. 2010).
These images were taken with the 
Photodetector Array Camera and Spectrometer (PACS) (Poglitsch et al. 2010)  
and the Spectral and Photometric Imaging Receiver (SPIRE) (Griffin et al. 2010) cameras; 
they are available online as of early 2011 
reduced to `level 2' by the {\it Herschel} pipeline 
but they need much more data reduction for bad pixels and other artefacts before 
they can be used to estimate fluxes.
It is beyond the scope of this paper to reduce {\it Herschel} data, 
but we can say that all our objects can be detected in the {\it Herschel} images, 
including the SPIRE 600~\micron\ images 
(except for G332.967, which is confused with G332.963), 
and thus the long wavelength parts of the SEDs seen in Fig. 19 are reasonable 
and the YSO envelopes must be large as the models predict (Table~3).

Unfortunately, the longest wavelength fluxes may include multiple sources 
(massive YSOs are, after all, born in clusters).
The result is that including longer wavelengths 
increases the wavelength of the SED maximum and also 
increases the total luminosity. 
In some cases it may be better to regard the luminosities in Table 3 
as possibly pertaining to a cluster rather than an individual YSO.
For example, in Table~3 we separate the YSO G333.466 from the 
total cluster IRAS 16175-5002, which has a much larger luminosity.

\subsection{Outflow YSOs with ice absorption features}

{\it All} YSOs defined by either their appearance on IRAC 3-colour images 
or by the presence of SiO emission to be outflow sources (Table 1) 
have ice absorption features at 6.0, 6.8, and 15.2~\micron.
These spectra are plotted in Fig. 3(b) and spectra 
with the PAH template subtracted in Fig. 7.
A number of icy compounds have been suggested as producing these features, such as
HCOOH and H$_2$O (6.0~\micron), CH$_3$OH and NH$_4^+$ (6.8~\micron) and CO$_2$ (15.2~\micron).
A review of the history of the detection of interstellar ices 
is beyond the scope of this paper -- for starters we refer the reader 
to the review by Boogert \& Ehrenfreund (2004) for a discussion of the 
high resolution spectra taken with the short-wavelength spectrometer on the 
{\it Infrared Space Observatory}.
A great number of spectra of YSOs have been taken with {\it Spitzer}, 
see for example, Boogert et al. (2008), Zasowski et al. (2009), and Oliveira et al. (2009).
With the high sensitivity of {\it Spitzer}, these authors have been able to show 
that there are many absorption features in addition to the main features 
at 6.0, 6.8, and 15.2~\micron. 
These new features enable the authors to identify additional molecules 
such as CH$_4$ at 7.67~\micron\ (\"{O}berg et al. 2008) 
and HCOOH at 7.25~\micron\ (Boogert et al. 2008),
thus showing details of the chemistry of the envelopes and discs of 
the YSOs.
There is a little dip at 7.67~\micron\ in most of the outflow sources 
that may be CH$_4$, 
but our spectra are so overladen with foreground PAHs and 
silicate absorption that it is not possible to estimate its reality or strength.

However, we do notice that the fit to the 9.6~\micron\ silicate feature is never good 
(Fig. 7) 
for the outflow sources whereas it is quite good for the red sources (Fig. 6).
The difference can be explained by additional absorption at around 9.0~\micron\ 
in the outflow source spectra;
an absorption feature at this wavelength has been attributed to NH$_3$ ice 
by Gibb et al. (2000).
Gibb et al. (2000), Boogert et al. (2008) and Bottinelli et al. (2010) 
also find absorption features due to CH$_3$OH ice at 9.7~\micron.
Unfortunately, our spectra have so much silicate extinction that the 
signal/noise is very poor at this wavelength and thus we cannot detect 
any such 9.7~\micron\ additional absorption.

We also do not detect the absorption features due to gaseous C$_2$H$_2$, HCN, and CO$_2$
that are found in massive YSOs in the Galactic Centre by An et al. (2009).
Because these features are narrow (and hence not confused by PAHs),
we should be able to detect them if they exist at similar strengths to 
the features measured by An et al. (2009).
We conclude our objects do not have amounts of warm gas similar to the Galactic Centre YSOs 
studied by An et al. (2009). 
Our objects could be cooler or at an earlier evolutionary stage. 
However, the three YSOs observed by An et al. (2009) could still be outflow sources 
since extended 4.5~\micron\ emission is detected in their vicinities 
(Chambers, Yusef-Zadeh, \& Roberts 2011).

\subsubsection{The 15~\micron\ CO$_2$ ice feature}

It is much easier to reliably detect the CO$_2$ ice feature than the features
at 6.0 and 6.8~\micron\ because the 15~\micron\ region is free
of the strong PAH emission features seen at the shorter wavelengths and at 16.3 -- 17.5~\micron.
As a result, the 6.0 and 6.8~\micron\ features are detected only at or near 
those YSOs that are IRAC band 3 point sources and have 
substantial 5.5~\micron\ continuum (Fig. 3b).
The 15~\micron\ CO$_2$ feature, however, is detected more widely, 
both in the extended region around the outflow YSOs and 
in two red sources (G333.163 and G333.375) that have no other ice features.

Because the CO$_2$ feature can have such good signal/noise and is so readily detectable, 
we have attempted to decompose its shape into polar and apolar CO$_2$ ice components 
as was described by Gerakines et al. (1999). 
The five components of the fit consist of the laboratory measurements by 
Ehrenfreund et al. (1997) for ices with ratios 
CO$_2$:H$_2$O=14:100, CO$_2$:CO=26:100, CO$_2$:CO=70:100, and pure CO$_2$ ice at 10 K. 
Since there is always smooth absorption in our spectra at shorter wavelengths 
than seen in any of Ehrenfreund et al.'s (1997) laboratory spectra, 
for the fifth component we shifted the CO$_2$:CO=70:100 component to shorter wavelengths by 
an arbitrary amount (calculated by the least-squares routine). 
There is no indication of any additional 15.4~\micron\ `shoulder' component as 
was fitted by Pontoppidan et al. (2008) and An et al. (2009), although 
it must be said that our continua are poorly estimated at the longer wavelengths 
of the CO$_2$ feature owing to the PAH emission from 16.3 -- 17.5 ~\micron. 
We estimated the component strengths by fitting the lab spectra of these five components 
plus a sloped linear continuum from 14.7 to 16.3~\micron\ 
by non-linear least-squares using {\sc mpfit} (Markwardt 2009). 
The polar CO$_2$:H$_2$O=14:100 component at $\sim 15.3$~\micron\ is by far the dominant, 
but there is always (if the S/N is adequate) 
a shorter wavelength component at $\sim 15.1$~\micron\ that can be attributed to 
one or all of the apolar components. 
The spectra of the six YSOs with the best S/N are plotted in Fig. 20. 
We list the CO$_2$ optical depths for these YSOs in Table~4, where we have summed the apolar 
components, since there probably is no real physical distinction between them.

%Figure 20
%\clearpage
\begin{figure}
\includegraphics[width=80mm]{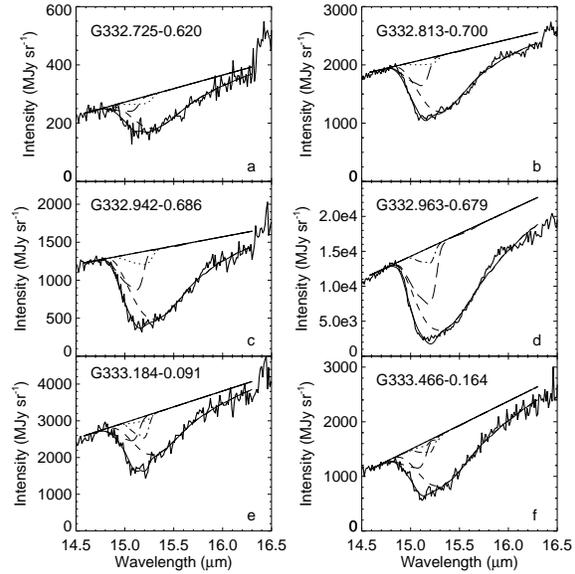}
\caption{Spectra, corrected for extinction, of the six YSOs that clearly show 
the CO$_2$ ice feature 
with fits to laboratory ice spectra from Ehrenfreund et al. (1997).
The dashed line is polar ice with CO$_2$:H$_2$O = 14:100.
The other ice features are all apolar ice: 
the dot-dashed line is CO$_2$:CO = 26:100,
the dotted line is CO$_2$:CO = 70:100, 
the dot-dot-dashed line is pure CO$_2$ ice, and 
the long dashed line is the CO$_2$:CO = 70:100 ice feature shifted to shorter 
wavelengths by an arbitrary (calculated) amount (see text).
The light solid line is the sum of all components and 
the heavy solid line is the fitted continuum, which stops short 
of the 16.4~\micron\ PAH feature. 
Only G333.466 (panel f) has any non-zero pure CO$_2$ ice component, and 
there it is small and not statistically significant.
}
\end{figure}

\setcounter{table}{3}
\begin{table*}
\centering
\begin{minipage}{130mm}
\caption{Maximum silicate and ice optical depths.
For each YSO in an outflow source, this table gives the optical depth of the silicate feature at 9.6 \micron, $\tau_{9.6}$,
the optical depth and error of the polar component of CO$_2$ ice, $\tau_{p{\rm -CO}_2}$, at $\sim~15.3$~\micron,
the optical depth and error of the sum of the apolar components of CO$_2$ ice, $\tau_{ap{\rm -CO}_2}$, at $\sim~15.1$~\micron\ (see text),
the ratio of the apolar to the polar components, $ap/p$, 
and the normalized central depths and errors measured by simultaneous Gaussian fits to the 6.0 and 6.8 \micron\ absorption features,
$T_{6.0}$ and $T_{6.8}$, respectively (see text).
}
%first line is central depth, second line is  optical depth
\begin{tabular}{@{}lcccccccccc@{}}
\hline
YSO & $\tau_{9.6}$ &  $\tau_{p{\rm -CO}_2}$ &  error  & $\tau_{ap{\rm -CO}_2}$ & error  & $ap/p$ &  $T_{6.0}$ &  error &  $T_{6.8}$ & error \\ 
\hline
G332.725$-$0.620 &  1.99 &   0.51 & 0.03  &   0.23 & 0.06  &  0.46 &   0.29 & 0.03  &    0.40 & 0.07 \\  
%  19   1.99    0.5073  0.0263     0.2317  0.0590    0.46    0.2922  0.0326      0.4001  0.0698   
%  19   1.99    0.5073  0.0263     0.2317  0.0590    0.46    0.3456  0.0461      0.5110  0.1164  
G332.813$-$0.700 &  2.94 &   0.60 & 0.01  &   0.29 & 0.03 &   0.48 &   0.89 & 0.04 &     0.68 & 0.02 \\
%  22   2.94    0.6023  0.0092     0.2915  0.0305    0.48    0.8880  0.0395      0.6815  0.0178 
%  22   2.94    0.6023  0.0200     0.2915  0.0305    0.48    2.1888  0.3526      1.1440  0.0559 
G332.942$-$0.686 &  4.45 &   1.08 & 0.03  &   0.58 & 0.12 &   0.54 &   0.95 & 0.05 &     0.77 & 0.02 \\
%  22   4.45    1.0770  0.0286     0.5830  0.1150    0.54    0.9502  0.0543      0.7690  0.0160 
%  22   4.45    1.0770  0.0286     0.5830  0.1150    0.54    2.9998  1.0899      1.4655  0.0693 
G332.963$-$0.679 &  6.90 &   1.51 & 0.02  &   0.89 & 0.14 &   0.59 &   0.96 & 0.07  &    0.90 & 0.01 \\
%  22   6.90    1.5098  0.0221     0.8920  0.1385    0.59    0.9634  0.0681      0.9019  0.0075
%  22   6.90    1.5098  0.0221     0.8920  0.1385    0.59    3.3076  1.8600      2.3221  0.0767 
G333.131$-$0.560 & 1.15  &  0.23 & 0.04  &   0.19 & 0.05 &   0.85  &  0.93 & 0.06  &    0.73 & 0.09 \\ 
%star1  1.15    0.2280  0.0409     0.1927  0.0517    0.85    0.9267  0.0596      0.7228  0.0910  
%star1  1.15    0.2280  0.0409     0.1927  0.0517    0.85    2.6134  0.8132      1.2831  0.3283  
G333.184$-$0.091 &  6.57 &   0.44 & 0.02 &    0.46 & 0.12 &   1.03  &  0.82 & 0.04  &    0.71 & 0.01 \\
%  26   6.57    0.4443  0.0193     0.4559  0.1218    1.03    0.8181  0.0399      0.7066  0.0110 
%  26   6.57    0.4443  0.0200     0.4559  0.1218    1.03    1.7045  0.2194      1.2261  0.0375 
G333.466$-$0.164 &  5.91 &   0.86 & 0.04  &   0.46 & 0.18  &  0.53  &  0.72 & 0.04  &    0.95 & 0.03 \\
%  49   5.91    0.8605  0.0429     0.4550  0.1751    0.53    0.7240  0.0443      0.9451  0.0302 
%  49   5.91    0.8605  0.0429     0.4550  0.1751    0.53    1.2873  0.1606      2.9015  0.5495  
\hline
\end{tabular}
\end{minipage}
\end{table*}

The pure CO$_2$ component has substructure with a feature at $\sim 15.28$~\micron; 
that component is identified by the least-squares routine 
only in those spectra that are noisy at that wavelength.
None of our outflow YSOs reliably show this structure in the 15~\micron\ feature, 
which structure can be ascribed to heat-processing of the CO$_2$ ice. 
Since such structure has been seen in certain of the low mass YSOs by Boogert et al. (2008) 
and in high-mass YSOs by Gerakines et al. (1999), 
we conclude that all of our outflow YSOs are at an earlier evolutionary stage 
such that the ice is still cold. 
The cold ice may be the reason why we do not see gaseous C$_2$H$_2$, HCN, and CO$_2$, 
as mentioned above.

%Figure 21
%\clearpage
\begin{figure}
\includegraphics[width=80mm]{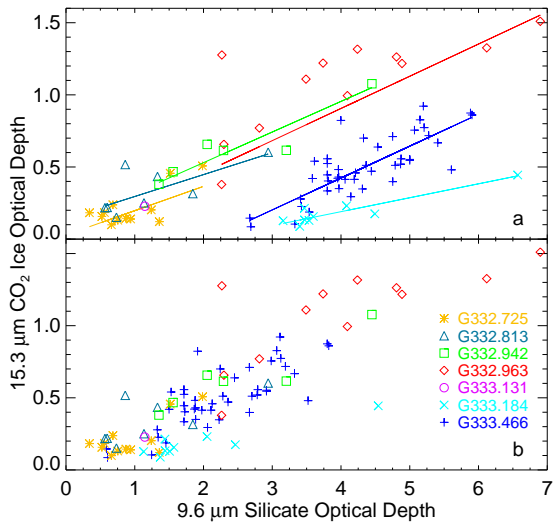}
\caption{Optical depths of the polar CO$_2$ component of the CO$_2$ ice feature 
plotted vs. the 9.6~\micron\ silicate optical depth.
The sources are identified by symbol and by colour as given in the figure.
Panel (a): a linear relation between the optical depths of the two compounds is 
given by the straight lines, computed by least-squares.
Panel (b): for the two sources with intercepts clearly not at the origin, 
the excess silicate optical depth is subtracted: 
$\Delta \tau_{9.6} = 2.03$ for G333.184 
and $\Delta \tau_{9.6} = 2.08$ for G333.466.
This excess silicate optical depth can be attributed to the intervening diffuse ISM, 
whose dust grains lack CO$_2$ ice (e.g., Whittet et al. 2007).
}
\end{figure}

In Fig. 21(a) we plot the optical depth of the 15.3~\micron\ CO$_2$ ice feature 
(the strong polar component) 
versus the optical depth $\tau_{9.6}$ of the 9.6~\micron\ silicate absorption feature
as estimated by our {\sc pahfit} computations (Section 3.1).
The plot shows that there is a correlation for individual sources, 
although there are differences between sources,
as shown by the fits to the CO$_2$ optical depths versus silicate optical depth.
In Fig. 21(b) we subtract the silicate optical depth corresponding to the 
intersection of the fitted line with zero CO$_2$ optical depth 
($\tau_{9.6} = 2.03$ for G333.184 and $\tau_{9.6} = 2.08$ for G333.466) -- 
that is, we assume that the differences between the sources is due 
to the foreground extinction through the diffuse ISM.
Now we see that there is much better agreement between sources.
We conclude that in massive YSOs with ices, the CO$_2$ ice absorption 
is proportional to the silicate absorption once the foreground ISM 
absorption, which lacks CO$_2$ ice, is removed.

\subsubsection{The 15~\micron\ CO$_2$ ice feature compared to the other ice features}

Because of the PAH emission present in every spectrum, 
the features at 6.0 and 6.8~\micron\ cannot be analysed with the same precision 
as the 15.2~\micron\ CO$_2$ ice feature.
In fact, we are not able to do any more than determine the normalized central depths, $T$, of 
the stronger features, which we estimated by fitting Gaussians to the two features.
We do not fit templates to the 6.0 and 6.8~\micron\ features because 
both are thought to be composites of at least two molecules each
(e.g., Boogert et al. 2008).
The measured central depths are given in Table~4 for the outflow YSOs.

From the table there is clearly a difference, source to source, between 
each of the ices: CO$_2$, the 6.0~\micron\ feature, and the 6.8~\micron\ feature.
The best correlation of the strength of the 6.0~\micron\ feature
relative to the  CO$_2$ optical depth is with the 
wavelength of the peak of the SED (Table~3 and Fig.~19):
colder SEDs have a larger 6.0~\micron\ feature relative to the CO$_2$ ice optical depth.
There is no apparent correlation with source morphology or luminosity.

\subsection{Sources with forbidden line emission}

\setcounter{table}{4}
\begin{table*}
\centering
\begin{minipage}{170mm}
\caption{Sources producing ionizing photons.
The spectrum of each source with localized forbidden lines after background subtraction 
was analysed as an \mbox{H\,{\sc ii}} region.
This table gives the electron density, $N_e$, estimated from the ratio of the \mbox{[S\,{\sc iii}]} 18.7 and 33.5 \micron\ lines;
ratios of neon, sulphur, and hydrogen ions, the ratio of the total neon/sulphur abundance; 
the radio flux at 5 GHz predicted from the strength of the H 7--6 recombination line, $S_5$;  
and the numbers of hydrogen-ionizing photons, $N_{\rm Lyc}$, estimated from the neon forbidden line fluxes (see text).
Exponents are written as follows: 1.8e48 equals $1.8 \times 10^{48}$.
}
\begin{tabular}{@{}lcccccccccc@{}}
\hline
Source   &   $N_e$(\mbox{S\,{\sc iii}})  &   Ne$^{++}$/Ne$^+$ &  S$^{3+}$/S$^{++}$ & Ne$^+$/H$^+$ &  Ne$^{++}$/H$^+$ &  S$^{++}$/H$^+$  &  Ne/S    &      $S_5$ & $N_{\rm Lyc}$ \\
       &    (cm$^{-3}$)  &          &              &    $\times 10^{-6}$   &  $\times 10^{-6}$   & $\times 10^{-6}$    &          &      (Jy)  & (s$^{-1}$) \\  
\hline
G332.800$-$0.595 &  $ < 100 $  &  $ 0.026\pm0.001$ & $ 0.0060\pm0.0003$ & $141\pm15$  &  $3.6\pm0.4$ & $10.8\pm1.1$ & $13.4\pm0.3$  &  $1.4$  &   1.8e48  \\
G333.131$-$0.560 &  $ 90\pm20$  &  $ 0.030\pm0.005$ &        -       &  - &  - & - &  $19.2\pm1.1$  &   -   &   3.4e45  \\
G333.163$-$0.100 &  $685\pm120$ &  $0.008\pm0.001$ &       -       &  $132\pm14$  &  $1.1\pm0.1$ & $5.2\pm0.7$ &  $25.9\pm2.2$  &  $0.30$ &   9.2e47  \\
G333.340$-$0.127 &  $150\pm80$  &  $0.039\pm0.006$  &  $ 0.107\pm0.049$ &  $ 52\pm12$  &  $2.0\pm0.6$ & $1.4\pm0.3$ &  $34.7\pm2.8$  &  $0.06$ &   3.1e46  \\
G333.375$-$0.202 &  $ < 100 $  &  $0.028\pm0.116$  &        -       &      -    &     -     &    -     &    -        &    -  &   5.4e44 \\
G333.466$-$0.164N & $190\pm55$  &  $0.059\pm0.002$  &   $0.011\pm0.002$ &  $147\pm19$  &  $8.6\pm1.1$ &  $9.1\pm1.2$ &  $17.0\pm1.3$  &  $0.09$ &   8.0e46  \\
G333.466$-$0.164W & $320\pm80$  &  $0.035\pm0.002$  &   $0.006\pm0.002$ &  $164\pm20$  &  $5.7\pm0.7$ & $11.4\pm1.4$ &  $13.8\pm1.2$  &  $0.10$ &   1.0e47  \\
\hline
\end{tabular}
\end{minipage}
\end{table*}

As was mentioned above, there are forbidden emission lines present in almost all the spectra.
In many of the spectra, for example in the background spectra, these forbidden lines 
clearly arise from the diffuse ionized ISM in the G333 cloud.
On the other hand, some of the outflow YSOs have minima of the forbidden lines 
at the YSO, probably because of their extreme extinction (Table~3).
Consequently, we identify sources producing ionizing photons only when 
there is a maximum of the forbidden lines uncorrected for extinction at the source location.
The sources producing ionizing photons are listed in Table 5 and 
contours showing the concentration of forbidden lines are plotted in Figs. 22 -- 26.

%Figure 22
%\clearpage
\begin{figure*}
\includegraphics[width=150mm]{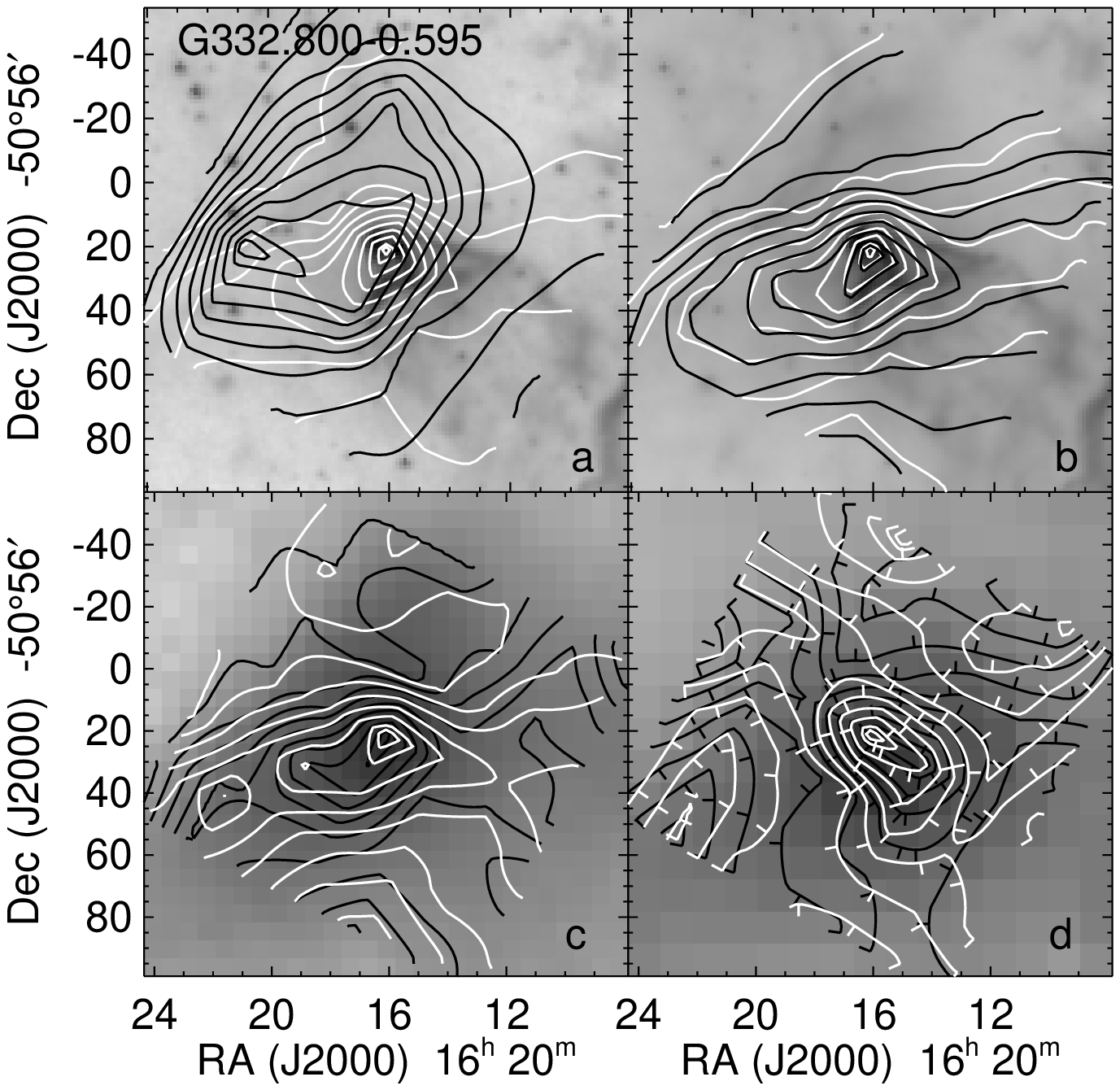}
\caption{Images of G332.800$-$0.595 with emission line contours.
Panel (a): the \mbox{[Ne\,{\sc ii}]} 12.8~\micron\ line contours 
are plotted in white 
and the \mbox{[Ne\,{\sc iii}]} 15.6~\micron\ line contours 
are plotted in black on the IRAC band 3 (5.8~\micron) image.
Panel (b): the \mbox{[S\,{\sc iii}]} 18.7~\micron\ line contours 
are plotted in white and 
the \mbox{[S\,{\sc iii}]} 33.5~\micron\ line contours 
are plotted in black on the IRAC band 4 (8.0~\micron) image.
Panel (c): the \mbox{[Ar\,{\sc ii}]} 7.0~\micron\ line contours are plotted in white 
and the \mbox{[Si\,{\sc ii}]} 34.8~\micron\ line contours 
are plotted in black 
on the MSX 21~\micron\ image.
Panel (d): the H$_2$ S(2) 12.3~\micron\ line contours  
are plotted in white and 
the H$_2$ S(1) 17.0~\micron\ line contours 
are plotted in black on the MIPS 70~\micron\ image.
}
\end{figure*}

%Figure 23
%\clearpage
\begin{figure*}
\includegraphics[width=150mm]{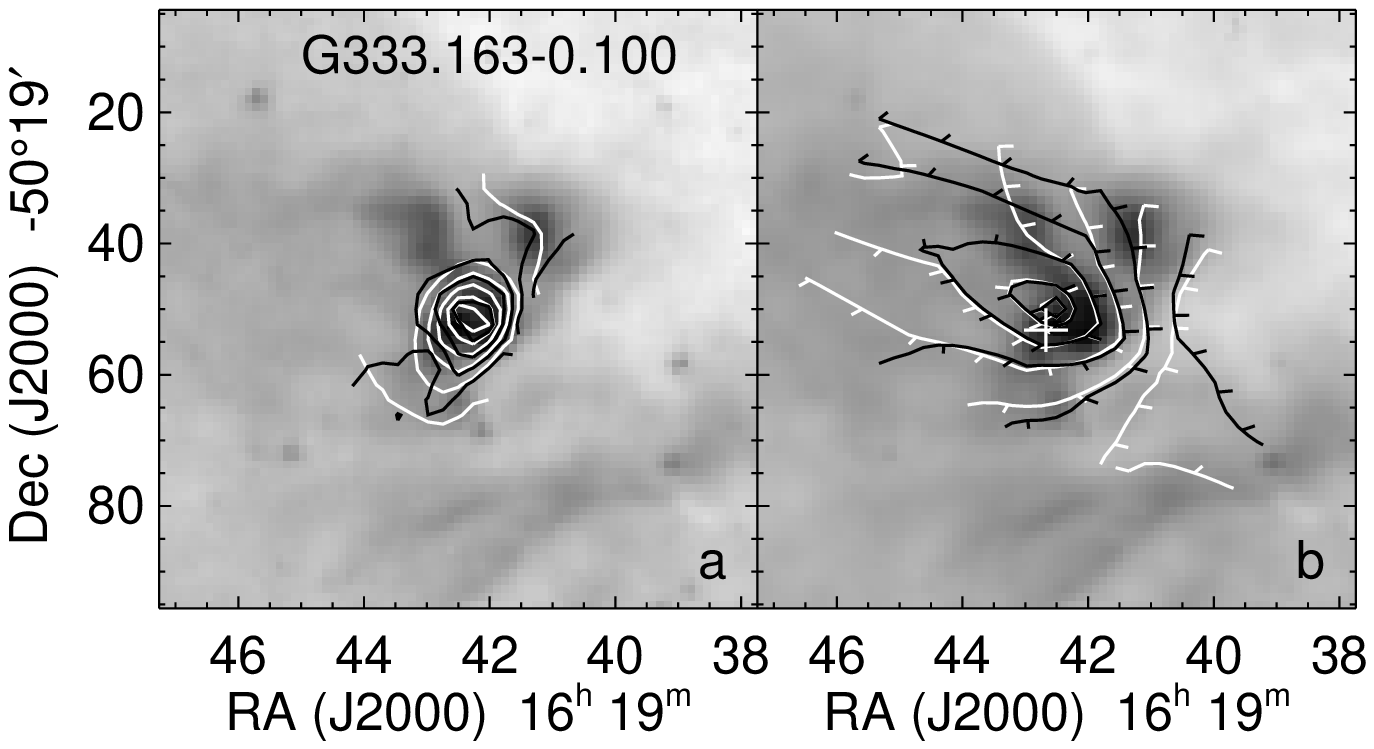}
\caption{Images of G333.163$-$0.100 with emission line contours.
Panel (a): the \mbox{[Ne\,{\sc ii}]} 12.8~\micron\ line contours 
are plotted in white and 
the \mbox{[Ne\,{\sc iii}]} 15.6~\micron\ line contours 
are plotted in black on the IRAC band 3 (5.8~\micron) image.
Panel (b): the \mbox{[S\,{\sc iii}]} 33.5~\micron\ line contours 
are plotted in white and 
the \mbox{[Si\,{\sc ii}]} 34.8~\micron\ line contours 
are plotted in black on the IRAC band 4 (8.0~\micron) image.
The white cross marks the location of the 6.7 GHz methanol maser (Caswell 2009).
}
\end{figure*}

%Figure 24
%\clearpage
\begin{figure*}
\includegraphics[width=150mm]{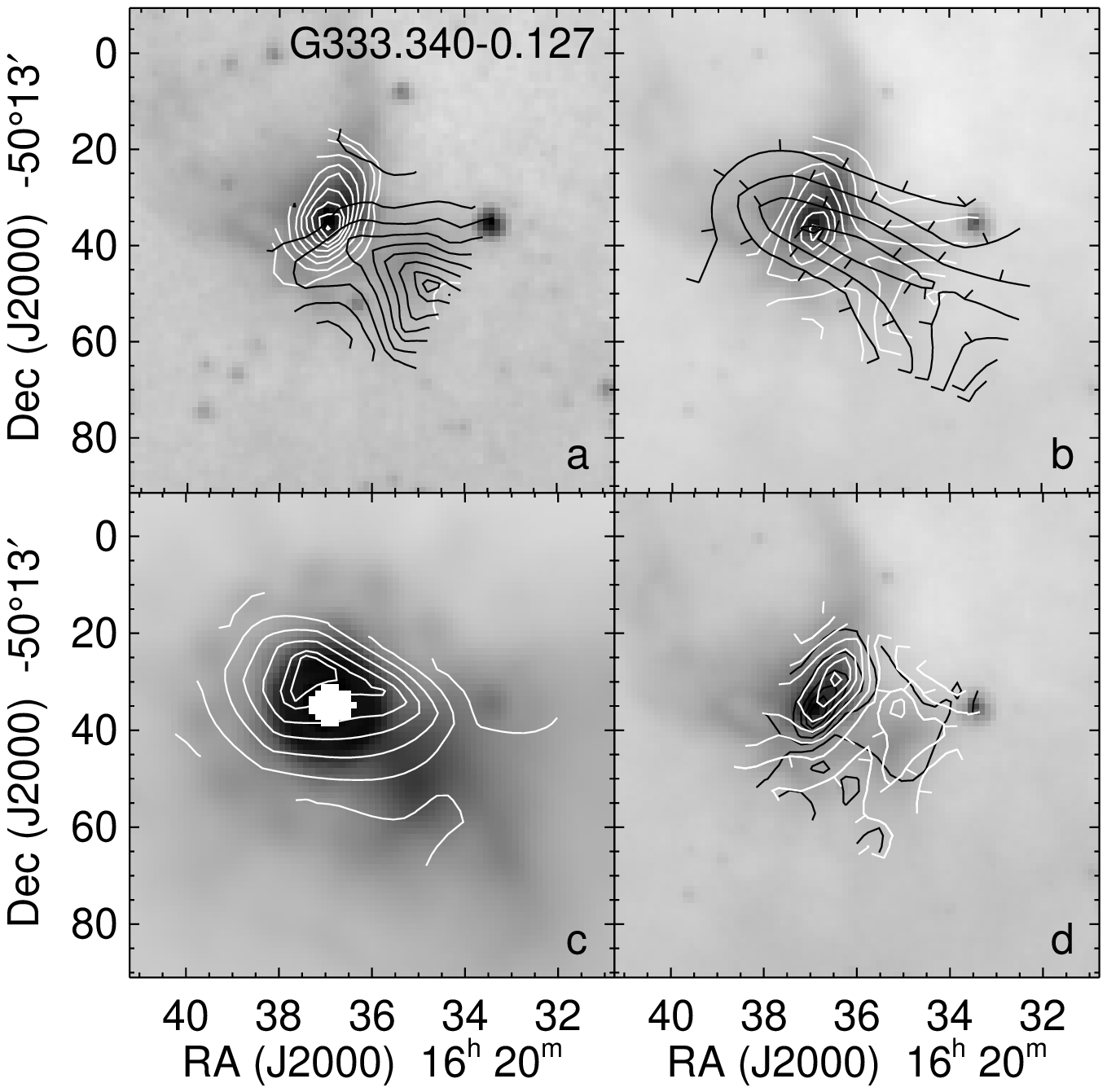}
\caption{Images of G333.340$-$0.127 with emission line contours.
Panel (a): the \mbox{[Ne\,{\sc ii}]} 12.8~\micron\ line contours 
are plotted in white and 
the \mbox{[Ne\,{\sc iii}]} 15.6~\micron\ line contours 
are plotted in black on the IRAC band 3 (5.8~\micron) image.
Contours of the \mbox{[S\,{\sc iv}]} 10.5~\micron\ line look similar to the \mbox{[Ne\,{\sc iii}]} line contours.
Panel (b): the \mbox{[S\,{\sc iii}]} 18.7~\micron\ line contours 
are plotted in white and 
the \mbox{[S\,{\sc iii}]} 33.5~\micron\ line contours 
are plotted in black on the IRAC band 4 (8.0~\micron) image.
Contours of the \mbox{[Fe\,{\sc iii}]} 22.9~\micron\ line look similar to the \mbox{[S\,{\sc iii}]} 33.5~\micron\ line contours.
Panel (c): the \mbox{[Si\,{\sc ii}]} 34.8~\micron\ line contours 
are plotted in white on the MIPS 24~\micron\ image.
Contours of the \mbox{[Fe\,{\sc ii}]} 26.0~\micron\ line look similar to the \mbox{[Si\,{\sc ii}]} line contours.
The centre of the MIPS image is saturated.
Panel (d): the H$_2$ S(2) 12.3~\micron\ line contours 
are plotted in white and 
the H$_2$ S(1) 17.0~\micron\ line contours 
are plotted in black on the IRAC band 4 (8.0~\micron) image.
}
\end{figure*}

%Figure 25
%\clearpage
\begin{figure*}
\includegraphics[width=150mm]{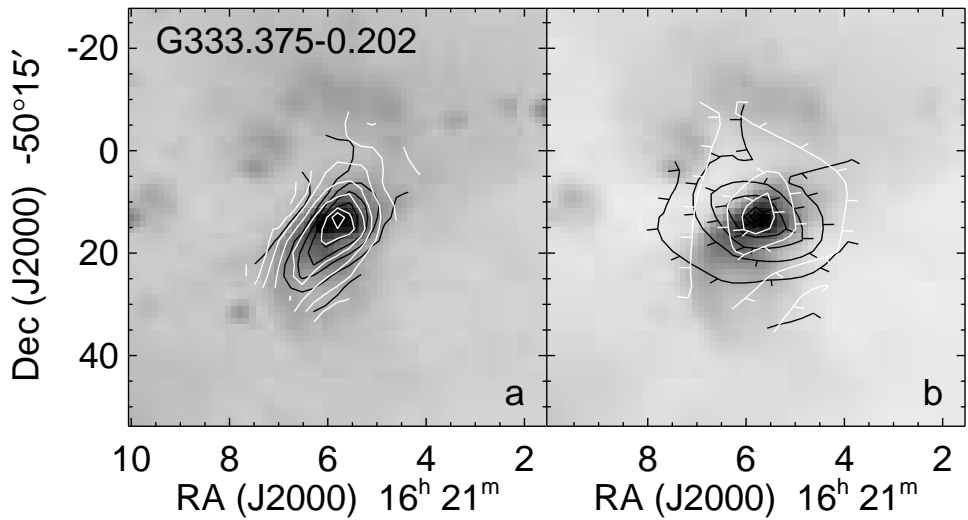}
\caption{Images of G333.375$-$0.202 with emission line contours.
Panel (a): the H$_2$ S(2) 12.3~\micron\ line contours 
are plotted in white and 
the \mbox{[Ne\,{\sc ii}]} 12.8~\micron\ line contours 
are plotted in black on the IRAC band 3 (5.8~\micron) image.
Panel (b): the \mbox{[Si\,{\sc ii}]} 34.8~\micron\ line contours 
are plotted in white and 
the \mbox{[S\,{\sc iii}]} 33.5~\micron\ line contours 
are plotted in black on the IRAC band 4 (8.0~\micron) image.
}
\end{figure*}

%Figure 26
%\clearpage
\begin{figure*}
\includegraphics[width=150mm]{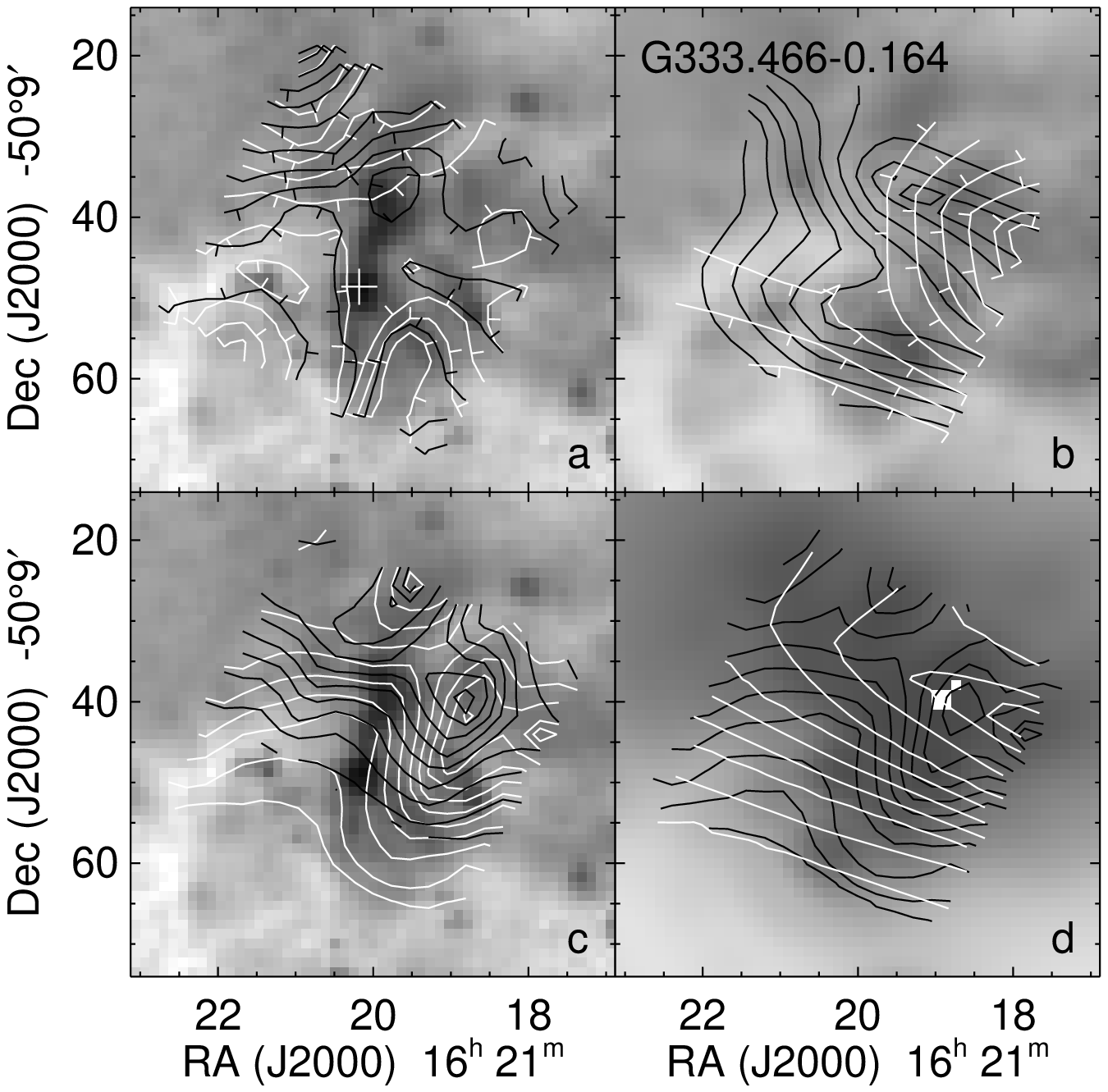}
\caption{Images of G333.466$-$0.164 with emission line contours.
Panel (a): the H$_2$ S(2) 12.3~\micron\ line contours 
are plotted in white and 
the H$_2$ S(1) 17.0~\micron\ line contours 
are plotted in black on the IRAC band 2 (4.5~\micron) image.
The white cross marks the location of the 6.7 GHz methanol maser (Caswell 2009);
this is also the location of the YSO with the ice absorption features.
Panel (b): the \mbox{[Si\,{\sc ii}]} 34.8~\micron\ line contours 
are plotted in white and 
the \mbox{[Fe\,{\sc ii}]} 26.0~\micron\ line contours 
are plotted in black on the IRAC band 4 (8.0~\micron) image.
Panel (c): the \mbox{[Ne\,{\sc ii}]} 12.8~\micron\ line contours 
are plotted in white and 
the \mbox{[Ne\,{\sc iii}]} 15.6~\micron\ line contours 
are plotted in black on the IRAC band 2 (4.5~\micron) image.
The \mbox{[Ar\,{\sc ii}]} 7.0~\micron\ line contours look similar to the \mbox{[Ne\,{\sc ii}]} contours.
Note the displacement of the ionized lines from the bright outflow emission, but there is a secondary contour maximum off the northern tip of the outflow.
Panel (d): the \mbox{[S\,{\sc iii}]} 33.5~\micron\ line contours 
are plotted in white and 
the \mbox{[S\,{\sc iii}]} 18.7~\micron\ line contours 
are plotted in black on the MIPS 24~\micron\ image.
Note that the \mbox{[S\,{\sc iii}]} line contours peak at the brightest (saturated) pixels of the MIPS image.
}
\end{figure*}

There are two outflow sources (G333.131 and G333.466) in Table 5 whose contours 
show the local production of forbidden lines. 
However, in neither of the outflow sources are the contours of maximum 
forbidden line emission located near the outflow source but are located at 
some distance, at least 20~arcsec but possibly further since the maps don't quite cover 
the region of forbidden line emission.
The G333.131 \mbox{H\,{\sc ii}} region is not imaged in this group of figures because 
the forbidden lines measured in most telescope pointings do not have good enough S/N -- 
the only telescope pointings with significant forbidden lines are in the bright fuzzy region 
north of Dec $-$50\degr~40\arcmin~35\arcsec\ (Fig.~2e).
These two sources both have the appearance of clusters in their 24~\micron\ MIPS images
(Figs. 2e and 26d) 
and thus the forbidden line emission may be associated with some star 
other than the YSO producing the outflow.
In the G333.466 source, the forbidden lines have their maxima at the west side of the image, 
but there is a secondary maximum at the northern tip of the outflow 
seen in the 4.5~\micron\ IRAC image (Fig. 26c).
In Table 5 and Fig. 27 these regions are described as 
G333.466$-$0.164W and G333.466$-$0.164N, respectively.

%Figure 27
%\clearpage
\begin{figure*}
\includegraphics[width=150mm]{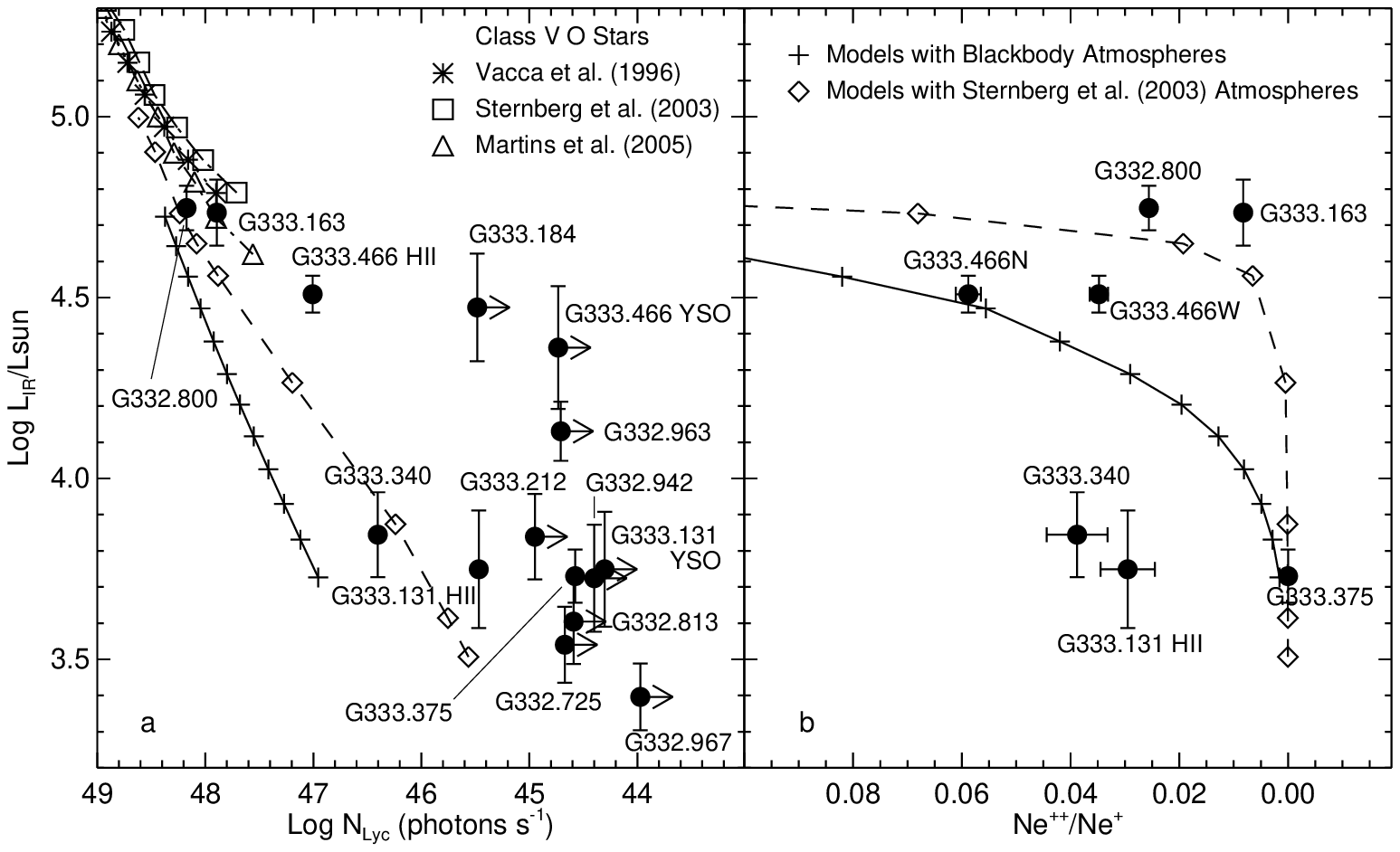}
\caption{
Panel (a): source luminosity plotted vs. the numbers of Lyman continuum photons per second 
for the six sources producing their own ionizing photons and 
3 $\sigma$ upper limits for the nine YSOs without measurable \mbox{H\,{\sc ii}} regions.
Estimated luminosities and $N_{\rm Lyc}$ for O stars of luminosity class V as determined by 
Vacca, Garmany, \& Schull (1996), Sternberg, Hoffmann, \& Pauldrach (2003), and Martins, Schaerer, \& Hillier (2005) are also plotted.
The connected diamonds and plus signs are models for \mbox{H\,{\sc ii}} regions ionized by ZAMS stars 
(lower luminosities than luminosity class V stars) 
with $T_{\rm eff}$ ranging from 25000 K to 36000 K for black body SEDs 
and $T_{\rm eff}$ ranging from 32000 K to 41000 K for SEDs from Sternberg et al. (2003).
The source luminosities are the average of the best fitting models from the SED fitter 
and the uncertainties are the standard deviations of these fits.
An additional uncertainty due to uncertainty in the distance $\sim 10$\% is not plotted.
Panel (b): source luminosity plotted vs. the neon ionic ratio for the seven \mbox{H\,{\sc ii}} regions 
(note that there are two positions for the G333.466 \mbox{H\,{\sc ii}} region, see text).
}
\end{figure*}

We analysed the forbidden lines and the H 7-6 recombination line 
using the code described by Simpson et al. (2007) with updates from Table~2 
for an assumed electron temperature, $T_e$, of 7000 K.
The line fluxes are given in Table A3.
The results are also in Table 5 for electron density, $N_e$, estimated from the 
ratio of the \mbox{[S\,{\sc iii}]} 18.7/33.5~\micron\ lines, 
and the ratios of various neon, sulphur, and H$^+$ ions.
The IR forbidden lines have very low temperature sensitivity.
However, the H$^+$ recombination line is sensitive to $T_e$ such that 
the ionic abundances with respect to H$^+$ are approximately proportional to $T_e^{-1}$.
Thus the fact that the Ne/H and S/H abundances are generally not very different from 
what one would expect for sources closer to the Galactic Centre than the Orion Nebula 
(Orion Nebula Ne/H = $1.01 \times 10^{-4}$ and S/H = $7.7 \times 10^{-6}$, Rubin et al. 2011),
given the Galactic abundance gradients (e.g., Simpson et al. 1995)
and the abundances in G333.6$-$0.2 
(Ne/H = $3.4 \times 10^{-4}$ and S/H = $1.3 \times 10^{-5}$, Simpson et al. 2004),
leads us to infer that $T_e = 7000$~K is a reasonable value.
Such an electron temperature is close to that expected for photoionized \mbox{H\,{\sc ii}} regions 
with these abundances and much lower than the temperatures predicted for 
gas ionized by shocks.

We also used the observed H 7-6 fluxes to predict the radio flux at 5 GHz, $S_5$, 
and the observed \mbox{[Ne\,{\sc ii}]} 12.8~\micron\ flux to estimate the number of 
ionizing photons, $N_{\rm Lyc}$, using equation 1 of Simpson \& Rubin (1990).
These parameters can also be estimated from the observed forbidden line fluxes, 
which is useful because the forbidden lines 
are much stronger than the radio flux or IR hydrogen recombination lines,
both of which are often not detectable or detectable only with very large uncertainties.

We start with equation 1 of Rubin (1968) but multiply the left side by 
$[1 + f_i<\rm He^+/(\rm H^+ + \rm He^+)>]$  (as was done by Simpson \& Rubin 1990), to get
\begin{equation}
N_{\rm Lyc} {\left[1 + f_i\left<{\rm He^+ \over (\rm H^+ + \rm He^+)}\right>\right]} = \int_0^{R_S} \alpha_{\rm B} N_e^2 dV,
%(1 + f_i<He^+/(H^+ + He^+)>) N_{\rm Lyc} = \int_0^{R_S} N_pN_e(\beta - \beta_1) dV.
\end{equation}
that is, the number of stellar photons in the Lyman continuum able to ionize the hydrogen
in an \mbox{H\,{\sc ii}} region, $N_{\rm Lyc}$,
augmented by the ionizing photons from He$^+$ recombination, is equal to 
the total number of recombinations to neutral H and He (ignoring heavy elements) 
integrated over the whole Str\"{o}mgren sphere of radius $R_S$ and volume $V$, 
where $N_e$ is the electron density and 
$f_i$ ($\approx 0.65$) is the fraction of helium recombination photons to excited states 
which are energetic enough to ionize hydrogen (Simpson \& Rubin 1990).  
For $T_e$ from 5000 K to 10,000 K and log~$N_e \sim 2.5$, 
the total recombination coefficient for Case B, $\alpha_{\rm B}$, 
equals $\sim 2.587 \times 10^{-13} t^{-0.81}$ cm$^3$ s$^{-1}$ for H 
(Storey \& Hummer 1995)
and $\sim 2.755 \times 10^{-13} t^{-0.78}$ cm$^3$ s$^{-1}$ for He (Hummer \& Storey 1998), 
where $t = T_e/10^4$.

We consider some heavy element X, with abundance with respect to hydrogen $N_{\rm X}/N_{\rm H}$,  
which has an ionized state with ionic density $N_i$ 
that emits a collisionally excited forbidden line with volume emissivity $\epsilon$,
$\epsilon = h \nu A (N_{upper}/N_i) N_i$,
where $h$ is Planck's constant, $\nu$ is the frequency of the transition, $A$ is 
the Einstein transition probability, and $N_{upper}$ is the density of the ions in 
the upper level of the transition.
For computational convenience we define
$\epsilon^\prime$
such that its only dependencies are $N_e$, $T_e$, and the atomic physics parameters (Table~2).
With that we have 
\begin{equation}
\epsilon = \epsilon^\prime {N_i \over N_{\rm X}} {N_{\rm X} \over N_{\rm H}} {N_p \over N_e},
\end{equation}
where $N_p$ is the proton density. 
In our code we divide $\epsilon^\prime$ by $N_e^2$ because the emissivity at low densities
is proportional to $N_e^2$ and the resulting $\epsilon^\prime/N_e^2$ has little density
dependence until the density is high enough to produce collisional de-excitation.
This collisional de-excitation allows estimation of $N_e$ from the ratios of 
two lines from the same element, such as S$^{++}$ (e.g., Simpson et al. 2004).

The flux from this line, observed over the whole emitting region, is given by
\begin{equation}
F = {1 \over {4 \pi D^2}} \int {\epsilon^\prime \over N_e^2} {N_i \over N_{\rm X}} {N_{\rm X} \over N_{\rm H}} {N_p \over N_e} N_e^2 dV
\end{equation}
where D is the distance to the \mbox{H\,{\sc ii}} region, which we assume is fully ionized ($N_p = N_{\rm H}$).
%Substituting, we get
Assuming uniform excitation and substituting the $\int N_e^2 dV$ derived from eq. (3) into eq. (1), we get 
\begin{equation}
N_{\rm Lyc} \gid {{2.587 \times 10^{-13} t^{-0.81} 4 \pi D^2 F} \over {\left[{\left<{\epsilon^\prime \over N_e^2}\right>}{\left<{N_i \over N_{\rm X}}\right>}{\left<{N_{\rm X} \over N_{\rm H}}\right>}{\left<{N_p \over N_e}\right>}\right]}{\left[1 + f_i\left<{\rm He^+ \over (\rm H^+ + \rm He^+)}\right>\right]}}
\end{equation}
where 
the brackets on $<N_i/N_{\rm X}>$, etc., indicate an average over the emitting volume 
(see Simpson et al. 1995, 2004, for further examples of the notation),
and the lower limit arises because of the possibility that the measured flux originates 
from less than the total ionized volume.
Here we have assumed the $\alpha_B$ for H since the correction 
for the different value for He as given above is so small. 

For use with forbidden lines 
(when no reliable hydrogen recombination line or radio flux is available), 
it is preferable to use a line that gives the least 
amount of uncertainty -- that is, low extinction, 
ionization correction factor ($<N_{\rm X}/N_i>$) close to unity, 
and relatively well-known abundance ($<N_{\rm X}/N_{\rm H}>$).
For the MIR observations of low-excitation nebulae (no He$^+$) discussed in this paper,
a good line is the 12.8~\micron\ line of Ne$^+$,
which has very little density sensitivity.
For the line parameters  and references in Table~2 we calculate 
$\epsilon^\prime/N_e^2 \approx 9.38 \times 10^{-22} t^{-0.275}$ erg cm$^3$ s$^{-1}$ 
for log~$N_e = 2.5$ ($N_e$ in cm$^{-3}$) and $T_e$ between 5000~K and 10000~K.
This gives 
\begin{equation}
N_{\rm Lyc} \gid 3.30 \times 10^{55} t^{-0.525} D_{\rm kpc}^2 F_{12.8} {\left<{N_{\rm Ne} \over N_{\rm Ne^+}}\right>}{\left<{N_{\rm H} \over N_{\rm Ne}}\right>}{\left<{N_e \over N_p}\right>}
\end{equation}
where $F_{12.8}$ is the observed flux of the \mbox{[Ne\,{\sc ii}]} 12.8~\micron\ line in W m$^{-2}$
and $D_{\rm kpc}$ is the distance in kpc.

In Table 5 we estimate $N_{\rm Lyc}$ from the observed neon fluxes 
and predict radio fluxes from the H 7-6 recombination line flux. 
Here we assumed a Ne/H ratio of $1.5 \times 10^{-4}$ 
(the weighted average of the Ne/H abundance ratios in Table 5), 
$T_e = 7000$~K, and $N_e = 316$ cm$^{-3}$, and used 
the volume emissivities computed by the forbidden line analysis program.
Since there are ionized lines in all spectra from the background, 
background fluxes were first subtracted.
These background fluxes may be overestimates 
since they were taken from positions relatively close to the sources 
where the intensity is at a minimum 
(see the plots of the SH slit locations in Figs. 8 -- 18), 
and consequently the estimates of $S_5$ and $N_{\rm Lyc}$ could be underestimates.
This is particularly true of the two sources with the smallest $N_{\rm Lyc}$, 
where the background intensity is a sizable fraction of the average source intensity.

Assuming, then, that the sources can be analysed as photoionized \mbox{H\,{\sc ii}} regions, 
we compare the estimated ionic ratios to ratios that we compute with 
{\sc cloudy}, version 08 (Ferland et al. 1998).
We computed models assuming that the ionizing sources can be described as 
zero-age main-sequence (ZAMS) stars, since such stars have the hottest 
effective temperatures, $T_{\rm eff}$ for their luminosities.
We took the stellar radius-$T_{\rm eff}$-luminosity relation for ZAMS stars from the 
`canonical' Z = 0.02 pre-main-sequence evolution tracks of Bernasconi \& Maeder (1996)
and used both black bodies and the log $g = 4$ atmospheres of 
Sternberg et al. (2003) 
for $T_{\rm eff} = 25000$ K to 36000 K or 40000 K. 
{\sc cloudy} was used to compute $N_{\rm Lyc}$ vs. $T_{\rm eff}$ for each atmosphere.

In Fig. 27(a) we plot the computed $N_{\rm Lyc}$ vs. luminosity for the ZAMS stars along with 
the results for our sources from Table 5.
Luminosity is plotted as the ordinate and $N_{\rm Lyc}$ as the abscissa because 
$N_{\rm Lyc}$ can be thought as a proxy for $T_{\rm eff}$.
Upper limits for the non-ionizing sources were estimated from the neon flux uncertainties.
We also plot the $N_{\rm Lyc}$ vs. luminosity relation for luminosity class V stars 
from Vacca et al. (1996), Sternberg et al. (2003), 
and Martins et al. (2005) in Fig. 27(a).
Because our sources lie in the region of the ZAMS stars on the plots,
we again conclude that it is likely that the sources are ionized by 
stellar photons and not shocks.

In Fig. 27(b) we plot the observed Ne$^{++}$/Ne$^+$ ratios vs. the FIR luminosities 
and the ratios predicted by the same models.
It is notoriously difficult to get models that predict the observed Ne$^{++}$/Ne$^+$ ratios 
because of the extreme sensitivity of the Ne$^{++}$ ionization level 
(which requires 41 eV) 
to details of the input stellar atmosphere models 
(e.g., Simpson et al. 2004; Rubin et al. 2008). 
In general, our observations indicate higher Ne$^{++}$/Ne$^+$ ratios 
than are predicted by the ZAMS star models 
(\mbox{H\,{\sc ii}} region models using luminosity class V star atmospheres would have even lower 
Ne$^{++}$/Ne$^+$ ratios at given luminosities because of the lower stellar $T_{\rm eff}$). 
Blackbody atmospheres were included for this project because 
their use in \mbox{H\,{\sc ii}} region models gives better agreement with observations 
than other model atmospheres. 
Most of the sources plotted in Fig. 27(b) have Ne$^{++}$/Ne$^+$ ratios lower than 
predicted by the blackbody models; 
however, the fact that two sources, G333.340 and the \mbox{H\,{\sc ii}} region north of G333.131, 
have higher Ne$^{++}$/Ne$^+$ ratios than even those predicted by the blackbody models 
may indicate an additional source of ionization, such as shocks. 
This will be discussed in Section 3.6.

\subsection{Molecular hydrogen lines}

It is not simple to derive the PDR or shock properties from the observed lines 
for several reasons: (1) The main indicators in a general sense are the ratios 
of the H$_2$ rotational lines, but only the first three [S(0), S(1), and S(2)] 
are readily detectable with the SH and LH modules' $\lambda/\delta \lambda = 600$ 
resolution, since the SL module's resolution is only $64 - 128$.
(2) The H$_2$ S(3) line falls in the bottom of the 9.6~\micron\ silicate absorption feature 
and the S(5) line is blended with the usually strong \mbox{[Ar\,{\sc ii}]} 6.98~\micron\ line.
(3) Other lines, such as \mbox{[Fe\,{\sc ii}]} 26.0~\micron\ and \mbox{[Si\,{\sc ii}]} 34.8~\micron, 
which are important PDR indicators (Kaufman, Wolfire, \& Hollenbach 2006), 
are also found in \mbox{H\,{\sc ii}} regions.
Moreover, one must assume some depletion factor onto grains.
(4) Finally, and most importantly, all the above mentioned lines also arise in shocks.
Considering the first two items, it is perhaps not surprising that we detect 
the S(3), S(5), and S(7) lines in only one source, G332.813$-$0.700, and there 
very poorly because of the strong continuum (see Fig. 3b), and no S(4) or S(6) at all.
We also detect the S(5) and S(7) lines in the G333.466$-$0.164 outflow, 
but no S(3).

The usual way of distinguishing shocks from PDRs is that shocks do not have much 
thermal continuum (e.g., Hollenbach, Chernoff, \& McKee 1989), unlike PDRs where the grains 
are heated by the stellar photons.
We cannot use this criterion here because of the strong FIR continuum from the YSOs.
Instead, we look at the morphology of the line maps, as we did for the ionized lines, 
and also compare the H$_2$ line ratios to theoretical models of both shocks and PDRs.
Contour plots of PDR/shock lines in outflow sources are shown in Figs. 28 - 30. 
We see no correlation of any of these lines with the outflows. 
We conclude that the morphology seen in the images is most likely due to 
the variations in extinction that are present in these sources. 
The best agreement is with the morphology of the PAH emission seen 
in the IRAC 8.0~\micron\ images, 
as would be expected if the H$_2$ lines are formed in PDRs like the PAHs. 

%Figure 28
%\clearpage
\begin{figure*}
\includegraphics[width=150mm]{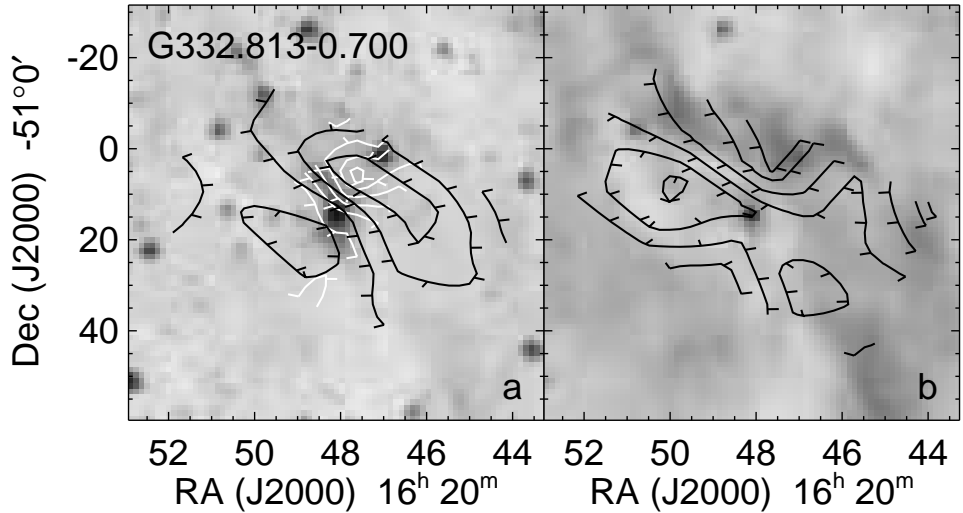}
\caption{Image of G332.813$-$0.700 with emission line contours.
Panel (a): the H$_2$ S(1) line contours are plotted in white and 
the H$_2$ S(0) line contours are plotted in black  
on the IRAC band 2 (4.5~\micron) image.
Panel (b): the \mbox{[Fe\,{\sc ii}]} 26.0~\micron\ line contours 
are plotted in black on the IRAC band 4 (8.0~\micron) image.
}
\end{figure*}

%Figure 29
%\clearpage
\begin{figure}
\includegraphics[width=84mm]{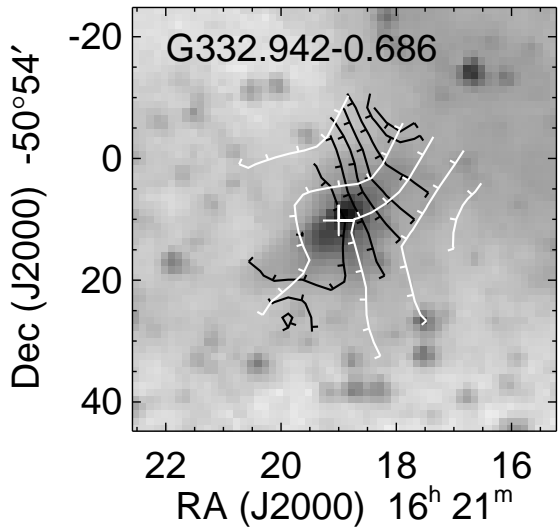}
\caption{Image of G332.942$-$0.686 with emission line contours.
The \mbox{[Si\,{\sc ii}]} 34.8~\micron\ line contours 
are plotted in white and the H$_2$ S(1) line contours 
are plotted in black on the IRAC band 2 (4.5~\micron) image.
It is curious that these contours appear orthogonal, 
because it usually expected that both lines are formed in PDRs, 
that is, in the same gas.
The white cross marks the location of the 6.7 GHz methanol maser (Caswell 2009).
}
\end{figure}

%Figure 30
%\clearpage
\begin{figure}
\includegraphics[width=84mm]{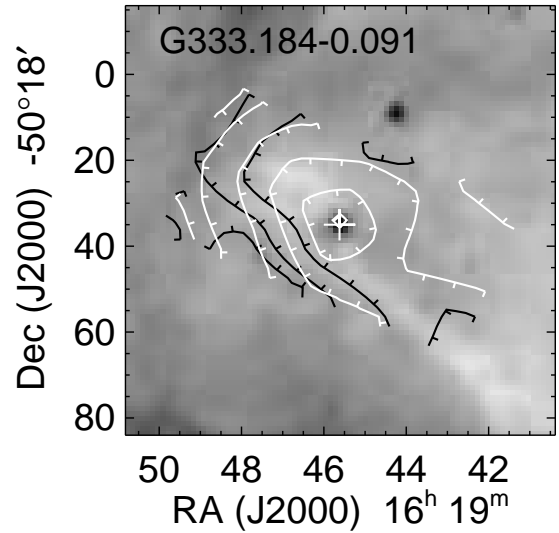}
\caption{Image of G333.184$-$0.091 with emission line contours.
The \mbox{[Si\,{\sc ii}]} 34.8~\micron\ line contours 
are plotted in white  
and the H$_2$ S(0) line contours are plotted in black on the IRAC band 4 (8.0~\micron) image.
The white cross marks the location of the 6.7 GHz methanol maser (Caswell 2009).
}
\end{figure}

Compared to the models, we typically see that the sources have
more H$_2$ S(0) relative to S(1) than the shock models of Hollenbach \& McKee (1989) 
and Flower \& Pineau des For\^ets (2010) 
and more H$_2$ S(2) relative to S(1) than the PDR models of 
Kaufman et al. (1999, 2006)\footnote{http://dustem.astro.umd.edu/index.html}.
We consider only the shock models with velocities less than 40 km s$^{-1}$ 
for the outflow YSOs with localized H$_2$ 
because at higher velocities, the models predict detectable lines of ionized elements, 
which are not observed (see the next two sections for possible exceptions).
The higher H$_2$ S(0)/S(1) ratio suggests that there is more lower temperature gas 
present (at $\sim 100$~K) than produced by shocks, which are generally 
much hotter.

The PDR models of Kaufman et al. (1999, 2006) 
all have S(2)/S(1)$< 1$ (the observed ratio is typically $\sim 1$)
except for the extreme model range of very high density, $n$ ($n \sim 10^{4.5}$ cm$^{-3}$), 
and either very high or very low $G_0$ 
(where $G_0$ is defined as the incident ultraviolet flux from 6.0 to 13.6 eV 
in units of $1.6 \times 10^{-3}$ erg cm$^{-2}$ s$^{-1}$).
The models with the S(1)/S(0) ratio in the range of $\sim 0.2$ to $\sim 3$ 
(the majority of the data) 
have either $n \sim 10^{3}$ cm$^{-3}$ and $G_0 \ga 10^{2.5}$ 
or $n \ga 10^{3}$ cm$^{-3}$ and $G_0 \la 10^{2}$.
We conclude that the observed H$_2$ ratios are in reasonable agreement 
with the high density, low $G_0$ models of Kaufman et al. (2006),
especially if we consider multiple components along the line of sight.
In fact, we should expect to see PDR H$_2$ line emission from our spectra, 
given the large amount of PAH emission seen in our spectra and in the IRAC 8.0~\micron\ images.
This PDR emission probably originates in the cloud as a whole and 
is not local to the outflow YSOs, which are all seen to be embedded in 
local dust lanes of higher extinction. 

\subsection{Shocks}

Both the red sources and the outflow sources show line emission indicative of shocks.
Although the peak of the low excitation forbidden lines is at the location of the YSO 
in the red sources G332.800 and G333.340, 
in both of these sources the high excitation forbidden lines 
peak well off the YSO (Figs. 22 and 24).
This is the opposite of what is expected for an \mbox{H\,{\sc ii}} region, 
where ions requiring higher energy photons for photoionization 
have much smaller Str\"omgren spheres immediately surrounding the ionizing star 
than ions needing the much more abundant photons with energies 
%$\gtrsim 13.6$ eV.
%$\ga 13.6$ eV.
only a little larger than 13.6 eV.
This is particularly apparent in the \mbox{[Ne\,{\sc ii}]} 12.8~\micron\ and the \mbox{[Ne\,{\sc iii}]} 15.6~\micron\ 
lines (panel a of both figures), which have similar extinction and critical densities.
Ions of intermediate ionization potentials, such as \mbox{[S\,{\sc iii}]} (panels b of Figs. 22 and 24), 
have intermediate morphologies.
We suggest that the offset high excitation lines are produced in 
an outflow that is shocking the surrounding envelope or ISM.
The alternative explanation, that a wind has blown out a cavity and 
the stellar photons are ionizing the cavity edges, is not in agreement 
with the observation that the highest excited lines (\mbox{[Ne\,{\sc iii}]} and \mbox{[S\,{\sc iv}]}) 
are found farther from the star than lower excitation lines like \mbox{[S\,{\sc iii}]}.

The various line ratios can be used with models to estimate the shock velocities, 
given the other necessary model parameters such as density and magnetic field,
neither of which we know.
For these two red sources, the most obvious shocked lines are \mbox{[Ne\,{\sc iii}]} 15.6~\micron\ 
and \mbox{[S\,{\sc iv}]} 10.5~\micron\ (similar morphology to the \mbox{[Ne\,{\sc iii}]} lines);
the lower excitation lines like \mbox{[Ne\,{\sc ii}]} 12.8~\micron, \mbox{[Ar\,{\sc ii}]} 7.0~\micron, 
\mbox{[Fe\,{\sc ii}]} 26.0~\micron, 
and \mbox{[Si\,{\sc ii}]} 34.8~\micron\ have extensions along the line of the \mbox{[Ne\,{\sc iii}]} emission.
The upper limit to the velocity is strongly constrained by the fact that there 
is no significant emission of the higher excitation lines of 
\mbox{[O\,{\sc iv}]} 25.8~\micron\ or [Ne V] 14.3 or 24.3~\micron.
Only H$_2$ S(2) appears to be affected by the shock in G333.340 (Fig. 24d) 
but no H$_2$ lines are affected in G332.800 (Fig. 22d). 

We find in the literature J-shock models for shock velocities 
of 30 - 150 km s$^{-1}$ (Hollenbach \& McKee 1989) 
and  100-1000 km s$^{-1}$ (Mappings III, Allen et al 2008). 
None of these models agrees with the data for G333.340 because 
all shock models with the \mbox{[Ne\,{\sc iii}]} 15.6~\micron\ line 
stronger than the \mbox{[Ne\,{\sc ii}]} 12.8~\micron\ line 
also have strong enough \mbox{[O\,{\sc iv}]} 25.85~\micron\ emission that we would have 
detected the line. 
Consequently, we must estimate shock velocities near the bottom of the range 
of model velocities, that is,  $\sim 100$ km s$^{-1}$, 
but emphasize that 
this value has a large uncertainty because we do not 
know the extremely important magnetic field parameter 
and because of the lack of fit. 
Further modelling is beyond the scope of this paper.

G333.163 and G333.375 also have the emission extended away from the YSO, 
although it is much less definite that the extension is due to an outflow 
(Figs. 23 and 25).

The shock velocities for the outflow sources are much lower. 
These sources have no forbidden line emission from any ions 
requiring as much as 13.6 eV for ionization
(although perhaps there is some emission beyond the end of the outflow in
G333.466, Fig. 26c). 
For these sources the outflows, 
as identified by the excess IRAC 4.5~\micron\ emission, 
are where we detect the \mbox{[S\,{\sc i}]} 25.25~\micron\ line. 
The \mbox{[S\,{\sc i}]} 25.25~\micron\ line is found only in shocks as the result of shock 
dissociation of molecules -- the ionization potential of S$^+$ 
is only 10.4 eV, such that any atomic sulphur is always 
ionized in the diffuse ISM. 
The shock models of Hollenbach \& McKee (1989) produce 
substantial \mbox{[S\,{\sc i}]} 25.25~\micron\ line emission, which 
is seen in both YSO outflows (e.g., Haas, Hollenbach, \& Erickson 1991) 
and in supernova remnants (e.g., Neufeld et al. 2009). 

\setcounter{table}{5}
\begin{table}
\centering
\begin{minipage}{70mm}
\caption{Maximum observed intensities of the \mbox{[S\,{\sc i}]} 25.25 \micron\ line.
The \mbox{[S\,{\sc i}]} lines in G332.725$-$0.620 and G332.813$-$0.700 
were measured with {\sc smart} for the centre positions in the LH map  
(LH aperture is 247.5-arcsec) because they needed defringing.
The \mbox{[S\,{\sc i}]} lines in the other sources were extracted with {\sc cubism}
with smaller extraction apertures in order to estimate the maximum intensities,
which occur in the outflows.
}
\begin{tabular}{@{}lcc@{}}
\hline
Source & Position & Intensity \\
%       &          & $\times 10^{-9}$ \\
       &          & (W m$^{-2}$ sr$^{-1}$) \\
\hline
G332.725$-$0.620 & Centre  & $8 \pm 1 \times 10^{-9}$  \\
G332.813$-$0.700 & Centre  & $4 \pm 1 \times 10^{-9}$  \\
G332.942$-$0.686 & Outflow  & $11 \pm 2 \times 10^{-9}$  \\
G333.131$-$0.560 & Outflow & $7 \pm 1 \times 10^{-9}$  \\
G333.466$-$0.164 & Outflow  & $18 \pm 4 \times 10^{-9}$  \\
\hline
\end{tabular}
\end{minipage}
\end{table}

Estimates of the maximum surface brightnesses 
of the five sources with detected \mbox{[S\,{\sc i}]} 25.25~\micron\ lines are given in Table~6. 
This line can be difficult to detect because of 
the small line/continuum ratio at the {\it Spitzer} resolution 
and the substantial ($\la 5$ per cent) fringing at this wavelength 
in the LH spectra with strong continua. 
For two of the sources, the line was detected only in the central position 
and only after the spectra were extracted and defringed with {\sc smart}.
Fortunately, the \mbox{[S\,{\sc i}]} 25.25~\micron\ line occurs in two separate orders of the LH array, 
and thus we can require that we measure the line in {\it both} orders 
to deem that we have a detection.
For the other three sources, the outflow regions 
have much less continuum than the YSO positions;  
consequently, fringing is less important and 
the lines are easily seen in spectra extracted with {\sc cubism}.

We do not see any increase in H$_2$ emission 
at the location of the outflows and \mbox{[S\,{\sc i}]} emission,
although generally 
the H$_2$ line ratios agree with neither the PDR nor the shock models.
If we assume that the H$_2$ S(0) line arises in PDRs and the S(1) and S(2) 
in shocks, we can estimate that the shock velocity is between $\sim 10$ km s$^{-1}$
and $\sim 40$ km s$^{-1}$ for the S(2)/S(1) line ratio $\sim 1$,
using the models of Flower \& Pineau des For\^ets (2010).
The velocities cannot be higher than this because the models of 
Hollenbach \& McKee (1989) produce lines of \mbox{[Ne\,{\sc ii}]} 12.8~\micron,
which we do not detect in the outflow sources. 

\setcounter{table}{6}
\begin{table*}
\centering
\begin{minipage}{160mm}
\caption{IRAC surface brightnesses and H$_2$ intensities. 
In response to the suggestion that the IRAC Band 2 extended emission could be due 
to emission from H$_2$ in outflows, this table lists the IRAC Band 2 intensities in 
both outflow regions and nearby background positions, 
and the difference that is attributable to the outflow
in units of both MJy sr$^{-1}$ and W m$^{-2}$ sr$^{-1}$.
The IRAC bandwidth for Band 2 is 1.015 \micron\ at effective wavelength 4.493 \micron.
The last column lists the measured H$_2$ S(7) intensities or $3 \sigma$ upper limits at the same positions.
Since the background was not subtracted, the H$_2$ S(7) intensities are upper limits 
to the H$_2$ S(7) intensities in the outflow region,
especially since there is no or little apparent increase in the H$_2$ intensity at the outflow positions in the table.
The H$_2$ intensites at 4.5 \micron\ [H$_2$ S(9) and S(10)] could be slightly but not substantially 
more than the H$_2$ S(7) intensity, given the flat extinction law at these wavelengths (Indebetouw et al. 2005).
Exponents are written as follows: 2.43e-6 equals $2.43 \times 10^{-6}$.
}
\begin{tabular}{@{}lcccccc@{}}
\hline
YSO and Position  &  RA       &       Dec &  IRAC Band 2 Intensity & Position$-$Bkgr  & Position$-$Bkgr &  H$_2$ S(7)  \\ %        tau_9.6   Band 2 corr   Band 2 Intensity  H2 S(1) corr
                  &  (J2000)   &  (J2000) &     MJy sr$^{-1}$      &   MJy sr$^{-1}$    & W m$^{-2}$ sr$^{-1}$ & W m$^{-2}$ sr$^{-1}$  \\ %               MJy sr-1      W m-2 sr-1     W m-2 sr-1
\hline
G332.725$-$0.620  &&&&& \\
E                &  16:20:03.4 & $-$51:00:35 &    27.32  &         16.09  &       2.43e-6 & $<$ 1.7e-7 \\ %    1.52       25.7      3.9e-6            1.4e-8
E background     &  16:20:03.9 & $-$51:00:37 &    11.23 &&& \\
NW               &  16:20:02.1 & $-$51:00:29 &    47.80  &         33.77  &       5.09e-6 &  $<$ 8.5e-8  \\ %   0.52       39.7      6.0e-6            2.4e-8
NW background    &  16:20:01.5 & $-$51:00:35 &    14.03 &&& \\
  &&&&& \\
G332.813$-$0.700     &&&&& \\
Peak             &  16:20:48.1 & $-$51:00:15 &          &                 &                &  ($1.22 \pm 0.03$) e-7 \\ %
NE               &  16:20:48.5 & $-$51:00:07 &    51.54 &          41.39  &       6.24e-6  &  ($7.72 \pm 1.34$) e-8 \\ %   0.73       51.9      7.8e-6            3.0e-8
NE background    &  16:20:48.0 & $-$51:00:02 &    10.15 &&& \\
S                &  16:20:48.0 & $-$51:00:20 &    58.42 &          52.17  &       7.86e-6  &  $<$ 7.9e-9 \\ %    0.60       62.8      9.5e-6            1.3e-8
S background     &  16:20:47.9 & $-$51:00:30 &     6.25 &&& \\
  &&&&& \\
G332.942$-$0.686      &&&&& \\
NW               &  16:21:18.6 & $-$50:54:08 &    50.28  &         43.78  &       6.60e-6  &  $<$ 5.6e-8 \\ %    1.77       75.7      1.1e-5            5.5e-8
NW background    &  16:21:17.9 & $-$50:54:14 &     6.50 &&& \\
SE               &  16:21:19.2 & $-$50:54:13 &   100.00  &         94.68  &       1.32e-5  &  $<$ 9.0e-8 \\ %    2.05      178.4      2.7e-5            2.5e-8
SE background    &  16:21:18.4 & $-$50:54:27 &     5.32 &&& \\
SE-end           &  16:21:20.2 & $-$50:54:19 &     8.41  &          5.12  &       7.72e-7  &  $<$ 9.4e-8 \\ %    2.05        9.6      1.4e-6            3.4e-8
end background   &  16:21:20.8 & $-$50:54:17 &     3.29 &&& \\
  &&&&& \\
G332.963$-$0.679   &&&&& \\
S                &  16:21:23.0 & $-$50:53:03 &    22.40   &        18.65  &       2.81e-6  &  $<$ 6.8e-8 \\ %    2.81       44.4      6.7e-6            4.4e-8
S background     &  16:21:23.9 & $-$50:53:06 &     2.79  &&& \\                    
  &&&&& \\
G333.184$-$0.091     &&&&& \\
NW               &  16:19:45.3 & $-$50:18:33 &    70.65  &         67.36  &       1.02e-5  &  $<$ 2.0e-8  \\ %   3.76      215.4       3.2e-5            3.6e-8
NW background    &  16:19:47.2 & $-$50:18:22 &     3.28 &&& \\
SE               &  16:19:46.1 & $-$50:18:39 &    90.61  &         85.78  &      1.29e-5  &  $<$ 5.0e-8  \\ %   3.60      261.0       3.9e-5            5.3e-8
SE background    &  16:19:44.6 & $-$50:18:54 &     4.83 &&& \\
  &&&&& \\
G333.466$-$0.164    &&&&& \\
N                &  16:21:19.9 & $-$50:09:38 &    78.07  &         72.43  &       1.09e-5  &  ($7.68 \pm 1.05$) e-8 \\ %  4.44      285.8       4.3e-5            3.3e-8
N background     &  16:21:20.9 & $-$50:09:43 &     5.64 &&& \\
\hline
\end{tabular}
\end{minipage}
\end{table*}

\setcounter{table}{7}
\begin{table*}
\centering
\begin{minipage}{130mm}
\caption{H$_2$ lines in G332.813$-$0.700.
The H$_2$ lines were well-detected only in G332.813$-$0.700.
This table gives the line intensities at the YSO centre and at the position of the strongest lines, `north' 
(the north position is at  
RA 16$^{\rm h}$ 20$^{\rm m}$ 47\fs6, Dec. $-$51\degr 00\arcmin 6\arcsec
and the centre position is at RA 16$^{\rm h}$ 20$^{\rm m}$ 48\fs1, Dec. $-$51\degr 00\arcmin 15\arcsec).
The line intensities were corrected for extinction ($I_{\rm corr}$) at both positions with the measured  
$\tau_{9.6}$, but the relative intensities for the centre position ($I_{\rm corr}$ centre) 
show that the correction is much too large for the line with the largest correction, S(3).
The last column gives the corrected intensities for a much lower value of $\tau_{9.6}$;
we infer from the more reasonable line ratios that the H$_2$ lines originate from a region foreground of 
the YSO in the Center.
PDR parameters, $n$ (density in cm$^{-3}$), and $G_0$ (incident ultraviolet flux from 6.0 to 13.6 eV in units of 
$1.6 \times 10^{-3}$ erg cm$^{-2}$ s$^{-1}$), were estimated from the corrected line ratios for these 
two positions (models of Kaufman et al. 2006).
Exponents are written as follows: 1.88e-8 equals $1.88 \times 10^{-8}$.
}
\begin{tabular}{@{}lccccc@{}}
\hline

Line &  North   &      Centre   &        $I_{\rm corr}$ north &    $I_{\rm corr}$ centre &      $I_{\rm corr}$ centre \\
     &  W m$^{-2}$ sr$^{-1}$ & W m$^{-2}$ sr$^{-1}$ &  W m$^{-2}$ sr$^{-1}$ &   W m$^{-2}$ sr$^{-1}$ & W m$^{-2}$ sr$^{-1}$  \\  
     &                     &        &  $\tau_{9.6} = 0.79$ &  $\tau_{9.6} = 2.94$ &  $\tau_{9.6} = 0.80$ \\
\hline
S(0) &    1.88e-8  &   8.84e-9 &         2.52e-8  &      2.61e-8  &         1.19e-8 \\
S(1) &    2.58e-8  &   1.20e-8   &       3.84e-8   &     5.23e-8     &      1.79e-8 \\
S(2) &    3.89e-8  &   1.71e-8   &       5.10e-8   &     4.70e-8     &      2.25e-8 \\
S(3) &    1.08e-7  &   4.30e-8   &       2.36e-7   &     8.07e-7     &      9.55e-8 \\
S(5) &    8.87e-8  &   9.01e-8   &       1.13e-7   &     2.23e-7     &      1.15e-7 \\
S(7) &     -       &   1.22e-7   &         -       &     3.03e-7     &      1.56e-7 \\
&&&&& \\ 
PDR Parameters \\
$n$ &              &             &         5.6e4   &                 &        1.0e4 \\
$G_0$ &            &             &         5.6e1   &                 &        1.0e1 \\
\hline
\end{tabular}
\end{minipage}
\end{table*}

When the excess IRAC 4.5~\micron\ emission was first discovered, 
an immediate suggestion was that it could be due to high-J H$_2$ emission 
from shocks in the outflows 
(e.g., Smith et al. 2006; Cyganowski et al. 2008; Chambers et al. 2009).
In fact, De Buizer \& Vacca (2010) found that this must be the case 
for G19.88$-$0.53, which has very little continuum at 4.5~\micron\ and substantial 
H$_2$ emission.
(Their other source, G49.27$-$0.34, has only continuum at 4.5~\micron\ and no H$_2$ emission.)
However, this cannot be the case for our sources.
In Table 7 we list the continuum intensities that 
we have measured for the excess 4.5~\micron\ emission positions 
in the IRAC band 2 images and 
measurements or upper limits for the H$_2$ S(7) line at 5.5~\micron. 
In Table 8 we list our measurements of the H$_2$ lines 
that we measured in G332.813, the only source with detectable S(3) - S(7) lines. 
Even though we have measured only the H$_2$ S(1) to S(7) lines 
and the lines in the IRAC band 2 bandpass are S(9) and S(10), 
the measured 4.5~\micron\ continuum is too large by a factor of at least 10 
to be due to H$_2$ line emission. 
We conclude that some other source of emission must be the 
dominant contributor to the IRAC band 2 intensities. 
This could be either CO emission (e.g., Marston et al. 2004) 
or perhaps scattered light from the large grains 
that are sometimes found in the cores of star-forming regions 
(Pagani et al. 2010). 
Some scattered light is almost certainly present, since 
the excess 4.5~\micron\ emission regions 
are clearly present in the IRAC band 1 images at 3.6~\micron\ and 
also in the 2MASS $Ks$ and sometimes $H$ images. 

\subsection{Individual sources}

In this section we comment on individual sources.

(i) {\it G332.725$-$0.620.} This source is missing the SL first order spectrum (SL1) 
from 7.4 -- 14~\micron\ because the predicted saturation of the IRS Peak-Up arrays, 
which share the same detector array as the SL spectrograph module, 
would render all spectral pixels in the same rows on the detector array uncalibratable.
This is unfortunate because the SL1 wavelength range provides the best 
estimate of the depth of the 9.6~\micron\ silicate feature. 
Hence our estimates of the extinction in this source are more uncertain than the 
estimates for the other sources.

(ii) {\it G332.800$-$0.595.} This source is too extended to map with the IRS slits touching. 
The result is that there probably is missed structure in the maps 
and the position of the ionized line outflow is poorly defined (see Fig. 22).
It is also not known which of the two bright sources seen in the centre of the IRAC 
images (see Fig. 22 and Fig. 31 of the Supporting Information) is the YSO.
The source on the left is extended and is bright in 2MASS $J$, $H$, and $Ks$ bands.
The source on the right is compact and much redder, being only barely visible 
in the 2MASS J band.
The sources are approximately 10~arcsec apart, which means they cannot be distinguished 
in the 19~arcsec resolution MSX 21~\micron\ image (Price et al. 2001) of Fig.~22(c)  
(the MIPS 24~\micron\ image was saturated) 
or the MIPS 70~\micron\ image. 
The MSX Catalog gives the coordinates for the source 
%as 16$^{\rm h}$ 20$^{\rm m}$ 16\fs6 $-$50\degr 56\arcmin 23\arcsec,
as RA 16$^{\rmn{h}}~20^{\rmn{m}}~16\fs6$, Dec $-50\degr~56\arcmin~23\arcsec$,
that is, the left-hand source.

(iii) {\it G332.813$-$0.700.} This source is the only outflow source without a methanol maser 
but it is also the only source with detectable H$_2$ S(3) lines (Table~8).
The H$_2$ lines peak in the PDR (region of strong PAH emission) to the north-west of the YSO
and not in the region of extended IRAC band 2 (4.5~\micron) emission as seen in Fig. 28(a).
The \mbox{[Fe\,{\sc ii}]} 26~\micron\ line also peaks in the PDR regions that show strong 
PAH emission at 8~\micron, as seen in Fig. 28(b).
We suspect that this PDR region is somewhat foreground of the main source 
(which lies in northeast-southwest lane of higher extinction) 
because the extinction-corrected S(3) flux is much higher 
than the fluxes of the S(1) and S(5) lines
(the S(3) lines is at 9.66~\micron).
Other YSO spectra may appear to have the S(3) 9.66~\micron\ line, 
but there are numerous bad pixels at this wavelength in the SL array;
since a resolution element in the IRS is 4 pixels and since the H$_2$ lines should 
be extended and not point-like, we do not believe we have identified a line 
unless we can see it on the array in a number of contiguous pixels, 
as we can for this source but not the others.

(iv) {\it G332.942$-$0.686.} In this outflow YSO the H$_2$ line emission peaks 
to the northwest,
in the direction of the PAH emission as seen in the IRAC 8.0~\micron\ image, 
but the \mbox{[Si\,{\sc ii}]} 34.8~\micron\ line emission increases to the southwest.
Perhaps in this case the Si$^+$ is found in the diffuse Galactic ISM 
but the H$_2$ is more associated with the G333 GMC, located towards the northwest.
These contours are plotted in Fig. 29.

(v) {\it G332.963$-$0.679.} This outflow YSO is the most luminous source of the little cluster 
that includes the red sources G332.967$-$0.683 and G332.960$-$0.681.
We see in Supporting Information Fig. 32(e) 
that it is the strongest source in the MIPS 24~\micron\ image,
and in fact, all the longer wavelength images (MIPS 70~\micron, and 
{\it Herschel} PACS and SPIRE 150--600~\micron\ images)
peak at its location.
The extinction map (Fig. 12) shows that much of the extinction in the region 
is due to the extended envelope ($\sim 20'' \approx 7 \times 10^4$ au).
The red sources are foreground, as seen by their lower extinctions.

(vi) {\it G332.967$-$0.683.} This YSO and the similar red YSO G332.960$-$0.681 
cannot be detected in images at wavelengths longer than the MIPS 24~\micron\ image,
where they are both much fainter than the outflow YSO, G332.963$-$0.679.
At least part of the reason for lack of detection 
is spatial resolution for the {\it Spitzer} 70~\micron\ image
and the {\it Herschel} long wavelength PACS and SPIRE images. 
As a result of lack of measurement at the longer wavelengths, the SED and total 
luminosity are poorly measured (Fig. 19).

(vii) {\it G333.125$-$0.562.} This is the name Garay et al. (2004) gave to 
the first detection of this massive, dense cold core.
We included it in our survey since its SiO emission (Lo et al. 2007) 
is thought to arise from shocks associated with an outflow. 
The YSO with the ice absorption features, G333.131$-$0.560, is GLIMPSE source 
SSTGLMC G333.1309$-$00.5601.
The other stars seen by the IRAC that are also bright at 24~\micron\ are 
SSTGLMC G333.1310$-$00.5658 and SSTGLMC G333.1295$-$00.5683.
Unfortunately, they lie beyond the range of our SL map (Fig. 2c), 
but since they have similar IRAC colours to 
SSTGLMC G333.1309$-$00.5601, we think it likely that they also have ice-coated grains 
in their surrounding envelopes.
The source G333.128$-$0.560 is probably the 70~\micron\ source since our spectral 
maps show that it is the brightest object at the longest wavelengths, 36~\micron, 
even though it is less bright than G333.131 at 24~\micron.

The shocked \mbox{[S\,{\sc i}]} 25.25~\micron\ emission is seen in LH telescope pointings 70 and 71; 
these slit positions are marked on Fig.~2(e) 
and correspond approximately to the extended IRAC 4.5~\micron\ emission
(seen more easily in the 3-colour image of Cyganowski et al. 2008).
It is not possible to identify which of the YSOs is the source of this outflow 
because it is faint and because both G333.131 and G333.128 lie along the same line,
as do the 6.7 GHz Class II methanol masers (Caswell 2009; Fig. 2b).
The SiO emission occurs at G333.131$-$0.560 
whereas the 4.5~\micron\  and \mbox{[S\,{\sc i}]} outflow region coincides with emission of HCO$^+$ 
(Lo et al. 2011).

We did not have enough flux data to model both G333.131 and G333.128 separately 
with the online SED fitter of Robitaille et al. (2007) -- 
the results in Table 3 are from a composite of both sources.
A model with a larger inclination angle, larger model age, and smaller envelope mass 
would have properties similar to G333.128$-$0.560 -- more flux at 35 -- 70~\micron\ 
but much less flux at 24~\micron\ (Fig. 2e) and 3 mm (Lo et al. 2011).

The fuzzy blob to the north of G333.131 appears to be an \mbox{H\,{\sc ii}} region, with forbidden lines 
and substantial 24~\micron\ continuum (Fig. 2e).
However, it is unusual for an \mbox{H\,{\sc ii}} region in that there is no obvious continuum emission 
at the longer wavelength maps ({\it Spitzer} MIPS and {\it Herschel}).
If it is actually associated with the two stars at its location, 
2MASS 16213573-5040306 and 2MASS 16213662-5040334,
it may be a foreground object since these stars are not very red and the former is 
seen on the red Digital Sky Survey image. 
The SL spectra show only continuum at 5.2~\micron\ and background PAHs.

(viii) {\it G333.163$-$0.100.} The spectral maps of this source show extensions to the north-east 
in the lines mapped by LH: \mbox{[S\,{\sc iii}]} 33.5~\micron\ and \mbox{[Si\,{\sc ii}]} 34.8~\micron, 
plotted in Fig. 23(b), and also \mbox{[Fe\,{\sc ii}]} 26.0~\micron\ and \mbox{[Fe\,{\sc iii}]} 22.9~\micron\ 
[this region was not covered in the contours of Fig. 23(a) taken from the much smaller SH map].
The SL map shows a similar morphology for the \mbox{[Ar\,{\sc ii}]} 7.0~\micron\ line.
This is somewhat orthogonal to the apparent direction of the red extended emission.
This is the only red source with a 6.7 GHz methanol maser (Caswell 2009; Fig. 23b).

(ix) {\it G333.184$-$0.091.} This outflow YSO has a distinct absorption lane visible 
in the IRAC images, especially the 8~\micron\ image, that is perpendicular 
to the apparent outflow direction (compare Figs. 14 and 30).
This morphology is still apparent in the 24~\micron\ MIPS image 
(Fig. 33e of the Supporting Information).
It is much too big to be a disc, but could be indicative of a toroidal structure 
in the envelope with a much larger optical depth.
The morphology of the H$_2$ S(0) and \mbox{[Si\,{\sc ii}]} 34.8~\micron\ lines (contours plotted in Fig. 30)
is probably due to absorption of background gas by the YSO envelope.

(x) {\it G333.212$-$0.105.} This red YSO does not appear to be present in the MIPS 70~\micron\ 
image (Supporting Information Fig. 34f), 
but that is due to the low {\it Spitzer} resolution 
and the high background from the other sources in its immediate vicinity.
The source is definitely visible in the higher resolution {\it Herschel} 
PACS and SPIRE images, even at the longest SPIRE wavelength of 600~\micron.

(xi) {\it G333.340$-$0.127.} This red source has a distinct outflow to the south-west as 
seen in the higher-excitation forbidden lines, particularly \mbox{[Ne\,{\sc iii}]} 15.6~\micron\ 
and \mbox{[S\,{\sc iv}]} 10.5~\micron\ (Fig. 24). 
These high excitation lines peak well away from the YSO central region, 
which does include the peak of the low excitation lines, like \mbox{[Ne\,{\sc ii}]} 12.8~\micron, 
\mbox{[Si\,{\sc ii}]} 34.8~\micron, and \mbox{[Fe\,{\sc ii}]} 26.0~\micron. 
Lines of intermediate excitation, like \mbox{[S\,{\sc iii}]} 18.7 and 33.5~\micron\ 
and \mbox{[Fe\,{\sc iii}]} 22.9~\micron, have intermediate morphologies. 
The iron lines are not plotted in these figures because they are much weaker and 
noisier, but the morphology is clear. 
There is a clump of warm dust in the outflow direction, but closer to the YSO, 
as seen in the MIPS 24~\micron\ image and also in the H$_2$ line contours.

(xii) {\it G333.375$-$0.202.} This red source is too low excitation to test for outflow 
morphology from high excitation lines. 
The \mbox{[Ne\,{\sc ii}]} 12.8~\micron\  lines are slightly elongated, similar to the structure 
seen in the PAH emission at 5.6 and 8.0~\micron\ and to the H$_2$ S(0) line 
(Fig. 25).

(xiii) {\it G333.466$-$0.164.} There are at least two sources in this apparent cluster: 
(1) the outflow YSO with the ice absorption features that is the peak 
at 70~\micron\ and longer wavelengths, and (2) an \mbox{H\,{\sc ii}} region 
located to the west of the YSO that is the continuum source seen 
in the MIPS 24~\micron\ image (Fig. 26). 
The forbidden lines have their maxima at the west side of the image, but there 
is a secondary maximum at the northern tip of the outflow seen in the 4.5~\micron\ IRAC image
(Fig. 26c).
In Table 5 and Fig. 27 these regions are G333.466$-$0.164W and G333.466$-$0.164N, respectively.
There is also some ionized gas in the outflow (Walsh et al. 1998).
The YSO with the ice features, G333.466$-$0.164, is the location of 
the 6.7 GHz methanol maser (Caswell 2009) and is at the south end of this outflow (Fig. 26a).
Fig. 26(a) shows that the H$_2$ lines follow the morphology of the PAH emission 
with a big extinction lane seen in the IRAC images.

\section{Discussion}

We review what we have learned about the YSOs in our survey of the G333 GMC. 
Even though the FIR SEDs (Fig. 19) and luminosities (Table 4) are similar, 
the spectra of the outflow sources are as different from the spectra of the red sources 
as their appearances are different in the IRAC bands: 
(1) The outflow sources have deep ice features at 6.0 and 6.8~\micron\ and 
substantial short-wavelength flux with distinctive colours. 
(Such sources could be easily searched for 
in three-colour images consisting of IRAC bands 1, 2, and 3, 
where they have a yellow colour in contrast to the blue of normal stars and the pink to red 
of the red sources, where bands 1, 2, and 3 are blue, green, and red, respectively.) 
The red sources are dominated by PAHs from 5 to 9~\micron. 
(It is conceivable that if there should be any 6.0 or 6.8~\micron\ ice features in the red sources, 
they would not be detectable because of the low signal/noise in the relatively low continuum 
and the contamination by the strong PAH features.) 
(2) In the 9 -- 20~\micron\ range, the outflow sources show substantial absorption 
from both silicates at 10~\micron\ and cold CO$_2$ ice at 15.2~\micron, 
whereas the red sources are extincted only by the diffuse ISM, with little 
CO$_2$ ice absorption. 
(3) The four red sources that are extended and the more luminous 
all ionize their own \mbox{H\,{\sc ii}} regions, 
whereas none of the outflow sources produce any ionizing photons
(except perhaps in shocks). 
(4) Five of the seven outflow sources produce \mbox{[S\,{\sc i}]} 25.25~\micron\ line emission 
from their outflow regions. 
This is indicative of low velocity shocks. 
The two red sources with forbidden-line morphology indicating shocks in outflows 
must have higher velocity shocks given the higher excitation of the lines. 
(5) Six of the seven outflow sources produce 6.7 GHz Class II methanol maser emission 
(as does one of the red sources). 
Such masers are commonly found in the earliest stages of massive star formation 
(e.g., Ellingsen 2006; Breen et al. 2010), and are 
particularly common among the EGOs studied by Cyganowski et al. (2009).

We suggest that the two groups of YSOs correspond to two separate evolutionary stages,
with the outflow stage being younger than the red stage.
(1) The outflow sources have large, thick envelopes containing cold dust grains coated with ices.
The sizes of the envelopes are some tens of arcsec across from the extinction seen in the images 
(Supporting Information Figs. 32, 33, 35 -- 38) 
and the optical depth maps, corresponding to a few times $10^4$ to 
$> 10^5$ au in radius. 
The envelopes of the red sources have smaller optical depths since there is little variation 
in extinction with distance from the YSO and the optical depth at the centre is not 
much larger than the optical depths due to intervening ISM.
(2) The central stars of the red sources have already contracted to the point that 
the protostars' surface temperatures are hot enough to produce photons that can 
ionize hydrogen and neon and doubly ionize sulphur and even some Ne$^{++}$.
The only ionized gas that could be associated with any of the outflow sources is found at the end 
of the outflow of G333.466 and is possibly shock ionized.

This separation by evolutionary stage is consistent with the results of fitting 
the SEDs with the models of Robitaille et al. (2006) as seen in Table~3 
(which indicates that the SEDs do have different shapes). 
For the most part (especially for the models with unconstrained interstellar extinction), 
the models of the outflow YSOs have younger ages 
and cooler effective temperatures for the central protostar, thereby requiring 
much larger protostar radii, 
since the luminosities and masses have similar ranges and distributions. 
This is in accordance with the models of Hosokawa et al. (2010), 
who find that as a protostar accretes mass, if the accretion rate is high enough  
that the protostar will end up as a massive star ($\sim 10^{-3}$ M$_\odot$ yr$^{-1}$), 
the protostar becomes very large -- `bloated' -- 
and as such is incapable of producing ionizing photons. 
Later, as the star contracts to the main sequence,  
it becomes hot enough to ionize an \mbox{H\,{\sc ii}} region. 
We conclude that it is likely that the optically thick, outflow sources are 
bloated protostars with large accretion envelopes.

The Australia Telescope Compact Array (ATCA) observations of G333.125$-$0.562 
are consistent with this,
finding weak thermal emission towards the likely protostar (G333.131$-$0.560),
but no evidence for an \mbox{H\,{\sc ii}} region of any type (Lo et al. 2011).

We return to Fig. 1. 
We did not observe a complete sample of all sources in the G333 cloud -- 
there are two more `possible' outflow sources (EGOs) in the region in 
the catalogue of Cyganowski et al. (2008) 
and a number of red sources can be found by looking more closely at the 
IRAC and MIPS images. 
Cyganowski et al.'s two EGOs and three more luminous red sources 
that are extended at 8~\micron\ are also plotted in Fig. 1.
The plotted sources all lie outside the main clusters and \mbox{H\,{\sc ii}} regions, 
which shine in the UV-excited PAH emission that dominates IRAC band 4.
It is likely that the nine tabulated outflow sources are a complete set at their level
of luminosity, because IRAC band 2 has good spatial resolution 
and is not limited by confusion. 
One might infer from the low inclination angles of the outflow sources in Table~3 
that there may be many more high-inclination, massive, cold YSOs 
invisible to IRAC or MIPS 
that should be included in our list of possible outflow YSOs. 
However, we do not find such in the {\it Herschel} PACS 130--210~\micron\ image
except possibly two sources confused with known \mbox{H\,{\sc ii}} regions at 24~\micron. 
We conclude from the scarcity of luminous, NIR- and MIR-obscured sources 
that additional work is needed modelling this stage of massive star formation.  
In addition, a more extensive comparison of the images and SEDs 
from 2MASS and IRAC through PACS and SPIRE wavelengths 
to identify YSOs from their colours will be a fine project for future work.

On the other hand, there are probably more red sources 
within the region of high IRAC band 4 PAH emission 
-- in fact, some FIR point sources can be detected in this region
on the {\it Herschel} SPIRE 250~\micron\ image.
However, the high band 4 emission regions are confused and often saturated at 
the longer MIR and FIR wavelengths, making it difficult to confirm 
whether clumpy structure consists of YSOs and not compact \mbox{H\,{\sc ii}} regions 
since the same regions also emit free-free radiation (Green et al. 1999). 
We conclude that it is not easy to identify YSOs that are past the 
bloated stage and are well on the way to contracting to the main sequence 
but have not contracted so far that they can ionize their own \mbox{H\,{\sc ii}} regions.
This probably means that the time-scale for the final contraction stage 
is very short.
The red sources that we did observe have molecular gas 
but our limited survey of the region including red sources
does not find as much molecular gas as is found in the outflow sources -- 
this may be another indicator of their more evolved nature.

We also see in Fig. 1 that there are tendencies for the YSOs to cluster, 
with at least three clusters if we include the YSO group in G333.125.
There can be both outflow and red sources in the same cluster, 
from which we infer that star formation occurs over a length of time 
longer than the contraction time needed for the smaller red sources like G332.967.
The outflow sources within the G333 cloud are all associated with dense molecular gas 
(see our previous surveys by Bains et al. 2006; Wong et al. 2008; Lo et al. 2009).
In these surveys we have used the Mopra radio telescope 
to map numerous molecular transitions
including  $^{13}$CO, C$^{18}$O, CS, HCN, HNC, HCO$^+$, N$_2$H$^+$,
C$_2$H, HC$_3$N, CH$_3$OH, SO, and SiO, 
all observed with an angular resolution of $\sim~35$~arcsec.
By comparing the tracers of higher density 
with the lower density $^{13}$CO tracer,
we are able to investigate the spatial and kinematic power spectra of the entire region.
While the analysis is ongoing, we find from a spatial power-spectrum analysis 
that all tracers have a similar power law distribution, 
suggesting well mixed gas with significant turbulent motions 
(M. Cunningham, in preparation).
We speculate that the observed clusters of YSOs are examples of star formation induced by
turbulent compression in the outer regions of GMCs (e.g., Elmegreen 2007).

\section{Summary and conclusions}

We have mapped 13 YSOs that were identified in the G333 GMC 
on {\it Spitzer} IRAC and MIPS images  
with the {\it Spitzer} IRS from 5-36~\micron.
We use these spectra plus available photometry and images to characterize the YSOs.
The objects are divided into two groups:
YSOs associated with extended emission in IRAC band 2 at 4.5~\micron\ (`outflow sources')
and YSOs that have extended emission in all IRAC bands peaking at the longest wavelengths
(`red sources').

We find the following results:
The source luminosities range from a few times $10^3$ L$_\odot$ to a few times $10^4$ L$_\odot$.
Modelling of the SEDs indicates that these YSOs have masses from 8 - 25 M$_\odot$.
The spectral energy distributions (SEDs) of all sources peak between 50 and 110~\micron,
thereby showing their young age, but the SEDs of the outflow sources peak at slightly longer
wavelengths than the SEDs of the red sources.

All spectral maps show ionized forbidden lines and
PAH emission features
(most of this is due to background emission from the G333 cloud and must be carefully
subtracted from the source spectra).
For four of the red sources, the line emission is concentrated
at the centres of the maps, from which we infer that these YSOs are
the source of ionizing photons.
The YSO luminosities and low excitation of the \mbox{H\,{\sc ii}} regions 
are both indicative of early B stars.

All the YSOs associated with outflows show evidence of massive envelopes surrounding the star. 
The extinction from these envelopes is characterized by deep silicate absorption features 
and by absorption by ices at 6.0, 6.8, and 15.2~\micron. 
Six out of seven of these YSOs also have 6.7 GHz Class II methanol maser emission. 
At least four of the objects are found in clusters of YSOs. 
Ionized lines are associated with two of the clusters containing outflow YSOs; 
however, the gas is more likely ionized by another star in the cluster 
since the lines peak at least 15~arcsec from the outflow YSO. 

For several objects of both types, the lines from the most highly excited ions (Ne$^{++}$) 
peak at some distance from the peak of the low excitation ion lines; 
we suggest that these ions indicate the presence of shocked gas. 
There is shocked gas associated with the 4.5~\micron\ emission, but this is seen in 
the presence of the \mbox{[S\,{\sc i}]} line in five of the seven outflow sources 
and not in the presence of any H$_2$ lines, as had been suggested by other observers.

We conclude that this spectroscopic survey has provided unique information
on this stage of the formation of massive stars including
the identification of the stars producing the IRAC 4.5~\micron\ outflows
as having envelopes cool enough to contain ice-coated grains
and the characterization of the only slightly more evolved YSOs as no longer exhibiting
ice features but already producing photons energetic enough to ionize a small \mbox{H\,{\sc ii}} region.

\section*{Acknowledgments}

This work is based on observations made with the {\it Spitzer Space Telescope}, 
which is operated by the Jet Propulsion Laboratory (JPL),
California Institute of Technology, under a contract with NASA. Support for
this work was provided by NASA through RSA 1376528 issued by JPL/Caltech.
The IRS was a collaborative venture between Cornell University
and Ball Aerospace Corporation
funded by NASA through the JPL and Ames Research Center.
{\sc smart} was developed by the IRS Team at Cornell University and is available
through the Spitzer Science Center at Caltech.
This research made use of Tiny Tim/Spitzer, developed by John Krist for the Spitzer Science Center. The Center is managed by the California Institute of Technology under a contract with NASA.
This research used observations with {\it AKARI}, a JAXA project with the participation of ESA.
{\it Herschel} is an ESA space observatory with science instruments provided by European-led Principal Investigator consortia and with important participation from NASA.
We thank S. Molinari and the Hi-GAL team for waiving the proprietary period 
such that we could look at their beautiful data in advance of publication.
We thank J. D. T. Smith for making the {\sc pahfit} program available such that users can 
modify it for other needs.
We thank the Spitzer operations team for managing the telescope schedule 
such that the cryogens lasted until mid-May, 2009.
Without this laudable effort, these observations would not have been possible.
We thank Sean Carey and Roberta Palladini for giving us the MIPSGAL 70~\micron\ images 
prior to publication and 
we thank the referee, Barbara Whitney, for her thoughtful and thought-provoking comments 
on the manuscript. 

\begin{appendix} 
\appendix

\section{Line flux tables}

The source fluxes, measured after the subtraction of the PAH template, 
are given in Table A1.
The contour levels for Figs. 22 -- 26 and 28 -- 30 are given in Table A2.
The measured line fluxes needed for Table 5 are given in Table A3.

\setcounter{table}{0}
\begin{table*}
\centering
\begin{minipage}{140mm}
\caption{Measured source continuum fluxes.
The continuum fluxes for each source after PAH subtraction are given in Jy.
G333.466--YSO is the point source at the head of the outflow that has the ice absorption features.
G333.466$-$0.164 is the whole source, IRAS 16175$-$5002.
}
\begin{tabular}{@{}lcccccccccc@{}}
\hline
Wavelength (\micron) &  5.48  &  7.99 & 9.52  &  12.02 & 14.02 & 18.02 & 20.01  &  25.03  &  30.00  &  35.02 \\
Source & &&&&&&&&& \\
\hline
% 5.48  7.40  10.18  12.02  14.02  18.02  20.01  25.03  &  30.00 & 35.02 \\
G332.725$-$0.620\footnote{For G332.725$-$0.620, the fluxes are measured at 7.40 and 10.18 \micron\ instead of 7.99 and 9.52 \micron.} &  0.966 & 0.487  &  0.810  & 2.76  & 2.82  & 3.88 & 6.09  & 16.5 & 30.7  & 55.2 \\
G332.800$-$0.595 &  3.93  & 9.76 & 14.4 &  59.5 &  97.5 &  200 &  293 &  526 &  709 &  946 \\
G332.813$-$0.700 &  0.770 &  1.04 & 0.635 &  2.07 &  2.68 &  3.36 &  4.92 &  12.0 &  19.8 &  31.2 \\
G332.942$-$0.686 &  1.24 &  1.66 & 0.384 &  1.68 &  1.50 &  1.24 &  1.79 &  6.91 &  17.9 &  28.8 \\
G332.963$-$0.679 &  0.897 &  1.77 & 0.056 &  1.09 &  2.36 &  2.81 &  5.98 &  36.8 &  83.2 &  96.5 \\
G332.967$-$0.683 &  0.279 & 0.640 & 0.241 &  1.17 &  1.24 & 2.18 &  3.31 &  8.61 &  16.3 &  24.3 \\
G333.131$-$0.560 &  0.211 & 0.264 & 0.183 & 0.757 & 0.832 &  1.22 &  1.86 &  4.59 & 8.68 & 15.9 \\
G333.163$-$0.100 &  0.683 &  2.16 & 0.235 &  5.42 &  9.59 &  11.8 &  20.9 &  71.7 &  136 &  206 \\
G333.184$-$0.091 &  1.13 &  1.42 & 0.046 &  1.46 &  2.33 &  1.77 & 3.32 &  12.5 &  26.1 &  46.3 \\
G333.212$-$0.105 &  0.164 & 0.857 & 0.088 &  1.23 &  1.92 &  1.92 &  3.51 &  10.7 &  18.0 &  27.6 \\
G333.340$-$0.127 &  0.548 &  2.81 & 0.815 &  5.82 &  9.85 &  17.2 &  25.5 &  57.1 &  82.6 &  105 \\
G333.375$-$0.202 &  1.65 &  1.69 & 0.318 &  2.22 &  3.33 &  5.52 &  9.30 &  24.9 &  40.8 &  55.1 \\
G333.466$-$0.164 &  1.84 &  4.52 & 0.201 &  6.26 &  11.2 &  13.2 &  22.2 &  78.9 &  157 &  225 \\
G333.466--YSO &  0.884 &  1.74 &  0.0039 &  1.29 &  2.40 &  3.02 &  5.77 &  26.8 &  60.1 &  88.1 \\
\hline
\end{tabular}
\end{minipage}
\end{table*}

\setcounter{table}{1}
\begin{table}
\centering
\begin{minipage}{84mm}
\caption{Contour levels for Figs. 22 -- 26 and 28 -- 30.
The figures noted are contour plots of various line intensities.
For each figure and line, the contours are usually 
0.95, 0.9, 0.8, 0.7, 0.6, 0.5, 0.4, 0.3, 0.2, 0.1, 0.05, and 0.025 times 
the maximum observed intensity (exceptions are footnoted),
where the line, maximum intensity, and minimum contour level are given in the table.
}
\begin{tabular}{@{}lcccc@{}}
\hline
Fig. &  Source     &      Line  &  Maximum  & Minimum \\
       &             &            &  Intensity & Contour \\
       &              &          & (W m$^{-2}$ sr$^{-1}$)  &   \\
\hline
 22 &   G332.800$-$0.595 & \mbox{[Ne\,{\sc ii}]} 12.8 \micron\ & $3.59 \times 10^{-6}$ & 0.1 \\ %  0.1, 0.2, 0.3, 0.4, 0.5, 0.6, 0.7, 0.8, 0.
    &                 & \mbox{[Ne\,{\sc iii}]} 15.6 \micron\ & $1.99 \times 10^{-7}$ & 0.05   \\ % 0.05, 0.1, 0.2, 0.3, 0.4, 0.5, 0.6, 0.7, 0.8, 0.9, and 0.95 
    &                 & \mbox{[S\,{\sc iii}]} 33.5 \micron\ & $1.09 \times 10^{-5}$ & 0.2    \\ %  0.2, 0.3, 0.4, 0.5, 0.6, 0.7, 0.8, 0.9, and 0.95 
    &                 & \mbox{[S\,{\sc iii}]} 18.7 \micron\ & $2.71 \times 10^{-6}$ & 0.1    \\ %  0.1, 0.2, 0.3, 0.4, 0.5, 0.6, 0.7, 0.8, 0.9, and 0.95 
    &                 & \mbox{[Ar\,{\sc ii}]} 7.0 \micron\ & $2.05 \times 10^{-6}$ & 0.05    \\ %  0.05, 0.1, 0.2, 0.3, 0.4, 0.5, 0.6, 0.7, 0.8, 0.9, and 0.95 
    &                 & \mbox{[Si\,{\sc ii}]} 34.8 \micron\  & $2.61 \times 10^{-6}$ & 0.3   \\ % 0.3, 0.4, 0.5, 0.6, 0.7, 0.8, 0.9, and 0.95 
    &                 & H$_2$ S(2) 12.3 \micron\ & $3.58 \times 10^{-8}$ & 0.025   \\ % 0.025, 0.05, 0.1, 0.2, 0.3, 0.4, 0.5, 0.6, 0.7, 0.8, 0.9, and 0.95 
    &                 & H$_2$ S(1) 17.0 \micron\ & $2.15 \times 10^{-8}$ & 0.1   \\ % 0.1, 0.2, 0.3, 0.4, 0.5, 0.6, 0.7, 0.8, 0.9, and 0.95 
 &&&& \\
 23 &  G333.163$-$0.100 & \mbox{[Ne\,{\sc ii}]} 12.8 \micron\ & $2.28 \times 10^{-6}$ & 0.15\footnote{Contour levels are 0.88, 0.75, 0.6, 0.45, 0.3, 0.15 times the maximum.}    \\ % .15,.3,.45,.6,.75,.88 
    &                 & \mbox{[Ne\,{\sc iii}]} 15.6 \micron\ & $2.60 \times 10^{-8}$ & 0.4\footnote{Contour levels are 0.85, 0.7, 0.6, 0.5, 0.4 times the maximum.}   \\ % 0.4, 0.5, 0.6, 0.7, and 0.85 
    &                 & \mbox{[S\,{\sc iii}]} 33.5 \micron\ & $3.32 \times 10^{-6}$ & 0.5    \\ % 0.5, 0.6, 0.7, 0.8, 0.9, and 0.95 
    &                 & \mbox{[Si\,{\sc ii}]} 34.8 \micron\ & $2.01 \times 10^{-6}$ & 0.5    \\ % 0.5, 0.6, 0.7, 0.8, 0.9, and 0.95 
 &&&& \\
 24 &  G333.340$-$0.127 & \mbox{[Ne\,{\sc ii}]} 12.8 \micron\ & $3.16 \times 10^{-7}$  & 0.15\footnote{Contour levels are 0.95, 0.9, 0.8, 0.7, 0.6, 0.5, 0.4, 0.3, 0.2, 0.15 times the maximum.}   \\ % 0.15, 0.2, 0.3, 0.4, 0.5, 0.6, 0.7, 0.8, 0.9, and 0.95 
    &                & \mbox{[Ne\,{\sc iii}]} 15.6 \micron\ & $3.30 \times 10^{-8}$ & 0.2    \\ % 0.2, 0.3, 0.4, 0.5, 0.6, 0.7, 0.8, 0.9, and 0.95 
    &                & \mbox{[S\,{\sc iii}]} 18.7 \micron\ & $5.66 \times 10^{-8}$ & 0.3    \\ % 0.3, 0.4, 0.5, 0.6, 0.7, 0.8, 0.9, and 0.95 
    &                & \mbox{[S\,{\sc iii}]} 33.5 \micron\ & $5.90 \times 10^{-7}$ & 0.4     \\ % 0.4, 0.5, 0.6, 0.7, 0.8, 0.9, and 0.95 
    &                & \mbox{[Si\,{\sc ii}]} 34.8 \micron\ & $3.51 \times 10^{-7}$ & 0.4    \\ % 0.4, 0.5, 0.6, 0.7, 0.8, 0.9, and 0.95 
    &                & H$_2$ S(2) 12.3 \micron\ & $2.82 \times 10^{-8}$ & 0.2    \\ % 0.2, 0.3, 0.4, 0.5, 0.6, 0.7, 0.8, 0.9, and 0.95 
    &                & H$_2$ S(1) 17.0 \micron\ & $2.22 \times 10^{-8}$ & 0.2   \\ % 0.2, 0.3, 0.4, 0.5, 0.6, 0.7, 0.8, 0.9, and 0.95 
 &&&& \\
 25 &  G333.375$-$0.202 & \mbox{[Ne\,{\sc ii}]} 12.8 \micron\ & $6.85 \times 10^{-8}$  & 0.4  \\ % 0.4, 0.5, 0.6, 0.7, 0.8, 0.9, and 0.95 
    &                 & H$_2$ S(2) 12.3 \micron\ & $2.73 \times 10^{-8}$ & 0.4   \\ % 0.4, 0.5, 0.6, 0.7, 0.8, 0.9, and 0.95 
    &                 & \mbox{[S\,{\sc iii}]} 33.5 \micron\ & $1.85 \times 10^{-7}$ & 0.7    \\ % 0.4, 0.5, 0.6, 0.7, 0.8, 0.9, and 0.95 
    &                 & \mbox{[Si\,{\sc ii}]} 34.8 \micron\ & $2.52 \times 10^{-7}$ & 0.7    \\ % 0.7, 0.8, 0.9, and 0.95 
 &&&& \\
 26 &  G333.466$-$0.164 & H$_2$ S(2) 12.3 \micron\ & $3.22 \times 10^{-8}$ & 0.2   \\ % 0.2, 0.3, 0.4, 0.5, 0.6, 0.7, 0.8, 0.9, and 0.95 
    &                 & H$_2$ S(1) 17.0 \micron\ & $2.20 \times 10^{-8}$ & 0.1   \\ % 0.1, 0.2, 0.3, 0.4, 0.5, 0.6, 0.7, 0.8, 0.9, and 0.95 
    &                 & \mbox{[Si\,{\sc ii}]} 34.8 \micron\ & $1.25 \times 10^{-6}$ & 0.2   \\ % 0.2, 0.3, 0.4, 0.5, 0.6, 0.7, 0.8, 0.9, and 0.95 
    &                 & \mbox{[Fe\,{\sc ii}]} 26.0 \micron\ & $3.26 \times 10^{-8}$ & 0.2   \\ % 0.2, 0.3, 0.4, 0.5, 0.6, 0.7, 0.8, 0.9, and 0.95 
    &                 & \mbox{[Ne\,{\sc ii}]} 12.8 \micron\ & $9.55 \times 10^{-7}$ & 0.05   \\ % 0.05, 0.1, 0.2, 0.3, 0.4, 0.5, 0.6, 0.7, 0.8, 0.9, and 0.95 
    &                 & \mbox{[Ne\,{\sc iii}]} 15.6 \micron\ & $7.28 \times 10^{-8}$ & 0.05   \\ % 0.05, 0.1, 0.2, 0.3, 0.4, 0.5, 0.6, 0.7, 0.8, 0.9, and 0.95 
    &                 & \mbox{[S\,{\sc iii}]} 33.5 \micron\ & $5.12 \times 10^{-6}$ & 0.1   \\ % 0.1, 0.2, 0.3, 0.4, 0.5, 0.6, 0.7, 0.8, 0.9, and 0.95 
    &                 & \mbox{[S\,{\sc iii}]} 18.7 \micron\ & $5.72 \times 10^{-7}$ & 0.025  \\ % 0.025, 0.05, 0.1, 0.2, 0.3, 0.4, 0.5, 0.6, 0.7, 0.8, 0.9, and 0.95 
 &&&& \\
 28 &  G332.813$-$0.700 & H$_2$ S(0) 28.2 \micron\ & $1.89 \times 10^{-8}$  & 0.4\footnote{Contour levels are 0.95, 0.85, 0.7, 0.55, and 0.4 times the maximum.}  \\ % 0.4, 0.55, 0.7, 0.85, and 0.95 
    &                 & H$_2$ S(1) 17.0 \micron\  & $2.58 \times 10^{-8}$ & 0.4   \\ % 0.4, 0.5, 0.6, 0.7, 0.8, 0.9, and 0.95 
    &                 & \mbox{[Fe\,{\sc ii}]} 26.0 \micron\ & $1.82 \times 10^{-8}$ & 0.5   \\ % 0.5, 0.6, 0.7, 0.8, 0.9, and 0.95 
 &&&& \\
 29 &  G332.942$-$0.686 & H$_2$ S(1) 17.0 \micron\ &  $3.07 \times 10^{-8}$ & 0.3   \\ % 0.3, 0.4, 0.5, 0.6, 0.7, 0.8, 0.9, and 0.95 
    &                 & \mbox{[Si\,{\sc ii}]} 34.8 \micron\ & $5.41 \times 10^{-7}$  & 0.35\footnote{Contour levels are 0.95, 0.9, 0.8, 0.65, 0.5, and 0.35 times the maximum.}  \\ % 0.35, 0.5, 0.65, 0.8, 0.9, and 0.95 
 &&&& \\
 30 &  G333.184$-$0.091 & H$_2$ S(0) 28.2 \micron\  & $4.10 \times 10^{-8}$ & 0.7  \\ % 0.7, 0.8, 0.9, and 0.95 
    &                 & \mbox{[Si\,{\sc ii}]} 34.8 \micron\ & $1.14 \times 10^{-6}$  & 0.5  \\ % 0.5, 0.6, 0.7, 0.8, 0.9, and 0.95 
\hline
\end{tabular}
\end{minipage}
\end{table}

\setcounter{table}{2}
\begin{table*}
\centering
\begin{minipage}{170mm}
\caption{Measured line intensities of the sources producing ionizing photons.
The intensities are integrated over a field of view comprising all or most of a source
(which can be the area emitting forbidden lines in an otherwise outflow source), 
with the field of view, or aperture, tabulated in square arcsec.
In addition to the line intensities and errors, 
the table gives the average optical depth at 9.6 \micron, $\tau_{9.6}$.
}
\begin{tabular}{@{}lcccccccc@{}}
\hline
%Name &    $\tau_{9.6}$ & Aperture & H 7-6  &  [S IV] 10.5 \micron\ & [Ne II] 12.8 \micron\ & [Ne III] 15.6 \micron\ & [S III] 18.7 \micron\ &  [S III] 33.5 \micron\ & [Si II] 34.8 \micron\ \\
Source &    $\tau_{9.6}$ & Aperture & H 7-6  &  \mbox{[S\,{\sc iv}]} 10.5 & \mbox{[Ne\,{\sc ii}]} 12.8 & \mbox{[Ne\,{\sc iii}]} 15.6  & \mbox{[S\,{\sc iii}]} 18.7 & \mbox{[S\,{\sc iii}]} 33.5 \\% & \mbox{[Si\,{\sc ii}]} 34.8 \\
     &              & (Square  &  $\times 10^{-8}$ &  $\times 10^{-8}$ &  $\times 10^{-7}$ &  $\times 10^{-8}$ &  $\times 10^{-7}$ &  $\times 10^{-7}$ \\%  &  $\times 10^{-7}$ \\   
     &              &  Arcsec)  &  (W m$^{-2}$ sr$^{-1}$) &  (W m$^{-2}$ sr$^{-1}$) &  (W m$^{-2}$ sr$^{-1}$) &  (W m$^{-2}$ sr$^{-1}$) &  (W m$^{-2}$ sr$^{-1}$) &  (W m$^{-2}$ sr$^{-1}$) \\%  &  (W m$^{-2}$ sr$^{-1}$) \\
\hline
%                     e-18               e-18          e-17       e-18          e-17           e-17           e-17  yso_lines.tab units
%becomes              e-8                e-8            e-7       e-8           e-7            e-7            e-7   W m-2 sr-1
G332.800$-$0.595\footnote{Intensities must be multiplied by a factor of 6.3 to compensate for the incomplete sampling.} &&&&&&&& \\% & \\
Background    &  0.7 & 3054 & $0.38 \pm 0.03$ &  $1.84 \pm 0.04$ &  $2.00 \pm 0.01$ & $ 3.88 \pm 0.04$ &  $1.73 \pm 0.02$ & $21.26 \pm 0.11$ \\%  & $11.16 \pm 0.17$  \\ 
Source        &  0.70 & 3054 & $1.47 \pm 0.04$ &  $3.43 \pm 0.05$ &  $9.48 \pm 0.10$ &  $7.60 \pm 0.05$ &  $8.42 \pm 0.05$ & $78.58 \pm 0.32$ \\%  & $18.90 \pm 0.60$ \\  
G333.131$-$0.560 &&&&&&&& \\% & \\
Background    &  0.6 & 184  & $1.05 \pm 0.17$  &    -         &  $4.50 \pm 0.04$ &  $0.23 \pm 0.14$ &  $2.35 \pm 0.03$ &  $5.68 \pm 0.04$ \\%  &   $6.26 \pm 0.05$ \\  
North \mbox{H\,{\sc ii}} Reg & 0.41 & 184  & $1.31 \pm 0.24$  &    -         &  $6.19 \pm 0.06$ &  $1.24 \pm 0.08$ &  $3.49 \pm 0.03$ &  $8.14 \pm 0.03$ \\%  & $6.22 \pm 0.08$ \\  
G333.163$-$0.100 &&&&&&&& \\% & \\
Background    &  3.1 & 654  & $1.33 \pm 0.11$  &    -         &  $6.51 \pm 0.05$ &  $2.87 \pm 0.10$ &  $2.48 \pm 0.01$ & $14.03 \pm 0.09$ \\%  &  $7.68 \pm 0.07$ \\  
Source        &  3.60 & 654  & $3.99 \pm 0.05$  &    -        &  $25.07 \pm 0.17$ &  $5.04 \pm 0.13$ &  $7.48 \pm 0.03$ & $24.64 \pm 0.37$ \\%  & $15.67 \pm 0.42$  \\ 
G333.340$-$0.127 &&&&&&&& \\% & \\
Background    &  1.3 & 654 &  $0.18 \pm 0.14$ &  $0.84 \pm 0.81$ &  $0.89 \pm 0.02$ & $ 2.31 \pm 0.19$ &  $0.49 \pm 0.02$ &  $2.02 \pm 0.04$ \\%  &  $1.50 \pm 0.04$ \\  
Source        &  2.00 & 654 &  $1.14 \pm 0.15$ &  $2.66 \pm 0.14$ &  $3.38 \pm 0.06$ &  $3.95 \pm 0.13$ &  $1.09 \pm 0.02$ &  $3.71 \pm 0.18$ \\%  &  $2.64 \pm 0.15$ \\  
G333.375$-$0.202 &&&&&&&& \\% & \\
Background    &  1.9  & 184  &     -          &   -         &  $1.31 \pm 0.02$ &  $0.35 \pm 0.11$ &  $0.30 \pm 0.02$ &  $1.23 \pm 0.04$ \\%  &  $2.06 \pm 0.05$  \\
Source        &  2.50 & 184  &     -          &   -         &  $1.42 \pm 0.01$ &  $0.40 \pm 0.18$ &  $0.22 \pm 0.01$ &  $2.50 \pm 0.20$ \\%  &  $3.34 \pm 0.14$ \\  
G333.466$-$0.164 &&&&&&&& \\% & \\
Background    &  1.6 & 153  & $0.70 \pm 0.21$  & $0.27 \pm 0.27$ &  $1.12 \pm 0.02$ &  $0.14 \pm 0.14$ &  $4.22 \pm 0.03$ &  $4.20 \pm 0.05$ \\%  &  $2.44 \pm 0.07$ \\  
North \mbox{H\,{\sc ii}} Reg &  3.50 & 153 & $4.13 \pm 0.22$ &  $2.56 \pm 0.27$ & $27.24 \pm 0.26$ & $22.41 \pm 0.16$ & $14.52 \pm 0.09$ & $42.75 \pm 0.33$ \\%  &  $4.63 \pm 0.09$ \\  
West \mbox{H\,{\sc ii}} Reg  &  4.00 & 153 &  $4.14 \pm 0.17$ &  $1.64 \pm 0.33$ & $30.94 \pm 0.33$ & $14.40 \pm 0.26$ & $17.51 \pm 0.19$ & $50.19 \pm 0.23$ \\%  & $ 9.68\ pm 0.28$ \\ 
\hline
\end{tabular}
\end{minipage}
\end{table*}

\end{appendix}

\section*{Supporting Information}

Additional Supporting Information may be found in the online version of this article:

{\bf Figures 31-41.}
In this Supporting Information, we show images from the IRAC (panels a -- d are IRAC bands 1 -- 4, respectively) 
and MIPS (panels e -- f are MIPS images at 24 and 70~\micron\ images) 
of each source.
An example of the Supporting Information figures is shown in Fig. 2 for G333.131$-$0.560. 
In each figure, on panels (c) -- (e), we overplot the locations of the IRS slits.
On panel (b) the locations of any 6.7 GHz methanol maser (Caswell 2009) are marked with white crosses.
All images are shown in logarithmic scaling.
For Fig. 31 of G332.800$-$0.595, the MIPS 24~\micron\ image is replaced by the 
{\it MSX} 21~\micron\ image because the MIPS 24~\micron\ image is badly saturated.
In Fig. 34, G333.212$-$0.105 is not visible in the MIPS 70~\micron\ image (panel f),
probably because of high background and low {\it Spitzer} resolution, 
but it is visible in all {\it Herschel} PACS and SPIRE 70 -- 600~\micron\ images.

\bsp

%\label{lastpage}

%\end{document}

\clearpage

\vspace*{\fill}
\begin{minipage}{155mm}
\centering{\huge{\bf Supporting Information}}
\end{minipage}
\vspace*{\fill}

\clearpage

\setcounter{figure}{30}

%Figure S2
%\clearpage
\begin{figure*}
\includegraphics[width=150mm]{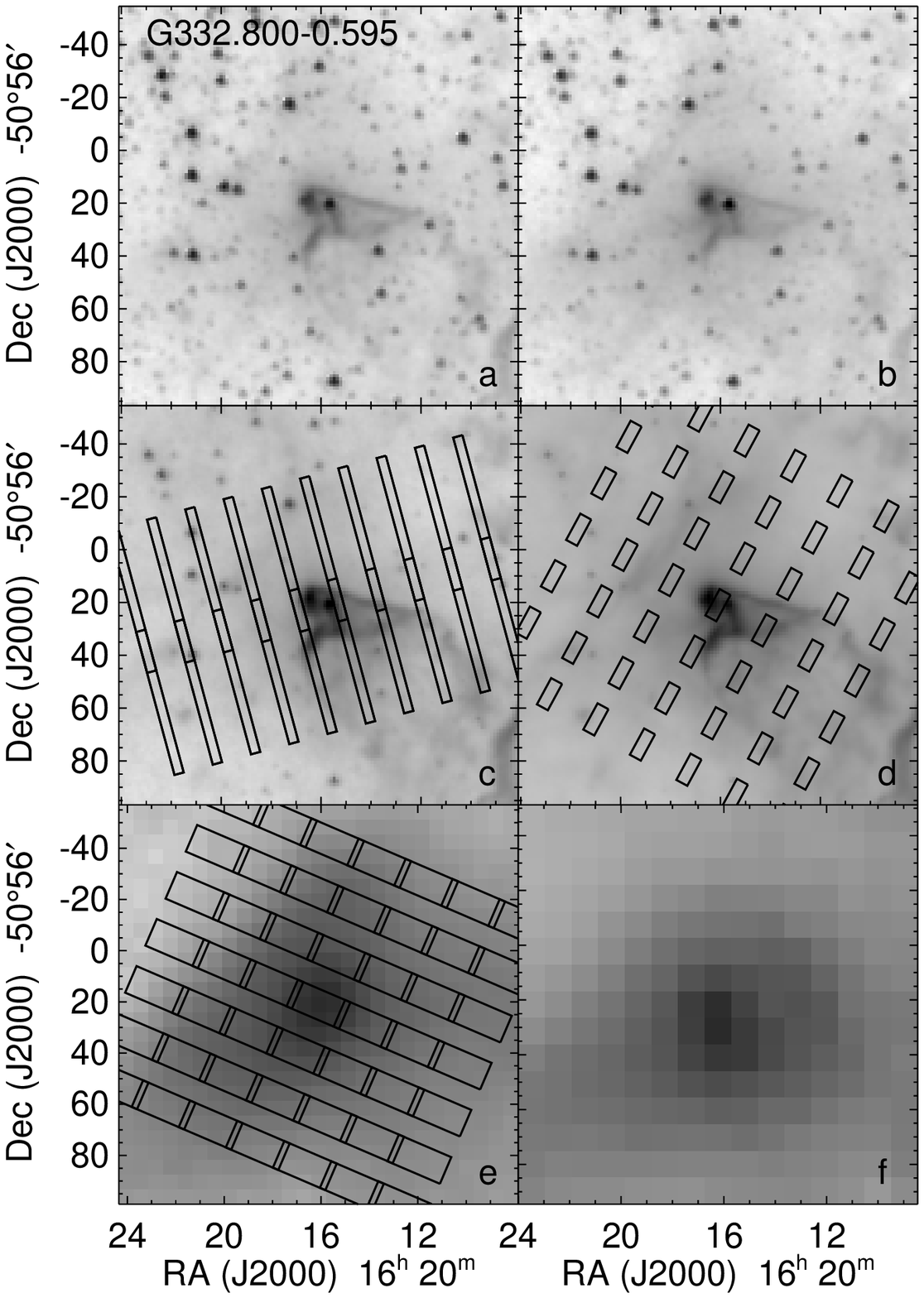}
\caption{IRAC, MSX, and MIPS images of G332.800$-$0.595.
Panel (a): IRAC band 1 (3.6~\micron) image. 
Panel (b): IRAC band 2 (4.5~\micron) image. 
Panel (c): IRAC band 3 (5.8~\micron) image with overlays of the SL slit positions.
Panel (d): IRAC band 4 (8.0~\micron) image with overlays of the SH slit positions.
Panel (e): MSX 21~\micron\ image with overlays of the LH slit positions. The MIPS 24~\micron\ image was not used in this figure because it is badly saturated.
Panel (f): MIPS 70~\micron\ image.
}
\end{figure*}

%Figure S5
%\clearpage
\begin{figure*}
\includegraphics[width=150mm]{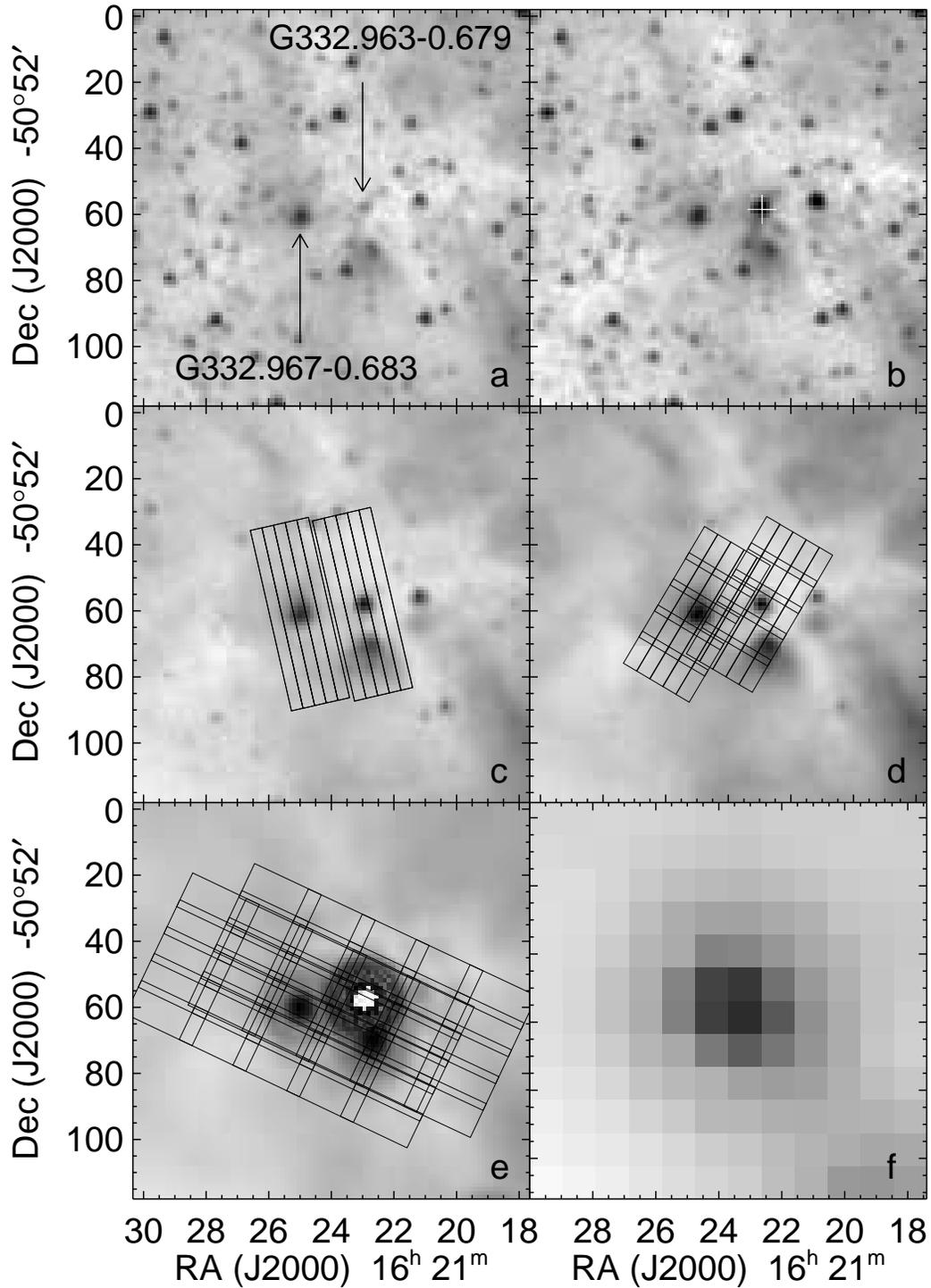}
\caption{IRAC and MIPS images of G332.963$-$0.679 and G332.967$-$0.683, which were observed 
in separate AORs.
Panel (a): IRAC band 1 (3.6~\micron) image. 
Panel (b): IRAC band 2 (4.5~\micron) image. 
The white cross marks the location of a 6.7 GHz methanol maser (Caswell 2009).
Panel (c): IRAC band 3 (5.8~\micron) image with overlays of the SL slit positions.
Panel (d): IRAC band 4 (8.0~\micron) image with overlays of the SH slit positions.
Panel (e): MIPS 24~\micron\ image with overlays of the LH slit positions.
Panel (f): MIPS 70~\micron\ image.
The centre of the MIPS 24~\micron\ image (panel e) is saturated at the location of G332.963$-$0.679.
}
\end{figure*}

%Figure S7
%\clearpage
\begin{figure*}
\includegraphics[width=150mm]{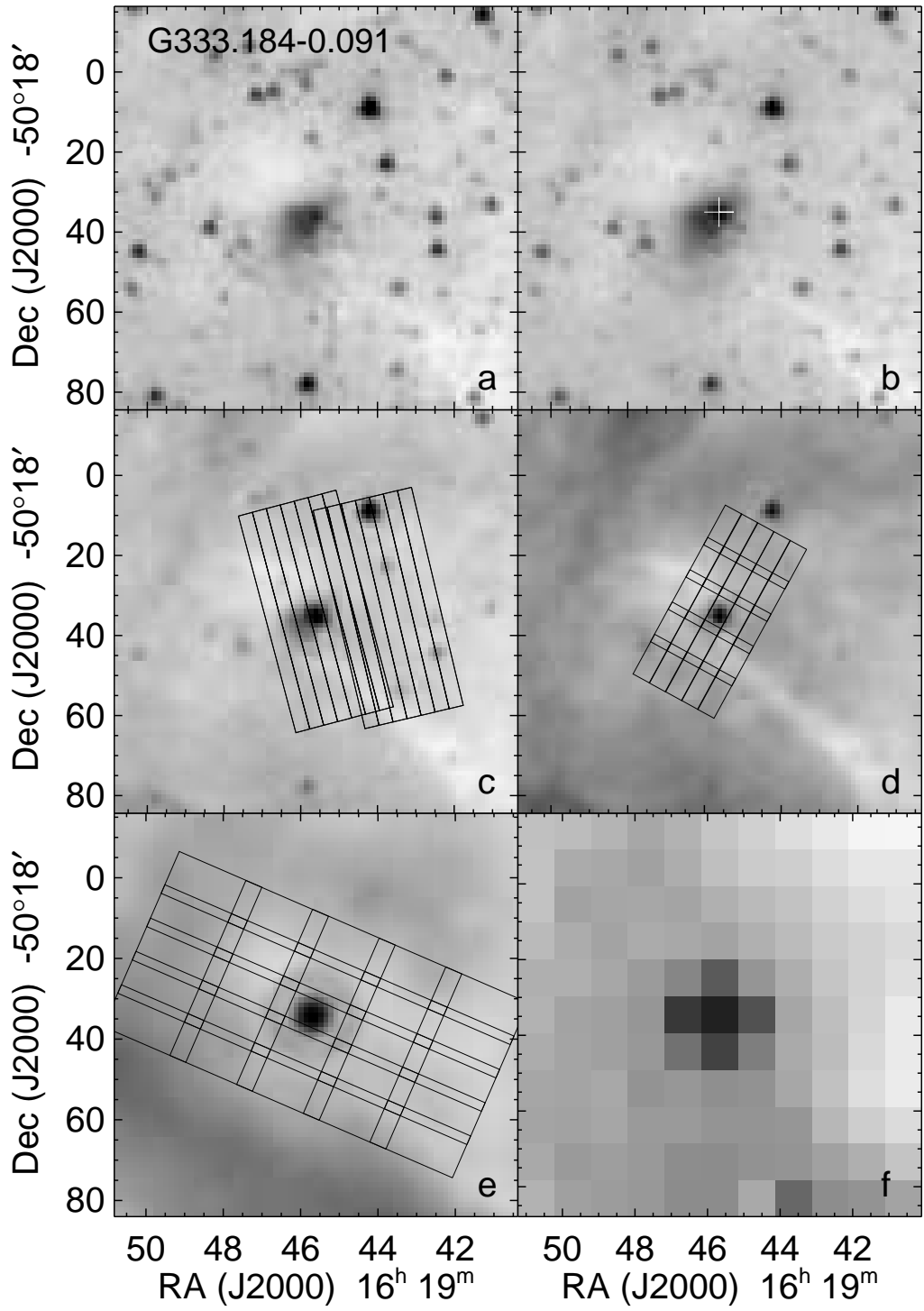}
\caption{IRAC and MIPS images of G333.184$-$0.091.
The images and overlays have the same description as Fig. 32 except that the MIPS 24 \micron\ image is not saturated.
Panel (c) also shows the location of the SL1 spectra from AOR 25915648.
}
\end{figure*}

%Figure S8
%\clearpage
\begin{figure*}
\includegraphics[width=150mm]{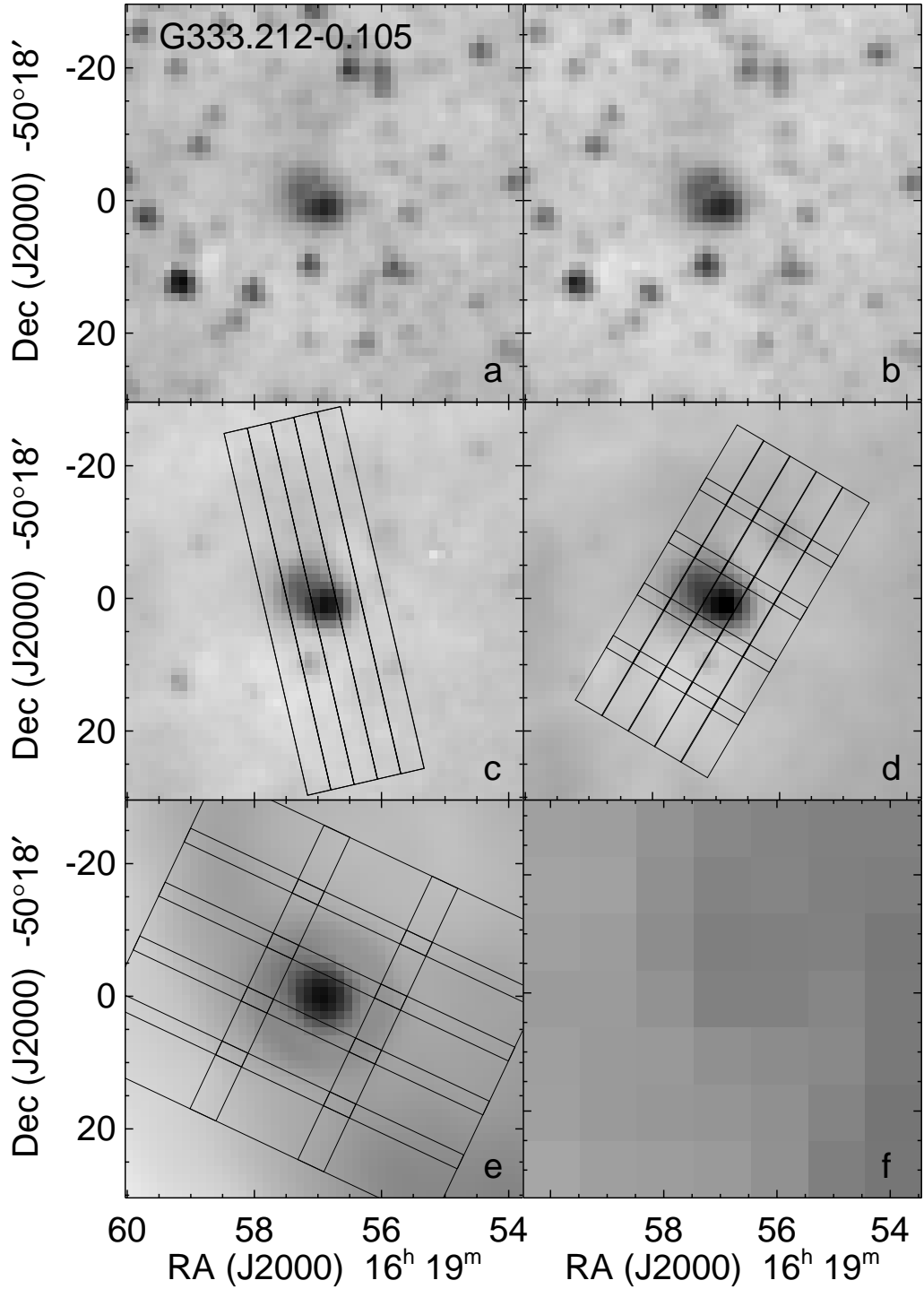}
\caption{IRAC and MIPS images of G333.212$-$0.105.
The images and overlays have the same description as Fig. 32 
except that there is no detected methanol maser and 
the MIPS 24 \micron\ image is not saturated.
G333.212$-$0.105 is not visible in the MIPS 70~\micron\ image (panel f),
probably because of high background and low {\it Spitzer} resolution, 
but it is visible in all {\it Herschel} PACS and SPIRE 70 -- 600~\micron\ images.
}
\end{figure*}

%Figure S1
\begin{figure*}
\includegraphics[width=150mm]{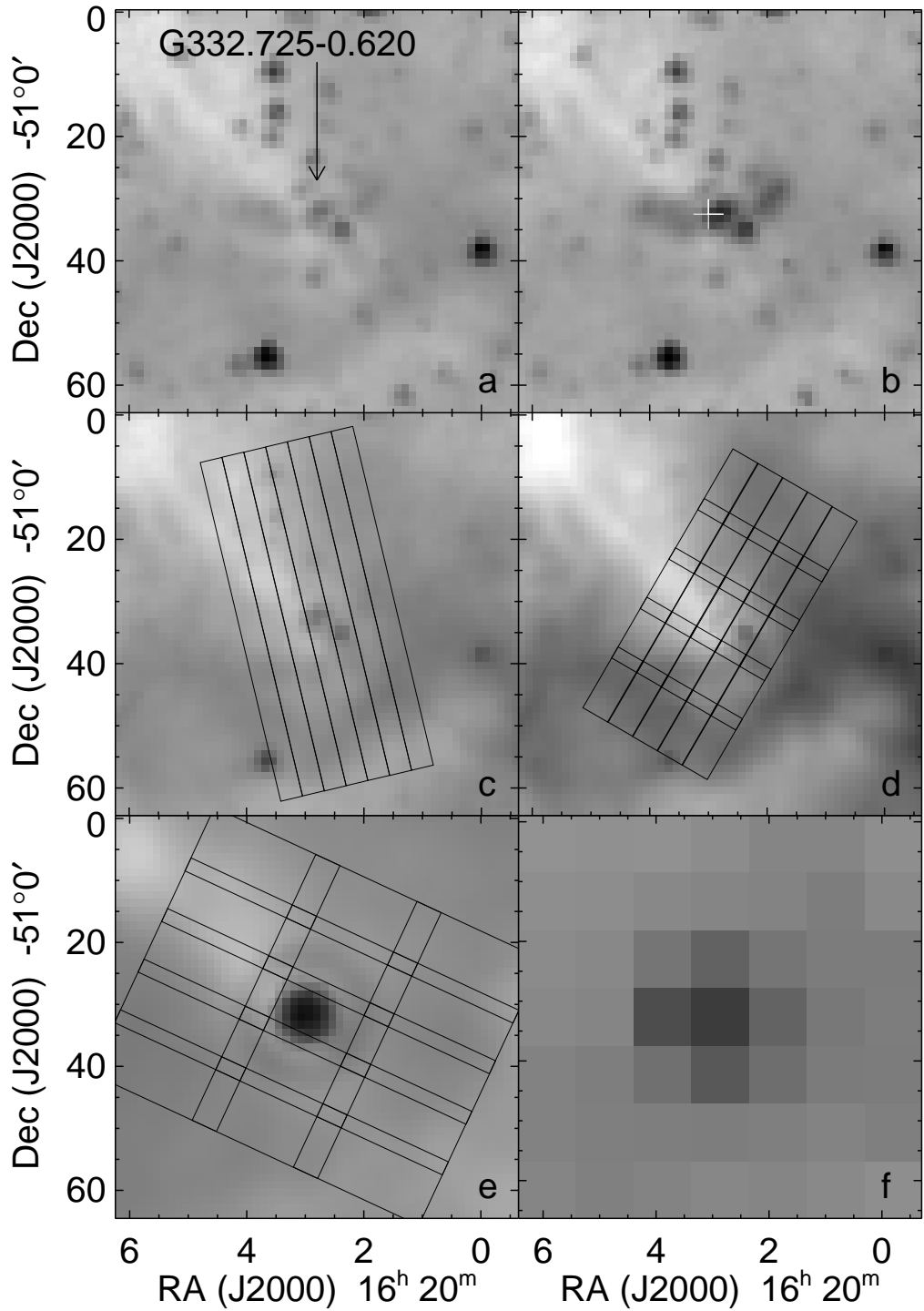}
\caption{IRAC and MIPS images of G332.725$-$0.620.
The images and overlays have the same description as Fig. 32 
except that the MIPS 24 \micron\ image is not saturated.
Panel (a): Note that the outflow region is visible on this image as well as on the IRAC band 2 (4.5~\micron) image.
}
\end{figure*}

%Figure S3
%\clearpage
\begin{figure*}
\includegraphics[width=150mm]{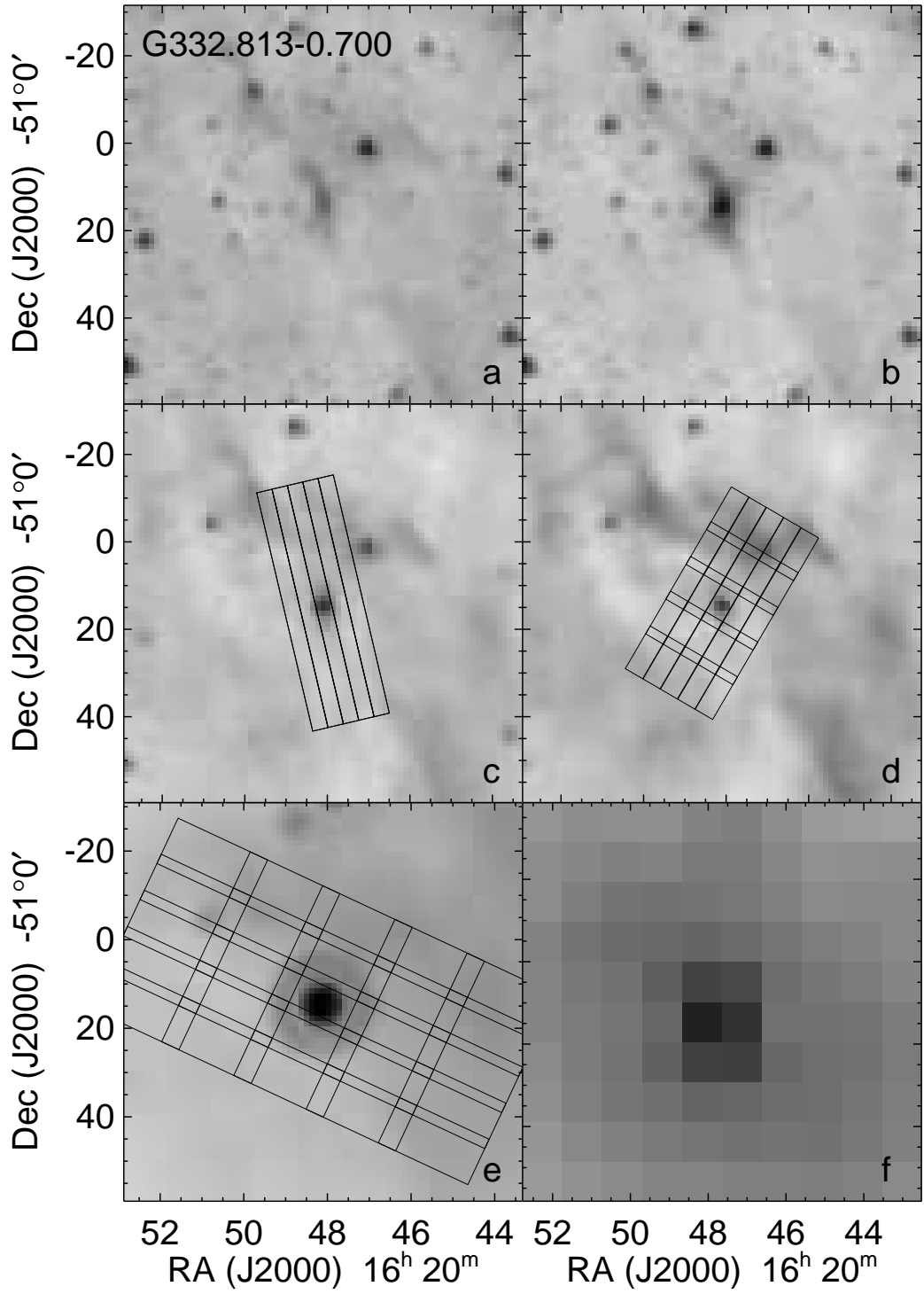}
\caption{IRAC and MIPS images of G332.813$-$0.700.
The images and overlays have the same description as Fig. 35.
}
\end{figure*}

%Figure S4
%\clearpage
\begin{figure*}
\includegraphics[width=150mm]{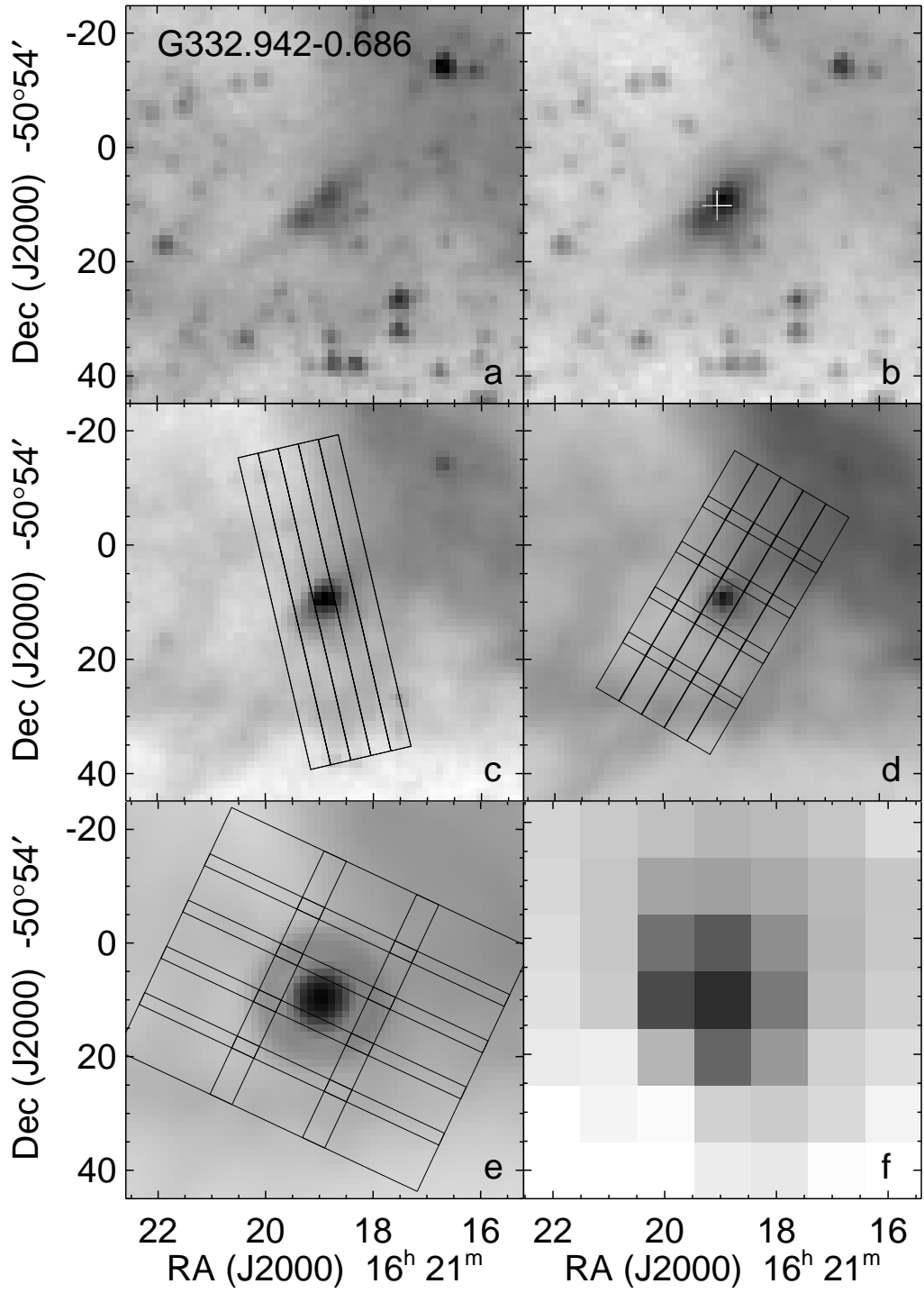}
\caption{IRAC and MIPS images of G332.942$-$0.686.
The images and overlays have the same description as Fig. 35.
}
\end{figure*}

%Figure S11
%\clearpage
\begin{figure*}
\includegraphics[width=150mm]{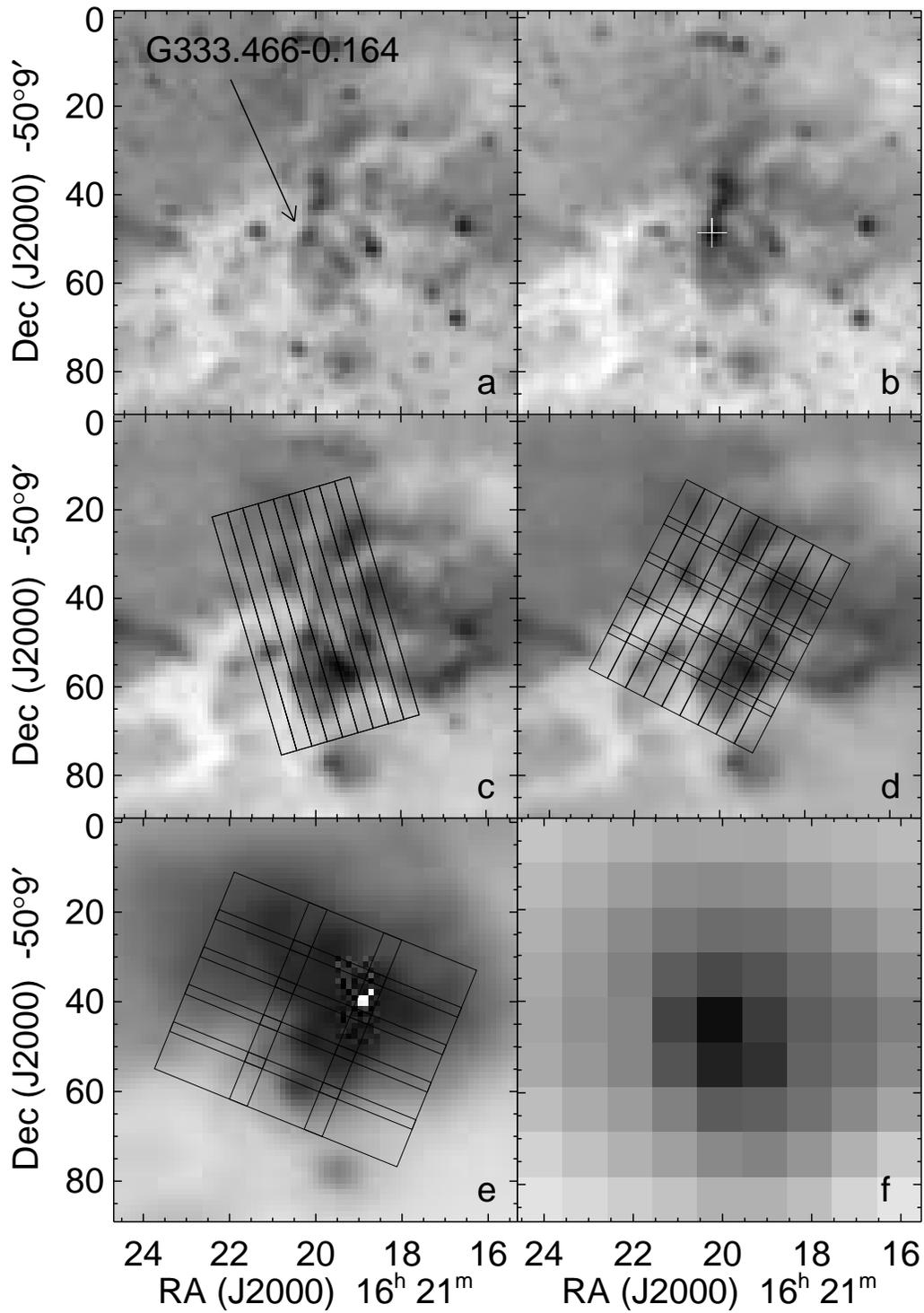}
\caption{IRAC and MIPS images of G333.466$-$0.164 (IRAS 16175-5002).
The images and overlays have the same description as Fig. 32.
The MIPS 24~\micron\ image (panel e) is saturated 
at the location of the maximum of the ionized forbidden line emission,
$\sim 15$~arcsec north-west of G333.466$-$0.164, the YSO at the head 
of the 4.5~\micron\ outflow (panel b).
}
\end{figure*}

%Figure S6
%\clearpage
\begin{figure*}
\includegraphics[width=150mm]{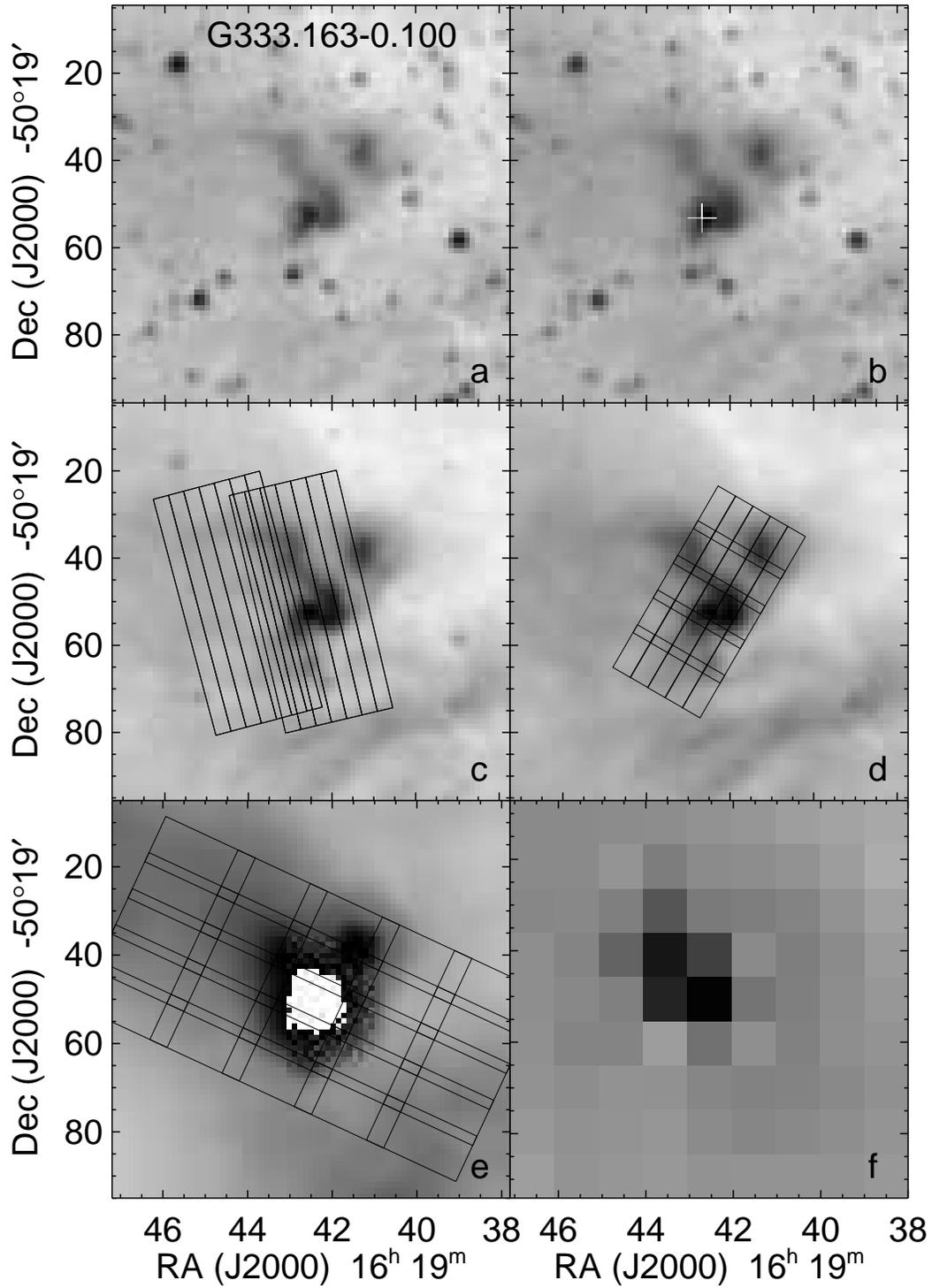}
\caption{IRAC and MIPS images of G333.163$-$0.100.
The images and overlays have the same description as Fig. 32.
Panel (c) also shows the location of the SL2 spectra from AOR 25913344.
The centre of the MIPS 24~\micron\ image (panel e) is saturated 
at the location of G333.163$-$0.100.
}
\end{figure*}

%Figure S9
%\clearpage
\begin{figure*}
\includegraphics[width=150mm]{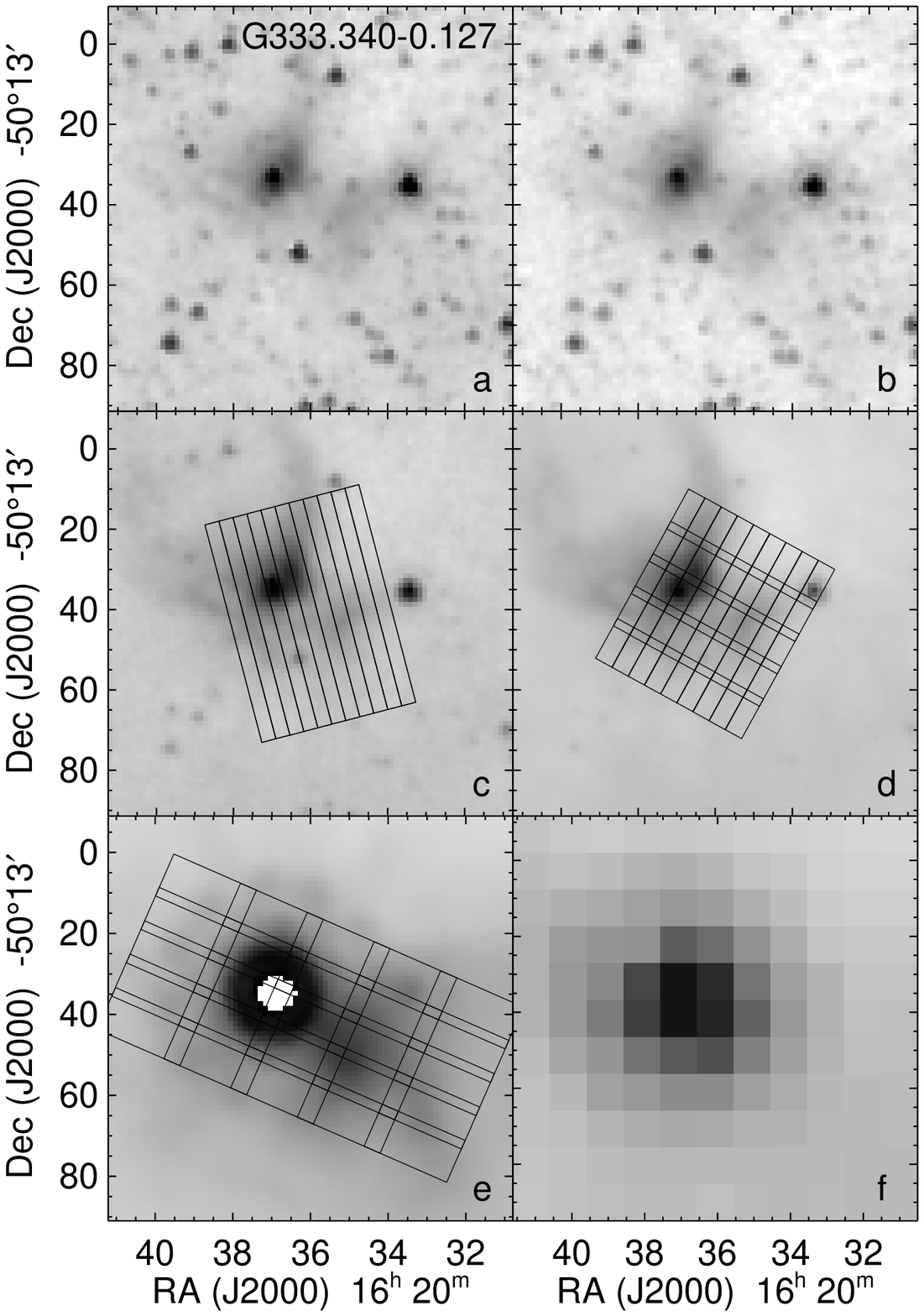}
\caption{IRAC and MIPS images of G333.340$-$0.127.
The images and overlays have the same description as Fig. 32 
except that there is no detected methanol maser.
The centre of the MIPS 24~\micron\ image (panel e) is saturated 
at the location of G333.340$-$0.127.
}
\end{figure*}

%Figure S10
%\clearpage
\begin{figure*}
\includegraphics[width=150mm]{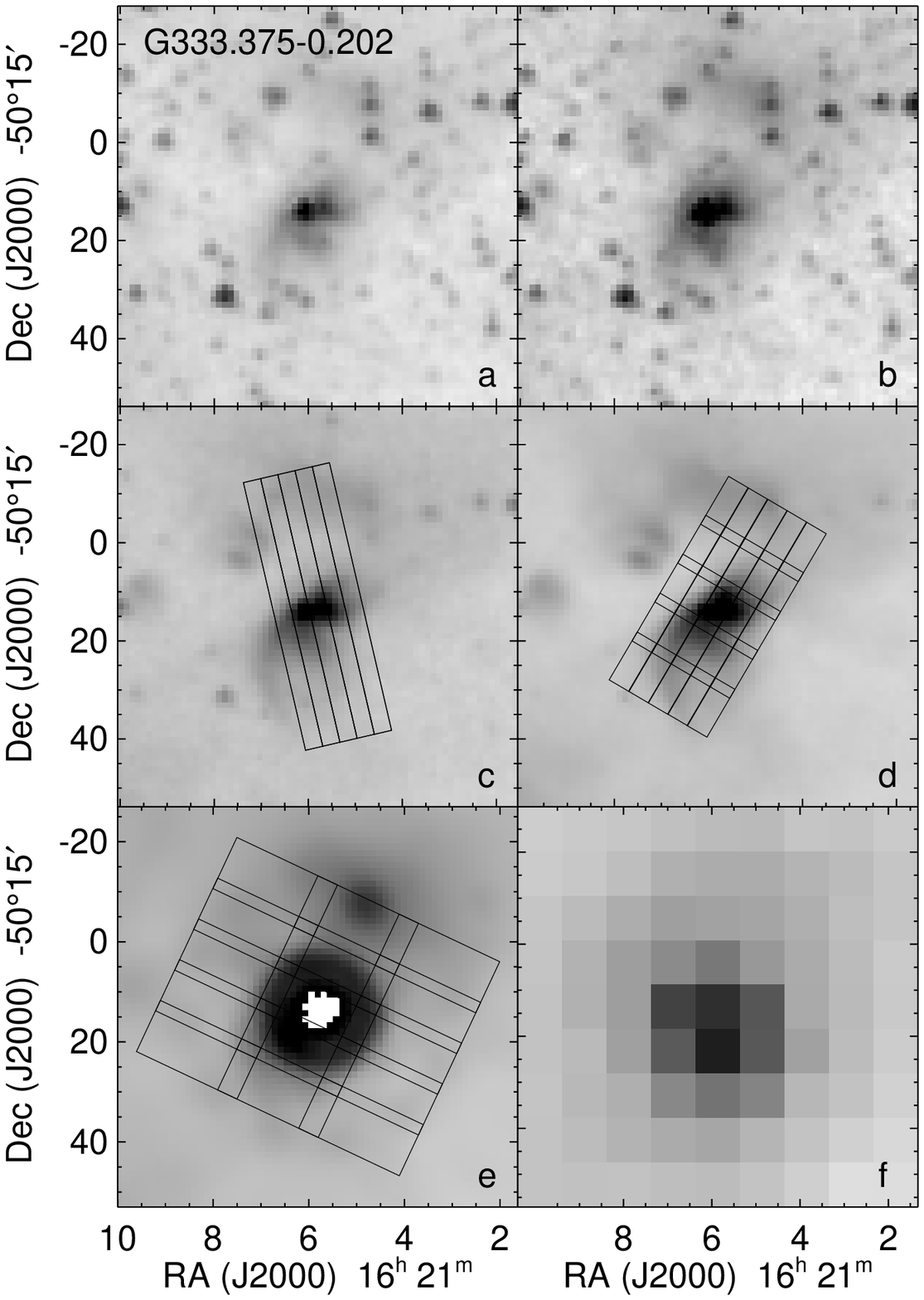}
\caption{IRAC and MIPS images of G333.375$-$0.202.
The images and overlays have the same description as Fig. 32 
except that there is no detected methanol maser.
The centre of the MIPS 24~\micron\ image (panel e) is saturated 
at the location of G333.375$-$0.202.
}
\end{figure*}

\end{document}